\documentclass[aps,twocolumn,prd,showpacs,showkeys,preprintnumbers,superscriptaddress,nobibnotes,floatfix,longbibliography]{revtex4-1}

\usepackage{graphicx}
\usepackage{bm}
\usepackage{times}
\usepackage{hyperref}
\usepackage{slashed}
\usepackage{color}
\usepackage{aas_macros}
\usepackage{nicefrac}
\usepackage{soul}

\usepackage{slashed}
\usepackage{lipsum}
\usepackage{subfigure}
\usepackage{multirow}
\usepackage{amsmath}
\usepackage{array}    
\usepackage{varwidth} 
\usepackage{comment}

\hypersetup{
    pdfnewwindow=true,    
    colorlinks=true,      
    linkcolor=blue,       
    citecolor=blue,       
    filecolor=blue,      
    urlcolor=blue         
}
\newcommand{\be}{\begin{equation}}
\newcommand{\ee}{\end{equation}}
\newcommand{\pfrac}[2]{\left(\frac{#1}{\rm #2}\right)}

\begin{document}

\title{Millisecond Pulsars Modify the Radio-SFR Correlation in Quiescent Galaxies}

\author{Takahiro Sudoh}
\affiliation{Department of Astronomy, University of Tokyo, Hongo, Tokyo 113-0033, Japan}

\author{Tim Linden}
\affiliation{Stockholm University and the Oskar Klein Centre, Stockholm, 10691, Sweden}
\affiliation{Center for Cosmology and AstroParticle Physics (CCAPP), Ohio State University, Columbus, OH 43210, USA}
\affiliation{Department of Physics, Ohio State University, Columbus, OH 43210, USA}

\author{John F.~Beacom}
\affiliation{Center for Cosmology and AstroParticle Physics (CCAPP), Ohio State University, Columbus, OH 43210, USA}
\affiliation{Department of Physics, Ohio State University, Columbus, OH 43210, USA}
\affiliation{Department of Astronomy, Ohio State University, Columbus, OH 43210, USA
\\
\footnotesize{\tt \href{mailto:sudoh@astron.s.u-tokyo.ac.jp}{sudoh@astron.s.u-tokyo.ac.jp},  \href{mailto:linden@fysik.su.se}{linden@fysik.su.se}, \href{mailto:beacom.7@osu.edu}{beacom.7@osu.edu}} \\
\footnotesize{\tt  \href{http://orcid.org/0000-0002-6884-1733}{0000-0002-6884-1733}, \href{http://orcid.org/0000-0001-9888-0971}{0000-0001-9888-0971}, \href{http://orcid.org/0000-0002-0005-2631}{0000-0002-0005-2631} \smallskip}}

\begin{abstract}
\noindent The observed correlation between the far-infrared and radio luminosities of galaxies illustrates the close connection between star formation and cosmic-ray production. Intriguingly, recent gamma-ray observations indicate that recycled/millisecond pulsars (MSPs), which do not trace recent star formation, may also efficiently accelerate cosmic-ray electrons. We study the contribution of MSPs to the galactic non-thermal radio emission, finding that they can dominate the emission from massive quiescent galaxies. This model can explain recent LOFAR observations that found a peculiar radio excess in galaxies with high stellar masses and low star-formation rates. We show that MSP-based models provide a significantly improved fit to LOFAR data. We discuss the implications for the radio-FIR correlation, the observation of radio excesses in nearby galaxies, and local electron and positron observations.
\end{abstract}

\maketitle

\section{Introduction}
\vspace{-0.1cm}

The radio--far-infrared (FIR) correlation is a cornerstone in our understanding of star-formation and cosmic-ray physics. Throughout their brief lives, massive stars produce bright radiation that is absorbed by interstellar dust and re-emitted in the FIR. In their violent deaths, these stars produce shocks that accelerate charged particles to GeV and higher energies. These cosmic rays lose energy via hadronic, inverse-Compton, and synchrotron interactions, producing a bright non-thermal radio flux, among other emissions. The close correlation between non-thermal radio and FIR emission has been found over a wide range of galactic masses and star-formation rates \cite{irc1973a,irc1973b, 1975ApJ...200L.127H, irc1984, 1984AJ.....89.1520R, irc1985, 1985A&A...147L...6D,1988A&A...199...91H,condon92, irc2001, irc2004, irc2010, irc15M, irc17Q, irc17T, irc18S, Read18, irc19F, irc19S}. A similar correlation has been found between the gamma-ray and FIR fluxes, providing additional support for the cosmic-ray origin of the radio emission~\cite{Ackermann:2012vca, Linden:2016, Ajello:2020zna}.

The increasingly high precision of radio and infrared measurements has isolated several confounding variables, including environmental effects~\cite{Murphy:2008dq} and active galactic nuclei (AGN) contributions~\cite{2002AJ....124..675C, 2010ApJ...724..779M}, and produced resolved analyses of the radio-FIR correlation within galaxies~\cite{1988A&A...191L...9B, 2006ApJ...638..157M, Paladino06, Murphy:2008ma, Heesen14, Heesen19}. Intriguingly, observations have detected dispersion in the radio-FIR correlation in the least luminous star-forming galaxies (SFGs). Early studies of low-luminosity galaxies found that both the FIR (due to ineffective dust absorption) and radio (due to ineffective cosmic-ray trapping) fluxes fall below predictions based on calorimetric models (which require that both ultraviolet photons and cosmic rays lose all their energy in the galaxy), implying the breakdown of calorimetry. Thus, a ``conspiracy'' of factors should exist to maintain the radio-FIR correlation over such a large dynamic range~\cite{Bell:2002kg, Lacki:2009mj}. 

Because the FIR flux may not always trace the SFR accurately, many studies have included optical and UV measurements to better probe the physical correlation between star formation and non-thermal emission~(e.g., Refs.~\cite{Hodge08, Brown17, Davis17, irc18H,  CalistroRivera17, Gurkan18, Wang19}). The radio-SFR correlation is expressed as $L_{\rm r}$~$\propto$~${\rm SFR}^\alpha$, where $L_{\rm r}$ is the radio luminosity, and $\alpha$ is the power-law index. Recent observations find $\alpha$ exceeding unity~(e.g., \cite{Hodge08, Brown17, Davis17, CalistroRivera17, Wang19}), which can be attributed to increasing cosmic-ray confinement and synchrotron radiation efficiency in rapidly star-forming systems~(e.g., \cite{Niklas97,Schleicher16}).

In these studies, the radio flux has been attributed to supernova remnants (SNRs) and normal pulsars, both of which trace recent star formation. However, recent gamma-ray observations suggest that recycled, millisecond pulsars (MSPs) can efficiently convert their power to cosmic-ray electrons and positrons~\cite{Hooper18}, possibly supplying additional power to the galactic non-thermal emission. Unlike SNRs and normal pulsars, MSPs first evolve through long-lived low-mass X-ray binary (LMXB) phases~\cite{2013ApJ...764...41F} and then slowly spin down over $\sim$~Gyr timescales~\cite{1984A&A...141...91C, Tauris06}. Thus, the MSP luminosity depends on the integrated SFR over the last $\sim$1--10~Gyr, and can be important for massive quiescent galaxies.

Interestingly, this result coincides with a recent observation by \citet{Gurkan18} (hereafter, G18), which found excess radio emission in galaxies with low star-formation rates compared to expectations from the radio-SFR correlation. Splitting their galaxy catalog into two components, they found that the trend is most pronounced in galaxies with total stellar masses that exceed 10$^{9.5}$~M$_\odot$, indicating that galaxy mass may play an important role in determining the total radio luminosity. While several effects, including contributions from AGN, pulsars, or Type-Ia SN were briefly mentioned, there is, at present, no clear explanation for this observation.

Here, we show that radio emission from MSPs may significantly contribute to (and even dominate) the radio luminosity of high-mass/low-SFR galaxies. We also produce the first quantitative fit to LOFAR data using models that include MSPs, finding that our model formally improves the fit. The paper is outlined as follows. In Sec.~\ref{sec:theory}, we present theoretical estimates for the radio flux from SNRs, normal pulsars, and MSPs. In Sec.~\ref{sec:data}, we explain our methodology for fitting the LOFAR data. In Sec.~\ref{sec:results}, we show the results of our analysis, and, in Sec.~\ref{sec:discuss}, we discuss the implications of our results.

\section{Theoretical Models for Radio Emission from SNRs and MSPs}
\label{sec:theory}

In ordinary galaxies, the dominant source of the diffuse non-thermal radio flux is due to the synchrotron emission of relativistic electrons in weak ($\sim\mu$G) galactic magnetic fields. Here we consider production within discrete sources, which could be important in quiescent galaxies. In Secs.~\ref{subsec:SNR1}--\ref{subsec:MSP}, we estimate the radio emission from each source class, showing that electrons from MSPs can be important in quiescent galaxies. In Sec.~\ref{subsec:radio}, we discuss the conversion of this electron power into synchrotron emission and summarize our radio emission model. In Sec.~\ref{subsec:qualitative}, we qualitatively describe the expected modification of the radio-SFR correlation by MSPs.

\subsection{Supernova Remnants (Primary)}
\label{subsec:SNR1}

Core-collapse supernovae inject $\sim$10$^{51}$~erg of kinetic energy into the interstellar medium (ISM), a subdominant fraction of which (roughly $\eta_{e}^{\rm SN}\sim10^{-3}$) is used to accelerate ambient electrons to relativistic energies~\cite{sn1993j, Park15, Sarbadhicary17}. To calculate the SNR flux in an SFG, we assume an SFR-dependent core-collapse supernova rate of $0.015\psi$~yr$^{-1}$~\cite{Lopez18}, where $\psi$ is the galactic SFR in $M_\odot$~yr$^{-1}$. This produces a steady-state electron injection power of:
\begin{equation}
\label{eq:Q_snr1}
    Q_e^{\rm SN,\ prim.} = 5\times10^{38}\psi\pfrac{\eta_{e}^{\rm SN}}{10^{-3}}~\rm erg~s^{-1}.
\end{equation}

\subsection{Supernova Remnants (Secondary)}
\label{subsec:SNR2}

SNRs also produce a significant population of non-thermal protons, which carry a much larger fraction ($\eta_{p}^{\rm SN}\sim0.1$) of the supernova kinetic energy. These protons can subsequently interact with the interstellar medium to produce pions, which promptly decay to produce secondary particles, including electrons and positrons (hereafter, electrons). The fraction of proton power transferred to pions is denoted $f_{pp}$, and depends on the mass, density, and diffusion properties of the specific galaxy. In the Milky Way, measurements of gamma-ray emission indicate $f_{pp}$ is approximately 0.03~\cite{Strong10}.

In each collision, approximately 1/6 of the initial proton energy is converted into relativistic electrons, while the rest is converted primarily into secondary protons, neutrinos and gamma rays. Therefore, the total electron power produced via these ``secondary" electrons is:
\begin{equation}
\label{eq:Q_snr2}
    Q_e^{\rm SN,\ sec.} = 8\times10^{37}\psi\pfrac{f_{pp}}{10^{-2}}\pfrac{\eta_{p}^{\rm SN}}{0.1}~\rm erg~s^{-1}.
\end{equation}
Thus, the conversion of SNR power to electron power has a total efficiency $\frac{1}{6}\eta_{p}^{\rm SN}{f}_{pp}$. If this exceeds $\eta_{e}^{SN}$, then synchrotron emission from secondary electrons dominates the galactic synchrotron emission. Because $\eta_{p}$ is unlikely to significantly vary between galaxies, the efficiency $f_{pp}$ determines the dominance of primary or secondary electrons. The efficiency $f_{pp}$ is higher for galaxies that can confine cosmic rays longer, and which have higher collision rates between cosmic rays and dense interstellar gas. It is generally expected that $f_{pp}$ eventually approaches unity (the calorimetric limit) in the strong magnetic fields and high densities of the most intensely star-forming galaxies~\cite{Thompson07,Lacki:2009mj}. 

This transition is consistent with gamma-ray observations of intensely star-forming galaxies, which indicate that the gamma-ray--FIR correlation exceeds unity, with $L_\gamma~\propto~L_{\rm IR}^{1.18}$~\cite{Linden:2016}. This suggests that $f_{pp}$ scales as $\sim\psi^{0.18}$. The value of $f_{pp}$ is also estimated for nearby galaxies: it is $\sim$1$\%$ for the Small Magellanic Cloud~\cite{Lopez18}, on the order of 10$\%$ for nearby starbursts M82 and NGC253~\cite{Lacki11}, and may reach unity for ultraluminous infrared galaxies like Arp220~\cite{Griffin16}. This indicates that secondary electrons are generally subdominant for quiescent galaxies, but can dominate in starburst sources~\cite{Lacki:2009mj,Lacki13}. 

In the following, we assume a scaling between $f_{pp}$ and $\psi$: 
\be
\label{eq:f_pp}
f_{pp} = \alpha_{pp} \psi^{\beta_{pp}}.
\ee

\subsection{Normal Pulsars}
\label{subsec:PSR}

Neutron stars are born as the remnants of core-collapse supernovae, with a rotational energy on the order of $10^{48}(P_i/150~{\rm ms})^{-2}$~erg, where $P_i$ is the initial rotational period of the pulsar. Over their lifetimes, these pulsars spin down, and their rotational energy is released as a relativistic wind of magnetized e$^+$e$^-$ plasma (the pulsar wind). This interacts with the ambient medium to create a shock where e$^+$e$^-$ are accelerated to very high energies to produce a pulsar wind nebula (PWN). Recent studies of non-thermal gamma rays around evolved pulsars (``TeV halos") have shown that pulsars convert a large fraction ($\eta_{e}^{\rm PSR}\sim 10-30\%$) of their spindown power into e$^+$e$^-$ pairs ~\cite{Hooper17halo,Linden17halo}. Assuming that the pulsar production rate is equivalent to the supernova rate, we obtain a steady-state electron power:
\begin{equation}
\label{eq:Q_psr}
     Q_e^{\rm PSR} = 5\times10^{37}\psi\pfrac{P_i}{150~ms}^{-2}\pfrac{\eta_{e}^{\rm PSR}}{0.1}~\rm erg~s^{-1}.
\end{equation}

From a comparison of Eq.~(\ref{eq:Q_psr}) and Eq.~(\ref{eq:Q_snr1}), the pulsar contribution is subdominant to the primary electron flux from supernovae. However, there are multiple uncertainties (most importantly in $\eta_{e}^{\rm PSR}$ and $P_i$) that may affect this conclusion. In particular, the average value of $(P_i)^{-2}$ is relatively unconstrained by pulsar statistics, which induce significant uncertainties in this estimate~(e.g.,\cite{FaucherGiguere:2005ny, deJager:2008ke}). 

It is important to note that the comparison between SNR and pulsar energetics is also energy-dependent. PWNe typically have a flat radio spectrum ($d\ln F_\nu/d\ln \nu \simeq-0.2$)~\cite{Gaensler06,Reynolds17}. This indicates that radio-emitting electrons have a hard spectrum ($d\ln N_e/d\ln E_e>-2$), i.e., that most of the energy is contained in higher-energy electrons that typically radiate X-rays. In contrast, SNRs are energetically dominated by low-energy electrons  ($d\ln N_e/d\ln E_e<-2$) that typically produce radio emission. Since our study focuses on LOFAR observations at 150~MHz, SNR contributions are likely more dominant in our study, compared to studies conducted at GHz frequencies. However, because we study only the integrated radio flux at a single frequency, our model cannot, in principle, differentiate these components.

Radio pulsars also directly produce pulsed and beamed radio emission. However, the fraction of the power carried by this emission is negligible, $\sim 10^{-4}$~\cite{Szary14}.

\subsection{Recycled/Millisecond Pulsars (MSPs)}
\label{subsec:MSP}

The time dependence of MSP cosmic-ray injection is different from every other source of galactic cosmic-rays. While emission from core-collapse SNe and normal pulsars (Eqs.~\ref{eq:Q_snr1},~\ref{eq:Q_snr2},~and~\ref{eq:Q_psr}) depends on the current star-formation rate ($\psi$), MSPs first evolve through long stellar-binary and LMXB phases, and inject cosmic-rays only after a significant time lag. Moreover, MSPs continue to accelerate non-thermal electrons over a long spin-down timescale, with a spin-down power that is relatively constant over  $\sim10(P_i/5~{\rm ms})^2(B_s/10^{8.5}~{\rm G})^{-2}~{\rm Gyr}$, where $B_s$ is the magnetic field strength~\cite{Lorimer13,Gonthier18}. Thus, the cosmic-ray injection from MSPs traces the average star-formation rate ($\bar{\psi}$) over the last $\sim$10~Gyrs. Indeed, MSPs are important sources of gamma-ray emission from globular clusters~\cite{Abdo09gc,Hooper16gc} and the Galactic bulge~\cite{Gonthier18,Macias19} , which indicates that they can power old stellar systems.

While $\bar{\psi}$ is not typically known for most galaxies, the total stellar mass ($M_*$) serves as an excellent tracer of star-formation over long timescales. Indeed, stellar mass is commonly employed as a tracer for the total population of low-mass X-ray binaries (LMXBs), which are the primary progenitors of MSPs~\cite{Gilfanov04,Lehmer10,Boroson11,Fragos13}. We assume that the total power from MSPs ($Q^{\rm MSP}$) also correlates with the mass.

Because there are significant uncertainties in the transition from the LMXB to MSP phase (and thus their relative rates), we normalize the MSP population using gamma-ray observations of Milky Way MSPs. Recent work by Ref.~\cite{Eckner2018} attempted to address the effect of incompleteness in the observation of dim MSPs, and estimated the total luminosity of galactic MSPs to fall between $(0.5-3)\times10^{38}~\rm erg~s^{-1}$, which is consistent with previous studies~\cite{Winter16,Ploeg17,Bartels18} (see, however,~Ref.~\cite{Bartels18b}, which finds a smaller value). Here, we normalize the total luminosity as $L_{38}^{\rm MW}=L_{\rm MSP}^{\rm MW}/10^{38}~\rm erg~s^{-1}$. The stellar mass in the Milky Way disk is $5\times10^{10}~M_\odot$~\cite{Licquia15}, which suggests the following relation:
\be
Q^{\rm MSP}_{\rm total} = 2\times10^{38}L_{38}^{\rm MW}\left( \frac{M_*}{10^{10}M_\odot} \right) \pfrac{\eta_\gamma}{0.1}^{-1}~\rm erg~s^{-1},
\ee
where $\eta_\gamma$ is the conversion efficiency from spindown power to gamma-ray luminosity, estimated to be $\sim10\%$~\cite{2013ApJS..208...17A}. 

These estimates do not include a contribution from galactic globular clusters, which might enhance the total gamma-ray luminosity from the galaxy. We also note that the Milky Way value may not be typical. Studies of the LMXB population by Ref.~\cite{Gilfanov:2003th} found that, while LMXBs are expected to trace stellar mass, the LMXB population of the Milky Way is roughly 2.5 times smaller than a chosen population of nearby Milky Way analogs. In particular, morphological analyses of the M31 galactic bulge indicate that the MSP population of M31 may be up to a factor of 4 larger than expectations based on Milky Way models~\cite{Ackermann17m31, Eckner2018}.

The power and spectrum of electrons produced by MSPs are highly uncertain, both theoretically and observationally. As in the case of normal pulsars, a substantial relativistic electron population is accelerated within the strong electric and magnetic fields of the pulsar magnetosphere. Notably, despite magnetic field strengths that are several orders of magnitude smaller than normal pulsars, the gamma-ray spectrum of MSPs and normal pulsars is almost identical, indicating that they may also accelerate similar electron populations. However, unlike normal pulsars, MSPs rarely produce bright PWNe~\cite{Stappers03,Hui06,Lee18}, and thus the relativistic electrons may not be subsequently accelerated by a termination shock. This also indicates that electron energy losses due to the adiabatic expansion of the nebula and synchrotron cooling inside it are much less important for MSPs, allowing a larger fraction of the injected power to be released into the ISM. Thus, it is likely that the ISM electron spectrum produced by MSPs differs substantially from that produced by normal pulsars. 

The conversion efficiency $\eta_{e}^{\rm MSP}$ is uncertain, and a wide range of values from a few percent to 90$\%$ have been tested in the literature. To date, the most stringent constraints on $\eta_{e}^{\rm MSP}$ come from observations at TeV scales. A recent study of the globular cluster M15 by the MAGIC collaboration suggests an efficiency less than $30\%$~\cite{MAGIC19gc} for a power-law injection, though it should be noted that frequent stellar interactions in the cluster may significantly suppress the particle production by MSPs~\cite{Cheng10}. Observational studies of TeV emission around Galactic MSPs suggests that the value of $\eta_{e}^{\rm MSP}$ might be $\sim$10$\%$~\cite{Hooper18}. Importantly, neither of these observations can strongly constrain the efficiency at the GeV scales that are most important for 150~MHz radio observations.

The lack of PWNe around MSPs makes it difficult to constrain their non-thermal electron spectra. Previous studies of non-thermal electron production in MSPs have used a diverse set of models with a wide range of parameters~\cite{Bednarek07, Cheng10, Harding11, Kisaka12, Venter:2015gga, Yuan15, Petrovic15, Bednarek16, Song19, Ndiyavala19, Bykov19}. For our analysis, which uses radio emission at only one frequency (150~MHz), changes in the electron spectrum and the electron acceleration efficiency are degenerate. Thus, we absorb the uncertainty in the MSP spectral shape into the parameter $\eta_{e}^{\rm MSP}$, writing the total electron power from MSPs as
\be
\label{eq:Q_msp}
Q^{\rm MSP}_{e} =
2\times10^{37}L_{38}^{\rm MW}\left( \frac{M_*}{10^{10}M_\odot} \right) \pfrac{\eta_{e}^{\rm MSP}}{\eta_\gamma}~\rm erg~s^{-1}.
\ee

\noindent While the contribution of MSPs is sub-dominant in typical galaxies, it becomes important whenever
\be
\label{eq:snr_to_msp}
L_{38}^{\rm MW}\pfrac{\eta_{e}^{\rm MSP}}{\eta_\gamma}\left( \frac{M_*}{10^{10}M_\odot} \right) \pfrac{\psi}{1~M_\odot~yr^{-1}}^{-1} \gtrsim 30.
\ee
For galaxies with low specific SFR (sSFR; SFR/Mass), the contribution of MSPs can be dominant. Intriguingly, this is the region ($M_*>10^{9.5}~{\rm M}_\odot$ and $\psi<10^{-2}~{\rm M}_\odot~{\rm yr}^{-1}$) where LOFAR has identified a radio excess.

\subsection{Modeling the Synchrotron Luminosity in SFGs}
\label{subsec:radio}

In previous subsections, we developed quantitative models for the total electron power from each source class, but thus far we have only qualitatively discussed the production of synchrotron radiation from these populations. There are three effects at play. The first is the energy dependence of the electron spectrum, which affects the fraction of the synchrotron power that is emitted at 150~MHz. The critical frequency for synchrotron radiation is given by
\be
\nu_c = 80~\left( \frac{E_e}{\rm GeV} \right)^{2}\left( \frac{B}{6~\rm \mu G} \right)~\rm MHz,
\ee
which indicates that GeV-scale electrons are most efficient at producing the 150~MHz radio emission studied here. The fraction of the total electron power that is stored in 150~MHz emitting electrons, $\chi_{150}$, strongly depends on the spectrum injected by sources.

The second effect pertains to competitive electron energy-loss mechanisms, including inverse-Compton scattering, bremsstrahlung, and ionization. The relative contribution of each component can be evaluated from their cooling timescales (e.g., Ref~\cite{Atoyan95}):
\begin{equation}
  \begin{split}
  \label{eq:cooling}
  &t_{\rm syn} = 2.6 \times 10^{8}~{\rm yr}~\nu_{150}^{-\nicefrac{1}{2}} \left( \frac{B}{\rm 6~\mu G} \right)^{-\nicefrac{3}{2}},\\ 
  &t_{\rm IC} = 2.3 \times 10^8~{\rm yr}~\nu_{150}^{-\nicefrac{1}{2}} \left( \frac{B}{\rm 6~\mu G} \right)^{\nicefrac{1}{2}} \left(\frac{w_{\rm ISRF}}{1~\rm {eV}~\rm{cm}^{-3}}\right)^{-1},\\ 
  &t_{\rm brems} = 1.1 \times 10^{8} ~{\rm yr} \left(\frac{n_{\rm gas}}{\rm 0.3~cm^{-3}}\right)^{-1},\\ 
  &t_{\rm ion} = 4.8 \times 10^{8} ~{\rm yr}~\nu_{150}^{\nicefrac{1}{2}} \left( \frac{B}{\rm 6~\mu G} \right)^{-\nicefrac{1}{2}} \left(\frac{n_{\rm gas}}{\rm 0.3~cm^{-3}}\right)^{-1},
  \end{split}
\end{equation}
where $\nu_{150}$ is the observation frequency in the unit of 150~MHz and the assigned galactic properties correspond to their average value over the cosmic-ray confinement volume. Also, we assume that inverse-Compton scattering proceeds in the Thomson regime, which is valid for GeV-scale electrons. The total cooling time, $t_{\rm cool}$, is estimated as
\be
\frac{1}{t_{\rm cool}} = \frac{1}{t_{\rm syn}} + \frac{1}{t_{\rm IC}} + \frac{1}{t_{\rm brems}} + \frac{1}{t_{\rm ion}}.
\ee

The relative contribution of each cooling process depends on the electron energy, as well as $n_{\rm gas}$, $B$, and $w_{\rm ISRF}$. If we adopt typical Milky Way parameters, e.g., $n_{\rm gas}$~$\simeq$~0.3~cm$^{-3}$,  $B$~$\simeq$~6~$\mu$G, and $w_{\rm ISRF}$~$\simeq$~1~eV~cm$^{-3}$, then the electrons that produce 150~MHz radio emission cool primarily via bremsstrahlung. In many galaxies, the magnetic field in synchrotron-emitting regions is found to be $B$~$\simeq$~10~$\mu$G under the assumption of cosmic rays and magnetic field equipartition~\cite{Beck19} (see, however, an arguments against equipartition models in starburst galaxies~\cite{Thompson06}), suggesting that synchrotron losses are important. Our focus on quiescent galaxies may motivate adopting target densities more consistent with massive elliptical galaxies that have lower gas densities, $\sim10^{-2}$~cm$^{-3}$~\cite{Mathews03}, so then bremsstrahlung and ionization losses may become unimportant. However, the magnetic fields of these galaxies are not tightly constrained.

The third effect pertains to cosmic-ray escape, which competes with each energy-loss process. In the Milky Way,  measurements of radioactive cosmic-ray nuclei indicate that GeV-scale cosmic rays are confined over a timescale of $t_{\rm esc}~\sim~10^{8}$~yr~\cite{Evoli20,Morlino20}, which indicates that GeV leptons lose most of their energy, although there are alternative models that suggest much shorter escape times~\cite{Cowsik16,Lipari17}. We note that in small galaxies that do not efficiently confine cosmic-rays within their bulk, self-confinement near sources may be important~\cite{Fujita10, Fujita11, Malkov13, Nava16, DAngelo:2017rou, Evoli:2018aza, Fang19}.

The competition between cooling and escape sets the lifetime of cosmic-rays in galaxies to be: 
\be
\frac{1}{t_{\rm life}} = \frac{1}{t_{\rm esc}} + \frac{1}{t_{\rm cool}},
\ee
which is related to the conversion efficiency of the injected electron power to the synchrotron radiation as 
\be
\label{eq:f_syn_def}
f_{\rm syn}= \frac{t_{\rm life}}{t_{\rm syn}},
\ee
which depends on the cosmic-ray confinement time, magnetic field strength, and radiation/gas densities. Most naively, $f_{\rm syn}$ is expected to be higher for more massive galaxies that confine cosmic rays for longer times. In the following, we assume a scaling between $f_{\rm syn}$ and $M_*$: 
\be
\label{eq:f_syn}
f_{\rm syn} = \alpha_{\rm syn} \left(\frac{M_*}{10^{10}~M_\odot}\right)^{\beta_{\rm syn}}.
\ee

In steady state, the radio luminosity of an SFG is the product of the injection rate of non-thermal electrons ($Q_e$) and $f_{\rm syn}$. The total 150~MHz luminosity can be expressed as the sum of contributions from different source classes:
\begin{equation}
\label{eq:L150}
    L_{150} = f_{\rm syn} \sum_{s} \chi_{150}^{s}Q_e^{s},
\end{equation}
where $s$ denotes the source class, $Q_e^{s}$ is a function of $\psi$ and $M_*$ (Eqs.~\ref{eq:Q_snr1}--\ref{eq:Q_snr2}, \ref{eq:Q_psr}, \ref{eq:Q_msp}), and $\chi_{150}^{s}$ depends on the source electron spectrum and the galactic magnetic field.

Combining Eqs.~(\ref{eq:Q_snr1})--(\ref{eq:Q_psr}), (\ref{eq:Q_msp}) and (\ref{eq:f_syn})--(\ref{eq:L150}), we represent the components of the radio luminosity with the following functional forms:
\begin{equation}
  \begin{split}
    L_{150} &\propto M_*^{\beta_{\rm syn}}\psi~:\rm \ SNR\ primary\ (and\ normal\ pulsars) \\
    L_{150} &\propto M_*^{\beta_{\rm syn}}\psi^{1+\beta_{pp}}~:\rm \  SNR\ secondary \\
    L_{150} &\propto M_*^{1+\beta_{\rm syn}}~:\rm \  MSP
   \end{split}
\end{equation}
In Sec.~\ref{sec:results}, we use these to fit the LOFAR data and constrain the free parameters in our model.

Finally, we note that low-frequency radio emission can be affected by free-free absorption by ionized gas. For typical galactic densities, the 150~MHz radio emission is not affected~\cite{Israel90,Hummel91,Basu15,Marvil15,Chyzy18}. However, in dense starburst galaxies, this can significantly reduce the 150~MHz luminosity~\cite{Torres04, Clemens10}.

\subsection{A Schematic Illustration of the Effect of MSPs on the Radio-SFR Correlation}
\label{subsec:qualitative}

\begin{figure}[b]
    \centering
    \includegraphics[width=0.9\columnwidth]{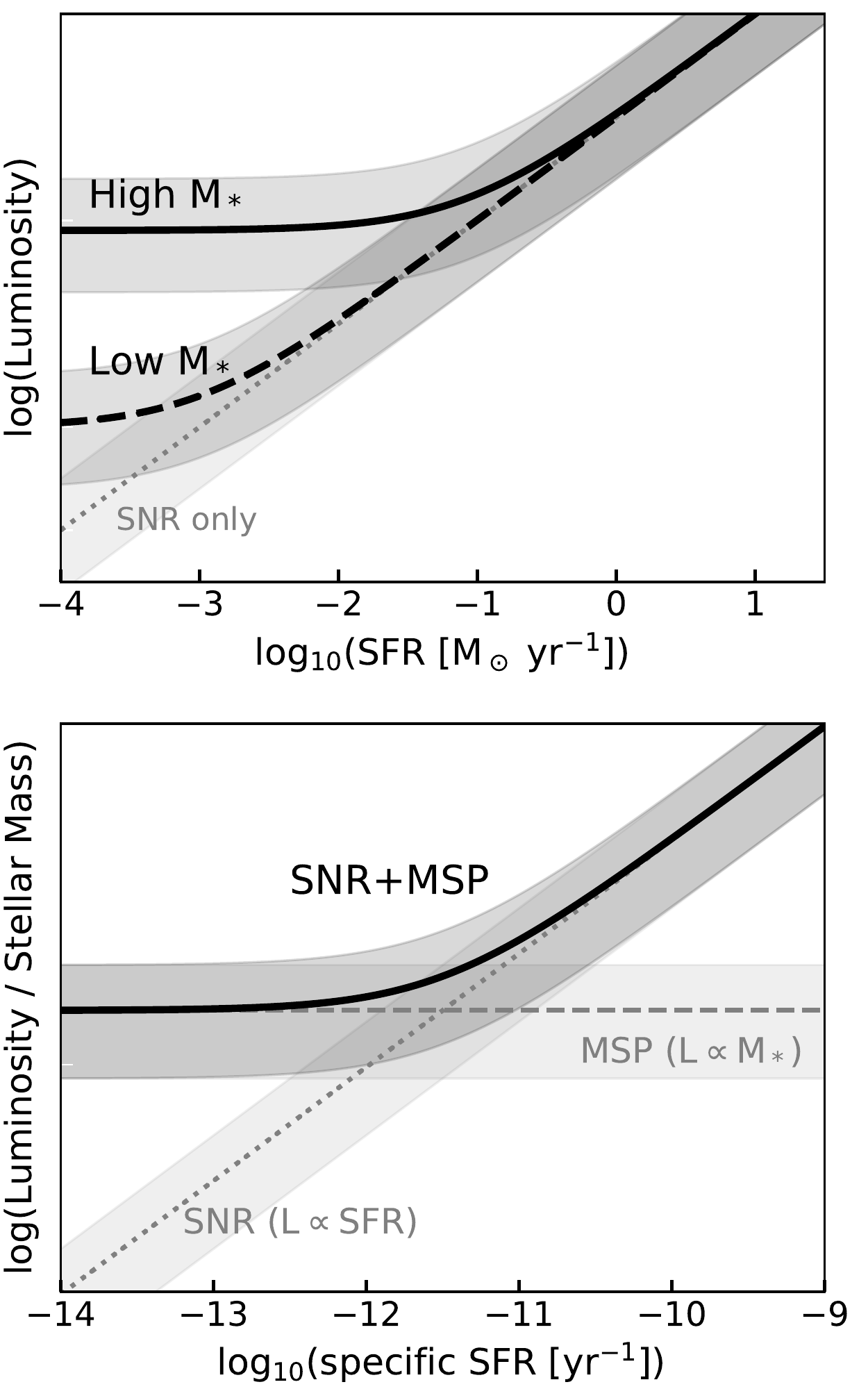}
    \caption{A schematic illustration of how MSPs can modify the radio-SFR correlation. While the radio-SFR correlation is dominated by the SNR in active galaxies, MSPs can dominate the flux of quiescent galaxies. This trend is particularly true for high-mass galaxies.}
    \label{fig:schematic}
\end{figure}

\vspace{-0.35cm}
In this section, we qualitatively describe the expected modification of the radio-SFR correlation induced by MSPs. In Fig.~\ref{fig:schematic} (top) we show the expected modification to the radio-SFR correlation in a scenario which includes MSP contributions. Specifically, the figure depicts the sum of two source terms, from SNR (Eq.~\ref{eq:Q_snr1}) and MSP (Eq.~\ref{eq:Q_msp}), with scatter that mimics source-to-source variation in $f_{\rm syn}$ (Eq.~\ref{eq:f_syn_def}). Note that we ignore the dependence of $f_{\rm syn}$ on $\psi$ and $M_*$ here. 

This figure highlights two aspects of our model. First, the effect of MSPs should be pronounced only in massive and low-SFR galaxies, as quantified in Eq.~(\ref{eq:snr_to_msp}). Therefore, MSP contributions can be clearly seen by splitting the sample into mass bins. Second, there should be source-to-source scatter due to galactic variations in the properties that affect the cooling of high-energy electrons~(see Eq.~\ref{eq:cooling}). However, theoretical modelling of the luminosity variation would require knowledge of the dispersion in the physical parameters of quiescent galaxies, which is largely unconstrained by observations.

Since the radio-SFR plot hides the masses of each galaxy, it may be useful to plot luminosities and SFRs scaled by stellar masses. Figure~\ref{fig:schematic} (bottom) illustrates a schematic expectation for luminosity/$M_*$ - SFR/$M_*$ plane. As the injection by MSPs is proportional to $M_*$, we would expect a plateau in this plane, if we ignore the dependence of $f_{\rm syn}$ on mass and SFR. In addition, in this projection the transition point from SNR to MSP domination is uniquely determined by the efficiency $\eta_{e}^{\rm MSP}$ (see Eq.~\ref{eq:snr_to_msp}) without any degeneracy with the radiation efficiency $f_{\rm syn}$. Therefore, this plot would be useful to assess the contribution of MSPs to the galactic radio emission.

\section{Data Analysis}
\label{sec:data}
In this section, we develop a method for comparing our models with the LOFAR data. First, in Sec.~\ref{subsec:data}, we briefly describe the dataset used in this work, and then in Sec.~\ref{subsec:model}, we introduce our fitting methodology.

\subsection{Dataset}
\label{subsec:data}
We utilize the flux densities, SFRs, and stellar masses of 15088 galaxies analyzed by G18. We refer the reader to Ref.~\cite{Gurkan18} for critical information regarding search strategies, catalog choices, and instrumental systematics, but summarize the key features here. G18 obtained flux density measurements for these sources from the HATLAS/NGP field survey, spanning the redshift range $0 < z < 0.6$, and then utilized a multi-step process to isolate SFGs. 

First, they identified radio-loud AGN by utilizing the radio source catalog constructed by Ref.~\cite{BH12}. Then, they divided the remaining sources into SFGs, Composite Systems, Seyferts, LINERs, and Ambiguous sources, utilizing a modified BPT-diagram focused on four emission lines: [NII]$\lambda$6584, [SII]$\lambda$6717, H$\beta$, OIII$\lambda$5007, and H$\alpha$. The necessity of a clear detection for each emission line sets a flux threshold that weights the sample toward systems observed at $z~\lesssim~0.25$. They fit multi-wavelength photometric data with the {\sc magphys} code to derive SFRs (averaged over the last 100~Myr) and the galactic stellar mass. Sources with bad {\sc magphys} fits were removed from the analysis. In the end, 3907 SFGs were analyzed by Ref.~\cite{Gurkan18}, and we use the same population in the following.

We note that 6370 of 15088 sources analyzed by G18 cannot be classified by BPT-diagram due to the lack of clearly detected emission lines. Because these sources can be contaminated by AGN emission, we do not use these unclassified sources in the main analysis. However, G18 find that they typically have low SFRs and high masses, where we expect that the contribution from MSP can be important. In Appendix~\ref{app:unclass}, we use these unclassified sources later to test the robustness of our results.

\subsection{Model Comparison}
\label{subsec:model}
To examine the role that MSPs play in the production of 150~MHz radio emission, we produce several models utilizing the source classes described in Secs.~\ref{subsec:SNR1} through~\ref{subsec:MSP}. First, we follow G18 and utilize a straightforward model for the radio-SFR correlation:
\be
\label{eq:s1}
L^{\rm model} = \alpha\psi^{\beta}, 
\ee
where $\beta$ is the index of the correlation, $\alpha$ is a normalization factor, and $L^{\rm model}$ is the expected 150~MHz radio luminosity. For an alternative model, we add a mass-dependent term. 
\be
\label{eq:s2}
L^{\rm model} = \alpha\psi^{\beta}M_*^{\gamma}.
\ee

In addition to these two empirical models, we construct two physically motivated models based on the source classes discussed in Secs.~\ref{subsec:SNR1}~through~\ref{subsec:MSP}. The first has only terms depending on the prompt SFR, and thus has a functional form:
\be
\label{eq:m2}
L^{\rm model} = (a_1\psi + a_2\psi^{\beta_{pp}})M_*^{\beta_{\rm syn}},
\ee
In the second, we add a contribution from MSPs including a mass-dependent component:
\be
\label{eq:m1}
L^{\rm model} = [a_1\psi + a_2\psi^{\beta_{pp}} + a_3M_*]M_*^{\beta_{\rm syn}}.
\ee
To reduce the number of free parameters, we fix $\beta_{pp}=0.18$ based on gamma-ray observations, noting that this choice does not affect our conclusions.

Finally, multiple confounding variables may also affect the radio flux in any given galaxy, including variations in $f_{\rm syn}$, additional sources (e.g., sub-dominant AGN activity), or additional sinks (e.g., dense gas). Thus, we introduce an intrinsic dispersion into our model. Specifically, we assume a probability distribution for the radio luminosity that follows a Gaussian distribution defined as: 
\begin{equation}
    \label{eq:dispersion}
    P_i(L) = \frac{1}{\sqrt{2\pi\sigma^2}}\exp\left(-\frac{|L - L^{\rm model}|^2}{2\sigma^2}\right),
\end{equation}
where we define $\sigma$ to be a combination of the measured uncertainty for each source and a modeling error. Quantitatively, we set $\sigma^2 = ({c}L^{\rm model})^2 + L_{\rm err}^2$, where $c$ is a free parameter that accounts for the intrinsic model dispersion and $L_{\rm err}$ is the 1$\sigma$ measurement error. We obtain best-fit parameters by minimizing the negative of the log-likelihood, $-\ln\mathcal{L} = -\sum_i\log(P_i)$, where the summation is taken for all sample SFGs. We utilize the {\sc iminuit} code~\cite{minuit} to find the best-fit model and calculate the error matrix for each model parameter. To calculate the best-fit parameters and likelihood values, we use the units [10$^{23}$ W~Hz$^{-1}$] for the radio luminosity and [10$^{10}$~M$_\odot$] for the stellar mass. Notice that while this affects the total quoted likelihood, it does not affect the difference of the log-likelihood values, $\Delta\rm{LG}(\mathcal{L})$, among different models. We have verified this approach with Monte-Carlo simulations (see Appendix~\ref{app:mcsimulation}).

Many studies of the radio-FIR correlation have analyzed the logarithmic correlation between each luminosity, using logarithmic error bars that relate to the fractional flux of the signal. Here, however, we utilize the true luminosity, because about $20\%$ of the SFGs in our study have best-fit luminosities that are negative (due to instrumental or systematic issues). If a full likelihood profile were available for the measured radio luminosity of each source, either choice should give the same final results. However, as G18 quote only 1$\sigma$ error bars, the choice of calculating likelihood profiles in linear or logarithmic space can affect the final answer. In Appendix~\ref{app:log_luminosity}, we analyze the data by utilizing a fit to the log-luminosity and analyzing only sources with positive luminosity. We find that our main conclusions are unchanged.

Finally, in our best-fit models, we find that 11 sources have a value of $-\ln\mathcal{L}$ larger than 50, representing a 7$\sigma$ rejection of our models for these sources. In each case, this stems from a source that is significantly brighter than our model prediction. Because alternative methods of producing bright radio emission (such as undetected AGN and galaxy interactions) may be present, we remove such sources from our fit so that they do not bias the relationship. In Appendix~\ref{app:bright_sources}, we include these sources and show that this treatment does not alter our conclusions.

\section{Results}
\label{sec:results}
In this section, we present the results of our analysis. In Sec.~\ref{subsec:plmodel}, we fit the data with empirical power-law models, showing that the inclusion of a mass dependence is significantly preferred. In Sec.~\ref{subsec:phymodel}, we analyze physically-motivated models and show that the inclusion of MSPs significantly improves the fit to sources in low-SFR and high-mass regime. In Sec.~\ref{subsec:robustness}, we test the robustness of our results by carrying out an alternative analysis. In Sec.~\ref{subsec:eta}, we discuss the viability of MSP scenario based on the best-fit parameters. Finally, in Sec.~\ref{subsec:systematics}, we discuss important uncertainties in our results.

\begin{figure*}[t]
\includegraphics[width=1.5\columnwidth]{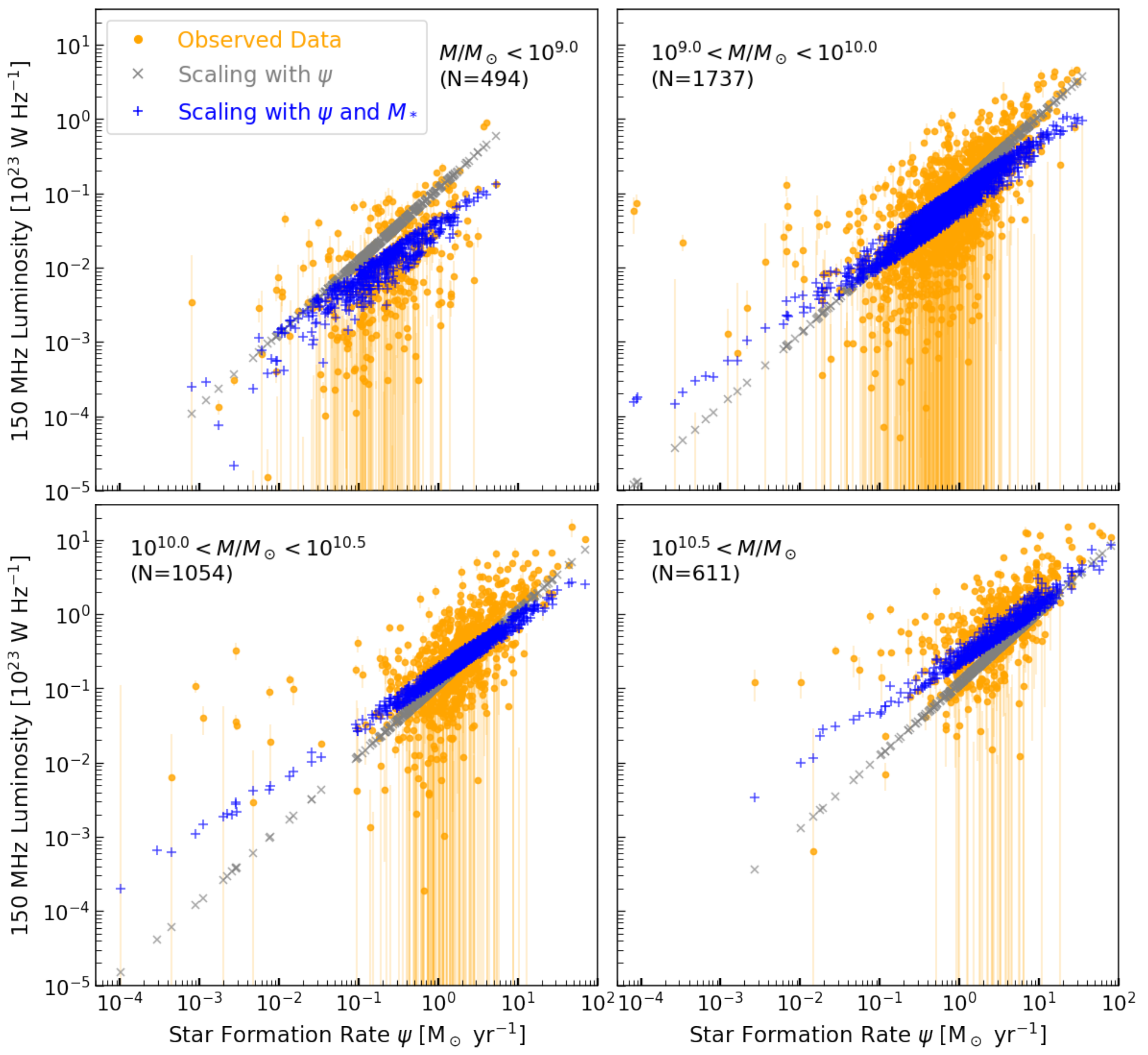}
\caption{The 150~MHz luminosity distribution of SFGs as a function of SFR (this and $M_*$ for each source are derived with {\sc magphys}).  The sample is split into four mass bins as labeled.  Orange (circle) points show the observed LOFAR 150~MHz luminosities and 1$\sigma$ errorbars. Gray (cross) and blue (plus sign) points are the best-fit luminosities predicted by the scaling relations~(Eqs.~\ref{eq:s1} and \ref{eq:s2}). The model with mass dependence (blue) fits better than prediction by SFR alone (gray), though neither can sufficiently explain low-SFR bright sources.}
\label{fig:scale}
\end{figure*}

\subsection{Testing the Simple Scaling Models}
\label{subsec:plmodel}

We first analyze and compare our simple scaling models (Eqs.~\ref{eq:s1} and \ref{eq:s2}) with the LOFAR data. The goal of this portion of the analysis is to determine whether total galaxy mass plays an important role in determining the observed radio luminosity.

Figure~\ref{fig:scale} shows the correlation between the radio luminosity and SFR, splitting our results into four mass bins. The orange points and error bars show the measured radio fluxes and uncertainties for individual SFGs in our sample. The SFR and stellar mass for each SFG are determined from {\sc magphys} fits~\cite{magphys,Gurkan18}. The gray and blue points correspond to the {\it predicted best-fit} values for these same SFGs, based on our models (theoretical uncertainties are discussed below). 

This analysis shows that a simple scaling between the SFR and the observed radio luminosity (Eq.~\ref{eq:s1}) systematically underpredicts the radio luminosity in galaxies with star-formation rates smaller than $\sim 0.1 M_\odot$~yr$^{-1}$. Adding a dependence on mass (Eq.~\ref{eq:s2}) adds a large scatter to the radio-SFR correlation and improves the fit.

We note that our likelihood function includes a significant dispersion (see Eq.~\ref{eq:dispersion}) that is not represented on this plot. This implies that it would be possible for the model with only $\psi$-scaling to provide an equally good fit to the LOFAR data, even though the $\psi + M_*$ model appears to better match the data in the figure. This could happen in a scenario where the dispersion in the LOFAR data is not related to the galactic mass. We examine this scenario as follows.

In Table~\ref{table:logL}, we calculate the mass dependence in the \mbox{LOFAR} data by comparing the log-likelihood fits of each model. We find that the addition of a mass-dependent term improves the quality of fit to the radio data by $\Delta\rm{LG}(\mathcal{L})$ = 838. If we restrict our analysis to only quiescent galaxies (132 have sSFR less than $10^{-11}$~yr$^{-1}$), we still improve the fit to the data by $\Delta\rm{LG}(\mathcal{L})$ = 212. This is notable, because this cut includes only 3\% of the galaxy counts (and is often biased towards galaxies with the largest radio uncertainties), but contributes nearly 25\% of the total improvement to the log-likelihood. This indicates that the mass dependence of the radio-SFR correlation is most pronounced in galaxies with the lowest current SFRs. 

These results are consistent with those of~G18, which also found a correlation between luminosity and stellar mass (see their Fig.~9). However, our results indicate that even in models that include a mass-dependent term, the predictions of scaling models tend to underestimate the radio luminosity of quiescent galaxies in a systematic fashion.

In Table~\ref{table:scaling}, we show the best-fit parameters for both simple scaling models. In our default model, we obtain $\beta$~=~0.98, close to the value of unity predicted from the radio-SFR correlation. In our mass-dependent model, $\beta$ drops to 0.70, an indication that there is degeneracy between the mass and star-formation rate, as expected. The model dispersion, $c$, is found to be $\sim$1.5 for both models, which suggests that the data has an intrinsic variation that spans a factor of $\sim$5 at the $\sim$3$\sigma$ level. This provides additional evidence that simple scaling models cannot explain bright low-SFR sources. The best-fit parameters derived in our work are similar to those in G18. 

\begin{table}
\begin{center}
     \caption{Values of $-\ln\mathcal{L}$ for different models. We show the sum of all SFGs (middle) and of low-sSFR SFGs ($<10^{-11}~\rm yr^{-1}$, right).\\}
\begin{tabular}{c|c|c} \hline
     & All Sources & Low sSFR  \\
    & ($N=3896$) & ($N=132$) \\
    \hline
    Scaling ($\psi$; Eq.~\ref{eq:s1}) & -391.4 &  258.4 \\
    Scaling ($\psi$ and $M$; Eq.~\ref{eq:s2}) & -1229.4 &  46.6  \\ \hline
    Model (SNR only; Eq.~\ref{eq:m2}) & -894.5 & 209.5 \\
    Model (SNR + MSP; Eq.~\ref{eq:m1}) & -1419.1 &  -69.3 \\
\end{tabular}
     \label{table:logL}
\end{center}
\end{table}

\begin{table}
\begin{center}
     \caption{Best-fit parameters for our simple scaling models. In parentheses, we show the best-fit values obtained in G18.\\}
\begin{tabular}{c|c|c|c|c} \hline
    & $\alpha$ & $\beta$ & $\gamma$ & $c$\\ \hline
    $\psi$ & 0.115 & 0.976 & - & 1.51 \\
     &(0.115) &(1.07) & - & -\\ \hline
    $\psi$ and $M_*$ & 0.124 & 0.702 & 0.422 & 1.41 \\
     &(0.135) &(0.77) & (0.43) & - \\
\end{tabular}
     \label{table:scaling}
\end{center}
\end{table}

\subsection{Testing the Physical Models}
\label{subsec:phymodel}

\begin{figure*}
\includegraphics[width=1.5\columnwidth]{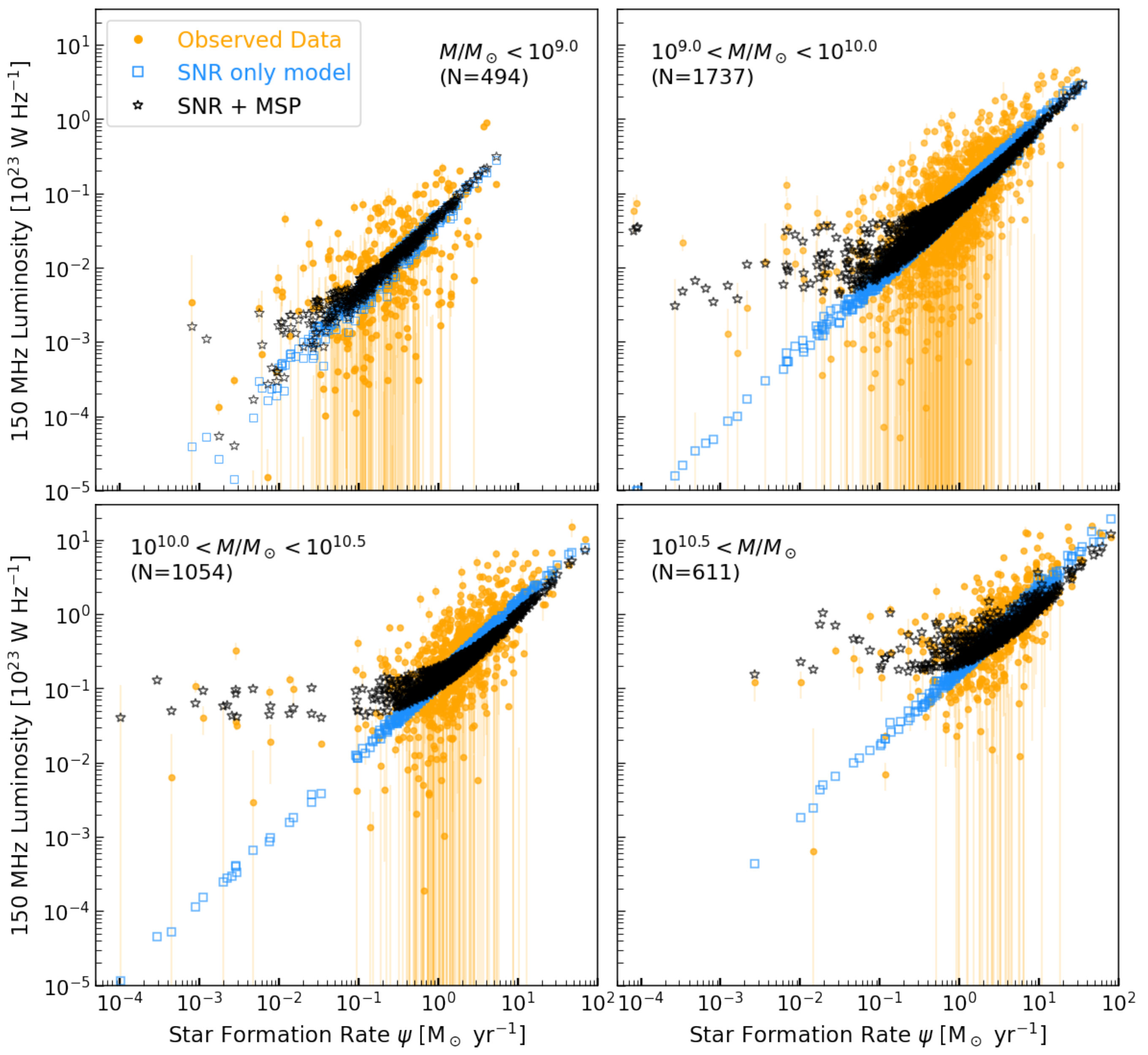}
\caption{Same as Fig.~\ref{fig:scale}, but for two physically-motivated models (Eqs.~\ref{eq:m2}~and~\ref{eq:m1}). Light blue (square) points show the model that only includes SNR contribution, which cannot fit the low-SFR data. Black (star) points show the predictions when MSPs are included, which significantly improves the fit to the LOFAR data.}
\label{fig:model}
\end{figure*}

Figure~\ref{fig:model} compares the two physical models for radio emission, described in Eqs.~(\ref{eq:m2})~and~(\ref{eq:m1}), with observations.  Compared to our simple scaling models, the SNR-only model (light blue, squares) has two additional degrees of freedom, including separate contributions from primary and secondary electrons (with different $\psi$-dependences), as well as a slight mass dependence relating to the efficiency of synchrotron production. However, this model is still incapable of explaining the bright radio emission from low-SFR sources, implying that the mass-dependent changes in the synchrotron prodcution efficiency are unlikely to explain the observed trend in the data. 

Our final model includes a contribution from MSPs, with a total flux that depends exclusively on the total galactic mass (black, star). Intriguingly, this significantly improves our fit to the radio data, particularly among among the population of quiescent galaxies. In Table~\ref{table:logL}, we find that this model improves the log-likelihood fit by 525, producing significant preference for this model compared to the SNR-only model. Restricting our analysis to only galaxies with low sSFR ($<10^{-11}$~yr$^{-1}$), we still find a log-likelihood improvement of 279, which indicates that the model is significantly favored even among only the 132 least luminous sources. 

We note that our physical SNR-only model provide a worse fit (by a log-likelihood of $\sim$306) than our simple scaling model that depends on $\psi$ and $M_*$, even though the SNR model has an extra degree of freedom. This is due to the fact that the scaling model prefers a power of $\psi$ that is smaller than one. Such a scenario is incompatible with the assumption of our SNR-only model, because the power of $\psi$ is fixed to either 1.0 or 1.18 for primary (secondary) components. {\it This indicates that SN-driven physics (with an input power that is at least linearly dependent on $\psi$) is unlikely to drive the radio excess in quiescent galaxies.} Additional factors, such as a competitive energy-loss process (like a cosmic-ray escape component that depends strongly on $\psi$) would be necessary to explain this data. However, this is not observed in bright SFGs, where the radio-FIR relation is found to be steeper than linear.

In Table~\ref{table:model}, we list best-fit parameters for our physical models. The normalization for the secondary term, $a_2$, is found to be unphysically small for the SNR-only model. This can be understood based on the preference of our scaling model (with $\psi$ and $M_*$) for a best-fit value $\beta <$ 1. Among the two terms that scale as $\psi$ and $\psi^{1.18}$, the best-fit model would only require the first term. This result indicates that our standard SNR model may be unable to provide a good fit to the data. Interestingly, we note that the SNR+MSP model predicts a value of $a_2$ that is physically reasonable.

We note that there are also sources that are significantly less luminous than our model predictions. However, our models would also predict significant dispersion in the radio luminosity of individual SFGs, which may explain these sources. 

In particular, in some systems $f_{\rm syn}$ may be small due to either efficient escape, a strong radiation field, a high gas density, or a weak magnetic field, all of which can lower the synchrotron signal. In addition, free-free absorption may significantly reduce the radio flux in galaxies with high gas densities. Some LOFAR sources even have radio luminosities that are negative, a clear indication of systematic or instrumental effects that are not included in our model.  We also note that, contrary to very dim sources, bright sources are difficult to explain solely by a variation in $f_{\rm syn}$, because it cannot exceed 100$\%$. Finally, we speculate that the star-formation history of each galaxy could stochastically change the total energetics from the population of MSPs, although exact assessment of this effect is difficult.

\begin{table}
\begin{center}
     \caption{Best-fit parameters for our models.\\} 
\begin{tabular}{c|c|c|c|c|c} \hline
    & $a_1$ & $a_2$ & $a_3$ & $\beta_{\rm syn}$ & $c$\\ \hline
    SNR only & 0.110 & 2.00e-10 & - & 0.285 & 1.46 \\
    SNR+MSP & 0.035 & 0.031 & 0.036 & 0.106 & 1.39 \\
\end{tabular}
     \label{table:model}
\end{center}
\end{table}

\subsection{Testing the Robustness of MSP Models}
\label{subsec:robustness}
Thus far, we have carried out our analysis on unbinned data. Here, we test whether an alternative, binned analysis of our dataset also produces a statistical preferences for a mass-dependent term. We note that this method should have less statistical power, because it throws away a significant quantity of information. We use our default dataset (all SFGs from the G18 sample) and separate sources into seven SFR bins and ten mass bins (with a constant logarithmic width). We then iteratively merge the smallest bins into larger ones, such that each bin has more than five sources. This avoids numerical issues that arise in very small bins, where the source to source dispersion is difficult to calculate. For each bin, we calculate the mean value of the SFRs and masses, along with the recorded radio luminosities and their standard deviation. We carry out model fit by minimizing $\chi^2$ values. We have verified this approach with Monte-Carlo simulations (see Appendix~\ref{app:mcsimulation}). We find the following $\chi^2$ statistic on binned data:
\begin{eqnarray*}
    \chi^2_{\rm (binned)} &=& 53.7~({\rm SNR~only}, {\it dof}=24)\\
    \chi^2_{\rm (binned)} &=& 25.8~({\rm SNR + MSP}, {\it dof}=23).
\end{eqnarray*}

Thus, our model remains statistically significant at a level exceeding 5$\sigma$, even after being binned relatively coarsely (which decreases the total information and thus the statistical significance). We emphasize that these simple $\chi^2$ tests on binned sources, while indicative, are not the full story, which requires our full maximum-likelihood approach. 

In our default, unbinned analysis, we  examine differences in $-\ln\mathcal{L}$, which demonstrate with high statistical significance that the inclusion of a MSP term improves the fit over the SNR-only model. Here we also evaluate the overall fit of our models to the unbinned data by calculating the chi-squared statistic, $\chi^2_{\rm (unbin)} = \sum_{i}{|L_{i, \rm model} - L_{i}|^2}/{\sigma_{i}^2}$, where the summention runs for all sources and $\sigma_i^2$ is the sum of $(cL_{i,\rm model})^2$ and $L_{i,\rm err}^2$. We find $\chi^2_{\rm (unbin)} =$ 5141 for our MSP+SNR model (sample size 3896), whereas $\chi^2_{\rm (unbin)} =$6002 for the SNR-only case, the null hypothesis.  This goodness-of-fit is dominated by the many points at high SFR, where we expect that the SNR-only model should sufficiently fit the data. 

To examine the goodness-of-fit for low specific-SFR sources, we re-calculate $\chi^2_{\rm (unbin)}$ using only the 132 sources that have $\psi/M_*$ less than $10^{-11}$ yr$^{-1}$. For our MSP+SNR model, the $\chi^2_{\rm (unbin)}$ is 205, whereas it is 1063 for the SNR-only null hypothesis. The large $\Delta\chi^2_{\rm (unbin)}$ obtained from the addition of a mass-dependent term indicates that it is likely the most important parameter needed to model radio emission from low-specific-SFR sources. In fact, because the total $\chi^2$ decreases by nearly 80\%, it can be shown that any parameter that decreases the $\chi^2$ by a larger amount must be correlated with mass. We note that while our model is preferred over the SNR-only model, both models produce a rather poor overall goodness of fit to individual galaxies. Even for the MSP+SNR model, the fit still has a $\chi_{\rm (unbin)}^2/d.o.f$ of 1.32, which indicates that the fit still has a low p-value. However, such a deviation should be expected based on the simplicity of our model. Our aim is to point out that adding mass-dependent term would significantly improve the fit and that MSPs can naturally produce such a term.  In future work, more complete models of the radio emission can be considered.

\color{black}

Our analysis shows that current LOFAR data favor a physical model with mass-dependent cosmic-ray injection (as is clear from Fig.~\ref{fig:model}). Next, we discuss the validity of the MSP model based on our best-fit parameters.

\subsection{Interpretation of Results}
\label{subsec:eta}

In the previous section, we have shown that the LOFAR data strongly prefers a physical model that includes at least one emission term that depends only on the galaxy mass. In Sec.~\ref{subsec:MSP}, we noted that a model including MSP-accelerated electrons would predict such a feature. This does not, however, prove that MSPs are the physical source of the excess radio emission. In this section, we show that such a scenario is possible, and, in fact, that current data suggests that MSPs can power bright radio emission with an intensity that is consistent with the excess.

Combining Eqs.~(\ref{eq:Q_msp}, \ref{eq:L150}, \ref{eq:f_syn}) and the third term in Eq.~(\ref{eq:m1}), we can write the MSP radio intensity as:

\be
a_3 = \frac{4}{3}\chi_{150}^{\rm MSP}\alpha_{\rm syn}\eta_{e}^{\rm MSP}L_{38}^{\rm MW},
\ee

\noindent where $a_3$ is best-fit parameter of the MSP contribution in Eq.~(\ref{eq:Q_snr1}), $\chi_{150}^{\rm MSP}$ is the ratio of the 150~MHz-emitting electron power to the total electron power, and the factor 4/3 arises from the conversion from W~Hz$^{-1}$ to erg~s$^{-1}$ at 150~MHz. We note that the electron power in the 150~MHz window is calculated over $\Delta\ln E_e=0.5$, as the luminosity is calculated by integrating the flux density over $\Delta\ln\nu=1$.

The radio spectral index of galaxies is approximately $F_\nu~\propto~\nu^{-0.7}$ near GHz frequencies and flattens to $F_\nu\propto\nu^{-0.5}$ near 100~MHz, which is likely caused by cooling and propagation effects~\cite{Israel90,Hummel91,Basu15,Marvil15,Chyzy18}. This translates to a steady-state differential electron spectrum of $E_e^{-2.4}$ above a few GeV and $E_e^{-2.0}$ below that. Adopting this spectral shape for electrons, we obtain $\chi_{150}^{\rm MSP}\sim0.1$, a value that only weakly depends on the spectral break and minimum electron energy. 

The efficiency of synchrotron emission, $\alpha_{\rm syn}$, may also depend on galaxy properties~(see Eq.~\ref{eq:f_syn}). For simplicity, we adopt typical Milky Way parameters to estimate the energy-loss timescales. We also assume that massive galaxies are calorimetric to cosmic-ray leptons, as is the case in the Milky Way~\cite{Strong10}. Under these assumptions, we obtain $\alpha_{\rm syn}\sim0.2$, which gives us:
\be
\label{eq:eta1}
\eta_{e}^{\rm MSP}\simeq 1(L_{38}^{\rm MW})^{-1}.
\ee

Thus, we find that the best-fit normalization of the MSP contribution ($a_3$~=~0.036) does not violate the total power of the MSP population. However, since $L_{38}^{\rm MW}\sim1$, this relation implies that our model does require the majority ($\eta_{e}^{\rm MSP}\sim1$) of the MSP spindown power to be injected into electrons. This might initially appear worrisome, as some previous estimates have utilized efficiencies of $\eta_{e}^{\rm MSP}\sim\eta_\gamma\sim0.1$. However, there has (to date) been no study validating these assumptions. 

Additionally, there are a number of uncertainties in our modeling that may significantly affect this result. Most importantly, the energetics of galactic MSPs are unknown. In this study, we normalize the total gamma-ray luminosity of MSPs to Milky Way observations. However, our MSP models are expected to dominate only in galaxies with low-SFRs and high masses, which may have different star formation histories than the Milky Way. Notably, if we instead normalized our results to M31, which has properties more consistent with quiescent galaxies (a larger stellar mass and a smaller SFR~\cite{Yin09,Sick15}), the necessary MSP efficiency would decrease by up to a factor of $\sim$4. Also, because the gamma-ray emission from MSP magnetosphere may be beamed, only some fraction of Galactic MSPs, $f_b$, can be observed from the Earth. Although $f_b$ is often assumed to be unity for gamma-ray pulsars, the actual value could be smaller by a factor of $\sim$2~\cite{Johnson14}, which would decrease the efficiency $\eta_{e}^{\rm MSP}$ by a factor of $1/f_b$. These (among other) uncertainties could lower the necessary efficiencies to the $\sim$10\% level.

In addition to observational uncertainties that may make the MSP efficiency smaller than our model prediction, we note that a large MSP e$^+$e$^-$ efficiency is consistent with our understanding of pulsar physics. Observations indicate that roughly 10\% of the MSP spin-down power is converted into gamma-ray emission within the magnetosphere, a negligible fraction of the total spin-down power is converted to radio, and the remaining power is carried primarily by e$^+$e$^-$ pairs, the magnetic field, and possibly protons. Although we lack knowledge concerning the energetics of the MSP pulsar wind, it is established for young pulsars that more than $\sim$90$\%$ of the spindown power is converted to pulsar-wind electrons that power the PWNe~\cite{Coroniti90}.

Observationally, the constraints on GeV-scale MSP emission are not strong. Ref.~\cite{Yuan15} found that e$^+$e$^-$ efficiencies up to 90\% can be reconciled with MSP models of the galactic center excess (see, however, Ref.~\cite{Hooper18}). Intriguingly, studies of GeV emission from the Galactic bulge by Ref.~\cite{TheFermi-LAT:2015kwa} find that the inverse-Compton flux exceeds standard predictions by more than a factor of 20, requiring a bright new source of energetic electrons. At the TeV scale, a stacking analysis of 24 MSPs observed at TeV energies by the HAWC telescope provided 2.6--3.2$\sigma$ evidence of TeV MSP emission, a result which would require a high efficiency for TeV e$^+$e$^-$ pair production from MSPs~\cite{Hooper18}. We note that observations of globular clusters in very-high-energy gamma rays suggest efficiencies below $\sim10\%$~\cite{Bednarek16, MAGIC19gc}. However, this result assumes particularly optimistic models for particle propagation within globular clusters (a Bohmian diffusion model), which has yet to be verified. Extrapolating this result to GeV energies also depends sensitively on spectral assumptions.

In light of these points, we conclude that MSPs can be efficient e$^+$e$^-$ accelerators. The necessity of an $\mathcal{O}(1)$ e$^+$e$^-$ efficiency may stretch current modeling. However, multiple uncertainties in our models may significantly lower the efficiency necessary to fit the radio excess. Furthermore, no observation rules out efficiencies as high as $\sim$90$\%$.

\subsection{Systematic Uncertainties}
\label{subsec:systematics}

We have shown that MSP-based models explain the flattening trend observed and detailed by G18. Here we note several systematic uncertainties that could affect the plateau detected by G18. We stress that while our models were fit to the G18, the qualitative hypothesis that MSPs contribute to the radio-SFR correlation does not necessarily require a flattening of the data at the level observed by G18.

In particular, we note that the accurate determination of the SFR and radio flux in the dimmest quiescent galaxies pushes the limits of current observational data. One worrisome point concerns any potential flux-sensitivity limit in the radio data. Such a limit could induce a plateau-like feature by excluding a vast sea of ``missing" galaxies with smaller radio fluxes. However, the methodology applied by G18 specifically accounts for such a scenario --- reporting the best-fit flux (including negative best-fit fluxes) for all galaxies that are determined to be SFGs via multi-wavelength photometric fits. We note two other facts that diminish the risk of such a systematic error. On the observational side, the large redshift range of SFG studies would smear out simple flux-sensitivity limits. On the theoretical side, we note that our model predicts the existence of very dim radio galaxies, due to the significant dispersion induced by variations in the magnetic field, ISRF, and interstellar gas densities in each galaxy. 

While systematic uncertainties in the radio luminosity are likely controlled by the analysis methods of G18, a more pressing concern may be the accurate determination of the star-formation rate. Because only a small number of quiescent galaxies are classified as SFGs in the G18 sample (and spectral-line classification of SFG samples may depend on the SFR of the galaxy), a systematic bias that shifts some galaxies to abnormally low-SFRs independent of their radio flux may be interpreted as a plateau feature in the radio-SFR correlation. We stress that in G18, SFRs and masses are derived by {\sc magphys} fit based on multiwavelength photometric data from SDSS {\it u}-band to submillimeter wavelength. However, the sensitive dependence of our results on this fit deserves further investigation.

A complete re-analysis of the SFRs in quiescent SFGs lies beyond the scope of this theoretical paper. Here, we test the results by replacing the SFRs and masses with those contained in the GSWLC-2 catalog of Ref.\cite{Salim18} (hereafter S18, see also Ref.~\cite{Salim16}). The galactic properties in this catalog are derived by SED fitting to the UV, optical and mid-IR data with the {\sc cigale} code~\cite{cigale}. We refer the reader to Refs.~\cite{Salim16, Salim18} for detail, but stress that one notable difference from G18 is that S18 includes short-wavelength UV radiation, which may produce more accurate measurements for quiescent galaxies. S18 produces three separate catalogs, shallow, medium, and deep UV imaging surveys, and we use the medium (GSWLC-M2), which is recommended for quiescent galaxies. 

We cross-correlate the catalog of S18 with G18, noting that only 1094 out of 3907 SFGs in G18 are included in the S18 catalog because GSWLC-M2 does not cover all SDSS targets. This is potentially a significant concern --- as important selection effects in the join-observation probability of the catalogs may affect our results, and are difficult to quantitatively assess. Keeping this in mind, we repeat our analysis, utilizing the radio luminosities and source classifications of G18 but utilize the SFRs and masses determined by S18. In Appendix~\ref{app:sfr}, we detail our analysis procedure.

We obtain a somewhat concerning result, which is that S18 systematically derives higher SFRs for the low-SFR galaxies observed by G18. This potentially suggests that the choice of methods for SFR measurements can have a significant impact. 

\begin{figure}
    \centering
    \includegraphics[width=0.9\columnwidth]{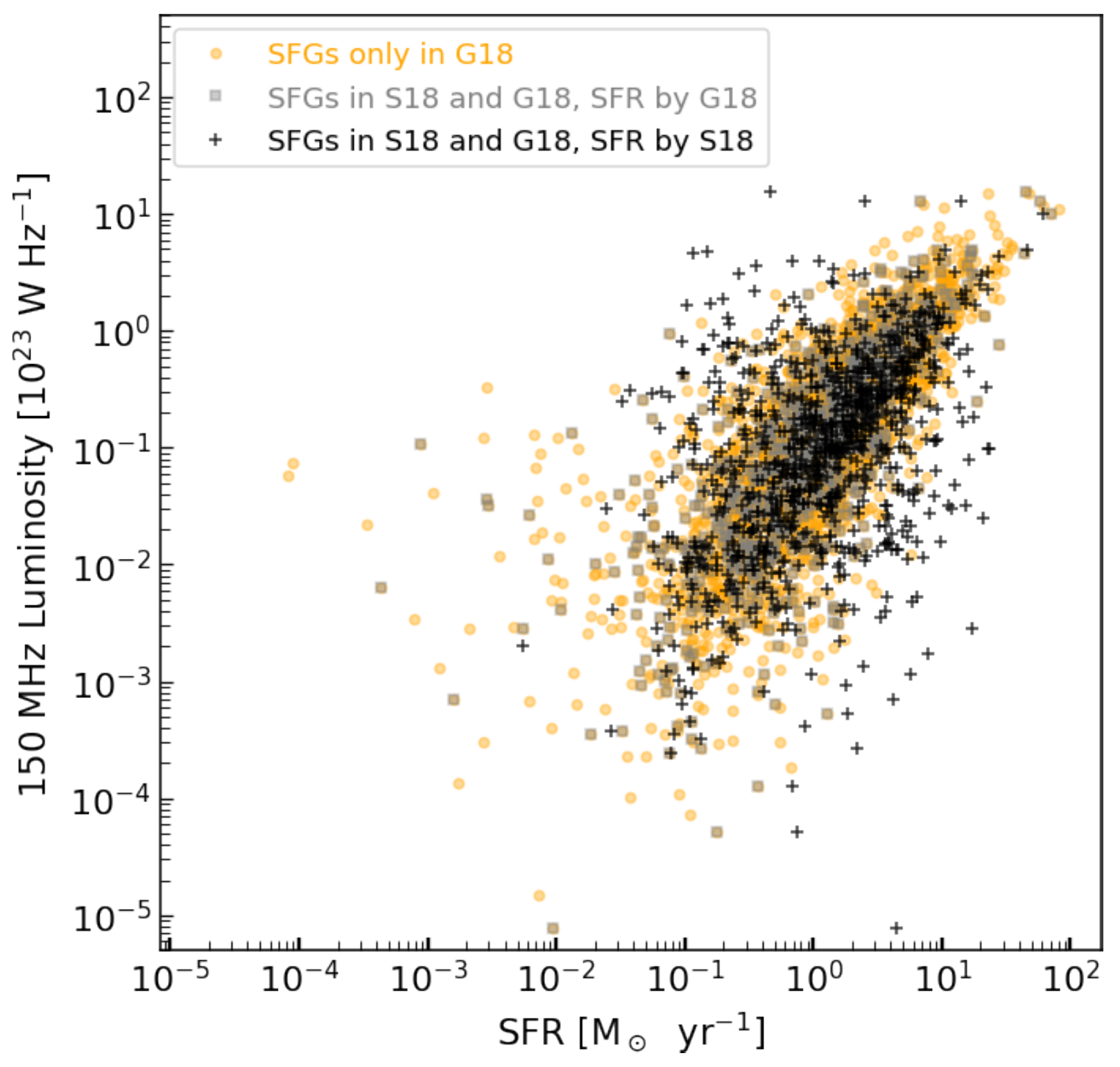}
    \caption{Comparison of the radio-SFR correlation for two different SFR estimations. We note that the SFR estimates of S18 significantly decrease the significance of the plateau feature observed in the radio-SFR correlation by G18.}
    \label{fig:s18:radio}
\end{figure}

We find that that this offset affects our results in two important ways. First (and perhaps most importantly), we find that the slope of the entire radio-SFR correlation becomes significantly flatter. Focusing our analysis only on galaxies with a \mbox{sSFR $>$ 10$^{-11}$~yr$^{-1}$}, where the radio-SFR correlation is thought to hold, we find that our cross-analysis model prefers a best-fit value ($L\propto\psi^{0.6}$), which is significantly flatter than standard radio-SFR measurements. Additionally, the scatter in the radio-SFR correlation increases. This is concerning, as no observation has previously reported a strongly sub-linear radio-SFR correlation --- and it calls into question whether systematic errors in the cross-correlation of these catalogs render the analysis untrustworthy. While the LOFAR analysis of the radio-SFR correlation takes place in a waveband that has not been extensively probed in other work, previous results by Ref.~\cite{1988MNRAS.235.1227C} found that the radio/FIR correlation was even steeper at 151~MHz than at higher-frequencies, contradicting the flattness of the results obtained here. 

The second impact is that the excess feature in low-SFR sources is much less pronounced, and a mass-dependent term (such as that from MSPs) is no longer statistically preferred. In Fig.~\ref{fig:s18:radio}, we show the radio-SFR correlation for our combined G18/S18 analysis, finding that the plateau feature produced in the G18 SFR calculation has disappeared. This is potentially worrisome, as it suggests that observations by G18 could potentially be explained by systematic uncertainties in SFR measurements. 

One alternative possibility is that the classification of ``SFG" sources by G18, which selects only a small fraction of the sources with low-SFRs, systematically biased the catalog towards sources that will have larger SFRs in S18. To test this possibility, we repeat cross-correlated study using radio luminosities from G18, but SFR measurements from S18. However, we loosen the restriction that the galaxy in question is labeled a SFG by G18, and instead also include ``unclassified" sources from G18. This increases our total population to 10277 sources, and adds a large population of sources that have low SFRs in both the G18 and S18 analyses. However, it potentially contaminates our result with galaxies that are radio bright due to low-luminosity AGN. 

In Appendix~\ref{app:unclass}, we detail this analysis, finding that this cross-correlation improves both the global and low-SFR agreement between G18 and S18. In particular, the radio-SFR correlation (for galaxies with sSFR$>$ 10$^{-11}$~yr$^{-1}$), is slightly hardened to $L\propto\psi^{0.8}$. Moreover, we find strong evidence in this dataset for a mass-dependent term compatible with our MSP hypothesis. We caution however, that AGN contaminated sources may also produce such a feature.

We thus conclude that a careful analysis of low-SFR galaxies is necessary in order to verify the contribution of MSPs to the radio emission in low-SFRs. Utilizing the results of the detailed study completed by G18, we find strong evidence in the data to support such a feature. The significance of this result decreases significantly if the results of S18 are instead used to calculate the SFRs of quiescent galaxies --- although the convolution of these studies also induces spurious effects into the main-bulk of the radio-SFR correlation that makes it difficult to interpret these results. One possible explanation may be systematic effects stemming from the interplay between the SFG-classification of galaxies in G18, and their total SFR utilizing the methods of S18. Futher analysis is thus necessary to confirm the plateau feature of G18 which is best fit by MSPs in our study. 

Noting that a significant MSP contribution is independently motivated by potential HAWC observations of gamma-ray emission from MSPs~\cite{Hooper18}, we now discuss the implications of our results based on the properties of quiescent galaxies.

\section{Discussion and conclusions}
\label{sec:discuss}

In this paper, we propose that MSPs can significantly contribute to the radio flux in quiescent galaxies, dominating the low-end tail of the radio-SFR correlation. We show that models including an MSP contribution significantly improve the fit to LOFAR radio data, explaining the observed excess in low-SFR, high-mass galaxies. We show that the energetics of our model are consistent with current observations and models of MSP activity. Finally, we discuss systematic uncertainties and future directions to test our model. Our results have several implications. 

\subsection{Radio-SFR (Radio-FIR) Correlation}

Previous studies of the radio-FIR correlation have found a striking continuation of power-law behavior over many orders of magnitude in galactic star-formation rate~(e.g., Refs.~\cite{irc1985, condon92, irc2001, Bell:2002kg}). 
This has been attributed to a variety of factors, the sum of which has been labeled a ``conspiracy" in the low-SFR behavior of the relationship. LOFAR data, on the other hand, appear to provide evidence for a break in that relationship among high-mass, low-SFR galaxies, and our interpretation offers yet another complicating factor that may shift this relationship from its linear functional form. Our model predicts that future observations of the radio-SFR correlation by LOFAR, as well as next-generation telescopes like SKA, will more clearly identify excess radio emission in high-mass galaxies that do not host AGN.

The tightness of the radio-FIR correlation has raised an expectation that the radio continuum emission can serve as a robust SFR tracer that is not affected by dust extinction. Our analysis suggests that the extrapolation of radio-SFR correlation to low sSFR sources may be insecure, and that future studies of radio emission in low-SFR galaxies should include MSP contributions. Alternatively, more detailed studies of the time-evolution of the MSP population and how it contributes to the galactic radio luminosity may allow radio measurements to inform measurements of star-formation histories in quiescent galaxies.

Finally, the LOFAR data have also been used to perform direct studies of the radio-FIR correlation in 150~MHz band~\cite{Read18}. Based on our results, one would expect excess radio emission for sources that have low FIR luminosity. This is not clearly seen in the data (though the uncertainties in the FIR luminosities for these sources are large). We speculate that this suggests another ``conspiracy" in the radio-FIR correlation. For massive galaxies, the radio luminosity is enhanced due to MSPs, and the FIR is also enhanced by the heating of interstellar dust by old stellar populations. Indeed, multiple studies have shown that intermediate and old stellar populations can produce significant IR emission even for galaxies with little star formation~\cite{Salim09,Calzetti10}. This new conspiracy might be important for future studies.

\subsection{Bright Radio/Gamma-Ray Emission from the Bulge of Disk Galaxies}

We have shown that the LOFAR data prefer a mass-dependent injection term, and have also shown that such a term is naturally produced by MSPs. However, there could be alternative explanations for the radio excess. Most importantly, although AGN have been removed from the LOFAR sample using BPT-diagram diagnostics, potential radio contributions from relatively-dim supermassive black holes cannot be ruled out. This scenario is particularly troubling, because supermassive black hole masses have been found to correlate with the total galaxy mass~\cite{1998AJ....115.2285M}, providing an alternative explanation for the mass dependence detected in our model (see also G18).

However, nearby, spatially resolved galaxies provide an excellent opportunity to differentiate these scenarios and study the contribution of MSPs to galactic radio emission. If the radio flux is dominantly from AGN, we would expect emission only from the galactic core, and would potentially expect variable emission. On the other hand, MSPs emission would be more extended (although it can be significantly enhanced in the bulge region) and should show no variability.

Intriguingly, there are several nearby galaxies that host large LMXB populations and also have bright diffuse radio excesses, most notably M31~\cite{2007A&A...468...49V,Ackermann17m31,McDaniel19}. Notably, Ref.~\cite{McDaniel19} determined the M31 bulge to be powered by an electron flux of $\sim10^{39}$~erg~s$^{-1}$, while SNRs are expected to injection only $\sim5\times10^{37}$~erg~s$^{-1}$. Utilizing a $M_* = 4\times10^{10}~M_\odot$ stellar mass for the M31 bulge~\cite{Eckner2018}, our model predicts that MSPs inject an electron flux of $\sim8\eta_e^{\rm MSP}\times 10^{38}$~erg~s$^{-1}$, explaining the majority of the electron power. Variations in the radio-FIR correlation are also seen across the M81 galaxy, with excess radio emission (compared to the Galactic average of~Ref.~\cite{irc2001}) found outside of active starbursts~\cite{Gordon:2004wn}. 

In addition, observations indicate that LMXBs and MSPs are highly overabundant in dense regions, such as globular clusters, compared to their average formation rate throughout the Milky Way plane~\cite{1984AdSpR...3...19G}. Therefore, cross-correlating diffuse radio emission with globular clusters may be useful to test an MSP origin of the radio excess, as it can constrain the energetics and spectrum of electrons. It may also be possible to detect radio emission around individual MSPs, if the particle diffusion around them is sufficiently suppressed. An alternative way to test and constrain our model is to observe gamma rays from globular clusters that are generated through inverse-Compton scattering. In this direction, a very recent study~\citep{2021arXiv210200061S} indeed detected such emission, providing direct evidences that MSPs can produce GeV-scale electrons. Interestingly, they find that injected electrons might have a steep spectrum. If we assumed such spectra, significant amount of electron energy should be contained in GeV-scale electrons, and thus the required efficiency $\eta_e^{\rm MSP}$ could be much smaller than estimate in Eq.~(\ref{eq:eta1}), although it would then highly depend on the minimum electron energy. On a similar ground, future observations by the Cherenkov Telescope Array will further constrain electron populations injected by MSPs~\citep{2021arXiv210205648M}.

\subsection{Implications for Gamma-Ray and Cosmic-Ray Astrophysics}

Finally, our results suggest that MSPs may efficiently convert a large fraction of their spin-down power into GeV-scale e$^+$e$^-$ pairs. Because MSPs do not include compact pulsar wind nebulae, these e$^+$e$^-$ pairs must escape into the ISM, where they subsequently cool via a combination of synchrotron (producing radio emission) and inverse-Compton scattering/bremsstrahlung (producing gamma-ray emission). The ratio of these components depends sensitively on the galactic environment. 

Recent observations have found a bright excess in GeV gamma-ray emission from the Galactic center of the Milky Way galaxy~\cite{Goodenough:2009gk, Daylan:2014rsa}. The most convincing explanations for this excess consist of dark matter annihilation~\cite{Goodenough:2009gk, Daylan:2014rsa} or the production of GeV gamma-ray emission within MSP magnetospheres~\cite{Abazajian:2010zy, Bartels:2015aea, Lee:2015fea}. Our model predicts that any such MSP population will be accompanied by a bright inverse-Compton emission in the Milky Way bulge. 

The impact of such a scenario on the interpretation of the Galactic center excess is unclear. At GeV energies, there is some evidence for excess inverse-Compton emission in the Milky Way bulge. In particular, models by the Fermi-LAT collaboration required that the normalization of the inverse-Compton scattering emission from the inner regions of the Milky Way was $\sim$20 times brighter than standard Galprop predictions (which, notably, do not include any cosmic-ray injection in the Galactic center region)~\cite{TheFermi-LAT:2015kwa}. Alternative models that do include significant hadronic cosmic-ray injection near the Galactic center include more modest enhancements to the leptonic emission~\cite{Carlson:2016iis}. 

Our results suggest that GeV-scale e$^+$e$^-$ from MSPs can significantly contribute to the background gamma-ray emission from the Galactic center, a scenario which may be compatible with MSP models for the Galactic center excess. On the contrary, if the MSP-induced ICS emission continues to TeV energies, the lack of bright TeV emission within the Galactic bulge would place a strong constraint on the contribution of beamed MSP emission to the Galactic center excess at GeV energies~\cite{Hooper18}.

If MSPs do produce bright TeV gamma rays via inverse-Compton scattering, a number of Milky Way MSPs are expected to be local and powerful enough to be seen by current and future TeV telescopes such as HAWC and CTA~\cite{Hooper18}. Such sources could contribute to the recently discovered population of ``TeV Halos" discovered by TeV gamma-ray observations around nearby pulsars like Geminga and Monogem~\cite{TeVhalo:HAWC17, Linden17halo}, now also observed at GeV energies~\cite{DiMauro19}. Importantly, unlike normal pulsars, MSPs lack associated SNRs and PWNe, which remain a confounding factor in assessing both the luminosity and morphology of TeV halos. The existence of TeV halo emission surrounding an MSP population would have important implications for our understanding of cosmic-ray propagation near bright TeV emission sources~\cite{Evoli:2018aza}.

As an efficient e$^+$e$^-$ accelerator, MSPs may produce a substantial contribution to the local e$^+$e$^-$ flux, potentially contributing to the positron excess observed by PAMELA and AMS-02~\cite{2010PhRvL.105l1101A, PhysRevLett.110.141102}. While some recent analyses, e.g., Ref.~\cite{2015ICRC...34..462V} argued that single MSPs explain only a few-percent of the excess, these results assumed electron production efficiencies of only a few percent. On the other hand, Ref.~\cite{Kisaka12} used an efficiency of 50$\%$ from spindown power to e$^+$e$^-$ pairs and found that MSPs can significantly contribute to the observed cosmic-ray electron and positron flux. As our analysis provides additional evidence supporting high e$^+$e$^-$ efficiencies in MSPs, it supports scenarios where MSPs significantly contribute to the positron excess.

Finally, even in low-SFR galaxies that are supposed to have little astrophysical emission, e$^+$e$^-$ pairs from MSPs may produce bright radio and gamma-ray emission. This can be an additional source of background emission for indirect searches of dark matters. In this context, the contribution from MSPs are evaluated in Ref.~\cite{Winter16}, but they only consider direct gamma-ray emission from the magnetosphere. Our results suggest that pulsar-wind e$^+$e$^-$ could significantly contribute to the background emission, potentially making additional factor of confusion for future dark-matter searches. Due to the small size of dwarf galaxies, the lumninosity of such a component might depend on the ability of MSPs to self-confine their own cosmic-ray electron population (as in, e.g., TeV halos)~\cite{Hooper18, Evoli:2018aza}.

\vspace{-0.3cm}
\section*{Acknowledgments}
\vspace{-0.3cm}
We thank G\"{u}lay G\"{u}rkan for providing us the data that are used in Ref.~\cite{Gurkan18} and also for helpful comments. We are grateful for helpful comments from Katie Auchettl, Yi-Kuan Chiang, Norita Kawanaka, Shaun Read, Lingyu Wang, and especially Rainer Beck, Samir Salim, and Todd Thompson. This research made use of {\sc astropy}~\cite{astropy0,astropy}, {\sc matplotlib}~\cite{matplotlib}, {\sc numpy}~\cite{numpy} and {\sc iminuit}~\cite{minuit}. T.S. is supported by a Research Fellowship of Japan Society for the Promotion of Science (JSPS) and by JSPS KAKENHI Grant No.\ JP 18J20943. T.L. is supported by Swedish Research Council Grant No.\ 2019-05135. J.F.B. is supported by NSF Grant No.\ PHY-1714479.

\newpage
\clearpage

\appendix

\section{Models using Log-Luminosity}
\label{app:log_luminosity}

In the main text, we fit our model against the luminosity values and uncertainties for each source using a linear fit to the data. This was due to the fact that some sources have negative best-fit values due to instrumental or systematic issues. Here, we re-analyze the data after taking the logarithm of the luminosity values, producing a probability model given by:
\begin{equation}
\label{eqA:p}
    P_i(L) = \frac{1}{\sqrt{2\pi \sigma_{\log_{10} L}^2}}\exp\left(-\frac{|\log_{10}(L) - \log_{10}(L^{\rm model})|^2}{2\sigma_{\log_{10} L}^2}\right),
\end{equation}
where $\sigma_{\log_{10} L}$ is a free parameter. In this analysis, we use only the 3215 sources that have positive best-fit luminosities. In Table~\ref{tableA:logL}, we calculate $-\ln\mathcal{L}$ for each model, verifying that the SNR+MSP model fits significantly better than other models.  These values cannot be directly compared with those in Table~\ref{table:logL} because the definitions of $P_i$ are different. In particular, while the 1$\sigma$ error in the uncertainty of each source is identical in both the linear and logarithmic constructions, the likelihood function for any other offset between the modeled and measured source flux will differ. 

In Table~\ref{tableA:bestfit}, we show the best-fit parameters, showing that they are also not significantly changed, and thus the main physical features of our model are robust to this choice.

\begin{table}[h]
\begin{center}
     \caption{Values of $-\ln\mathcal{L}$ for different models for the case when we use log-luminosity (Eq.~\ref{eqA:p}). \\} 
\begin{tabular}{c|c|c} \hline
    & All Sources & Low sSFR \\
    & ($N=3215$) & ($N=108$) \\
    \hline
    Scaling ($\psi$; Eq.~\ref{eq:s1}) & 2704 & 356.7  \\
    Scaling ($\psi$ and $M$; Eq.~\ref{eq:s2}) & 2193 & 193.3  \\ \hline
    Model (SNR only; Eq.~\ref{eq:m2}) & 2400 & 384.0 \\
    Model (SNR + MSP; Eq.~\ref{eq:m1}) & 2050 & 117.4  \\
\end{tabular}
     \label{tableA:logL}
\end{center}
\end{table}

\begin{table}[h]
\caption{\label{tableA:bestfit} Best-fit parameters when we use log-luminosity (Eq.~\ref{eqA:p}). \\}
\begin{ruledtabular}
\begin{tabular}{cccccc}
    & $\alpha$ & $\beta$ & $\gamma$ & $\sigma_{\log_{10} L}$ & \\ 
    Scaling($\psi$) & 0.108 & 0.973 & - & 0.561 & \\
    Scaling($\psi$ and $M$) & 0.127 & 0.665 & 0.530 & 0.479 & \\ \hline
     & $a_1$ & $a_2$ & $a_3$ & $\beta_{\rm syn}$ & $\sigma_{\log_{10} L}$\\ 
    Model (SNR only)& 0.119 & 1.06e-9 & - & 0.351 & 0.351 \\
    Model (SNR+MSP) & 0.031 & 0.046 & 0.026 & 0.199 & 0.458\\
\end{tabular}
\end{ruledtabular}
\end{table}

\section{The Effect of Removing Outliers}
\label{app:bright_sources}

In the main text, we removed from our analysis several outliers hat had radio luminosities that significantly exceeded model predictions. This is well justified, because other emission sources (e.g, AGN) or additional effects (e.g., galaxy interactions) may produce radio excesses that do not correlate with recent or historic star formation.

\begin{figure}[t]
\includegraphics[width=0.9\columnwidth]{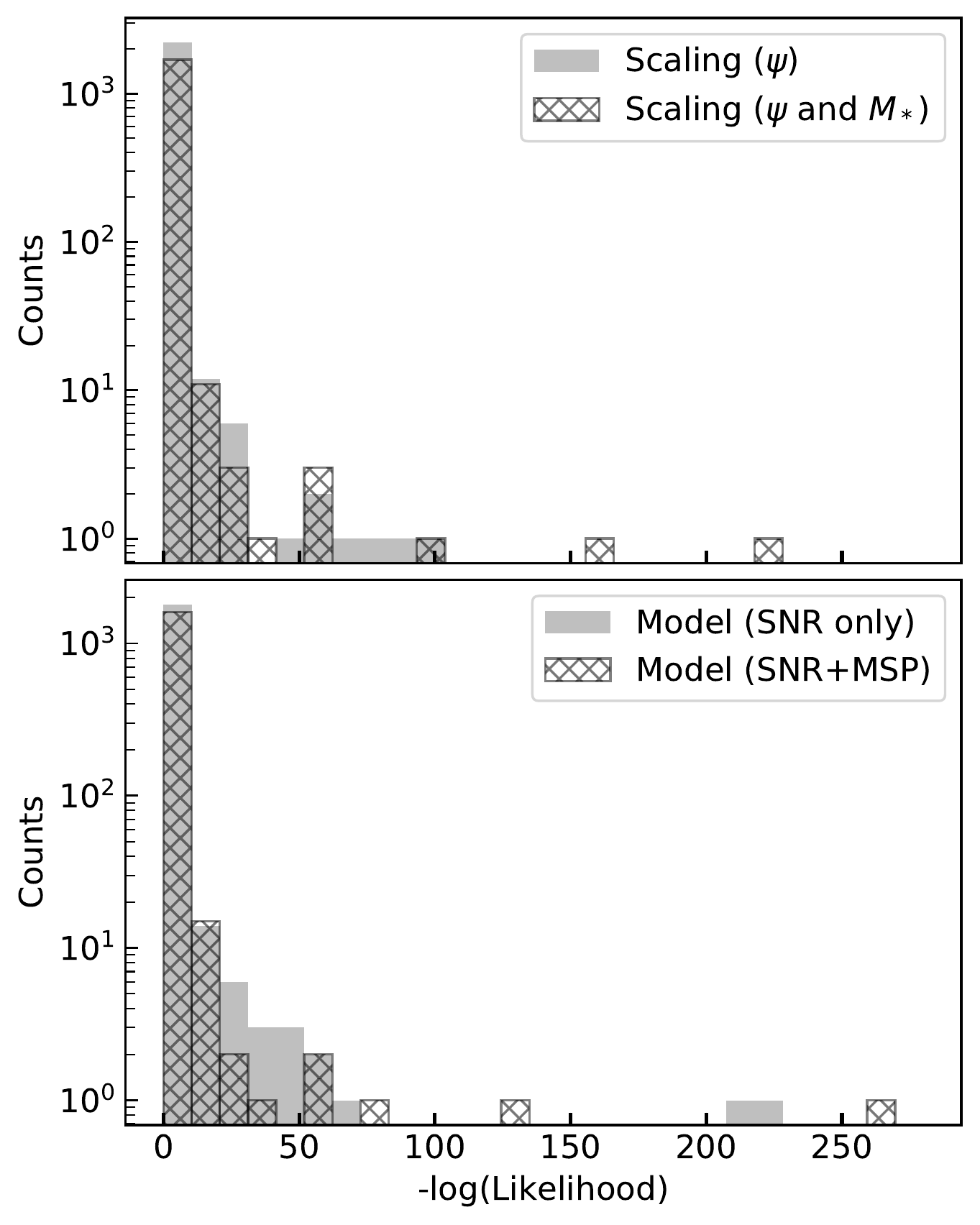}
\caption{The distribution of $-\ln{\mathcal{L}}$ for each SFG. For the scaling model with $\psi$ (top, gray), there is one source that has $-\ln{\mathcal{L}}=989$, which is not shown in this histogram.}
\label{figB:logL}
\end{figure}

In Table~\ref{tableB:logL}, we show the values of $-\ln\mathcal{L}$ for each model in a scenario where we do not discard these outliers. This confirms that the SNR+MSP models still provide the best fit. However, a comparison of these fits against those in Table~\ref{table:logL} indicates that our fits are highly affected by several very bright sources. In Fig.~\ref{figB:logL}, we show the distribution of the log-likelihood value for individual sources. While most of sources have $-\ln\mathcal{L}$ smaller than 10, some individual sources have $-\ln\mathcal{L}$ more than 50 or even 100. These sources dominate the sum of log-likelihood fit, which could potentially affect our results. 

Repeating our analysis, we have verified that our conclusions are unchanged if we set the upper limit for outlier removal to log-likliehood values of 100, 25, and 12.5. In all cases, the SNR+MSP model is favored over any other model by $2\Delta\ln\mathcal{L}>196$. The best-fit parameters remain largely unchanged.

\begin{table}[h]
\begin{center}
     \caption{Values of $-\ln\mathcal{L}$ for different models for the case when we include all 3097 sources in our analysis.\\} 
\begin{tabular}{c|c|c} \hline
    & All Sources & Low sSFR \\
    & ($N=3907$) & ($N=137$) \\
    \hline
    Scaling ($\psi$; Eq.~\ref{eq:s1}) & 2625.7 & 566.3 \\
    Scaling ($\psi$ and $M$; Eq.~\ref{eq:s2}) & -213.4 & 17.9 \\ \hline
    Model (SNR only; Eq.~\ref{eq:m2}) & 375.2 & 312.4 \\
    Model (SNR + MSP; Eq.~\ref{eq:m1}) & -580.6 &  -32.0 \\
\end{tabular}
     \label{tableB:logL}
\end{center}
\end{table}

\section{The Effect of SFR Modeling}
\label{app:sfr}

Here, we present an alternative analysis produced by replacing SFRs and masses from G18 with those obtained in S18~\cite{Salim18}. S18 produces three separate catalogs for three different exposure times for UV imaging. While the shallowest catalog contains the largest dataset (about 90$\%$ of SDSS sources are contained), it can be inaccurate for quiescent and passive galaxies. On the other hand, the deepest catalog covers only a small field, and thus includes only $\sim$7$\%$ of SDSS sources. Therefore, we choose to use the catalog of medium exposure time, which can be used for off-main-sequence galaxies and contains about 50$\%$ of SDSS sources. We utilize SFRs and masses from this catalog, but continue to utilize the radio luminosities and galaxy classifications determined by G18.

\begin{figure}[b]
    \centering
    \includegraphics[width=0.9\columnwidth]{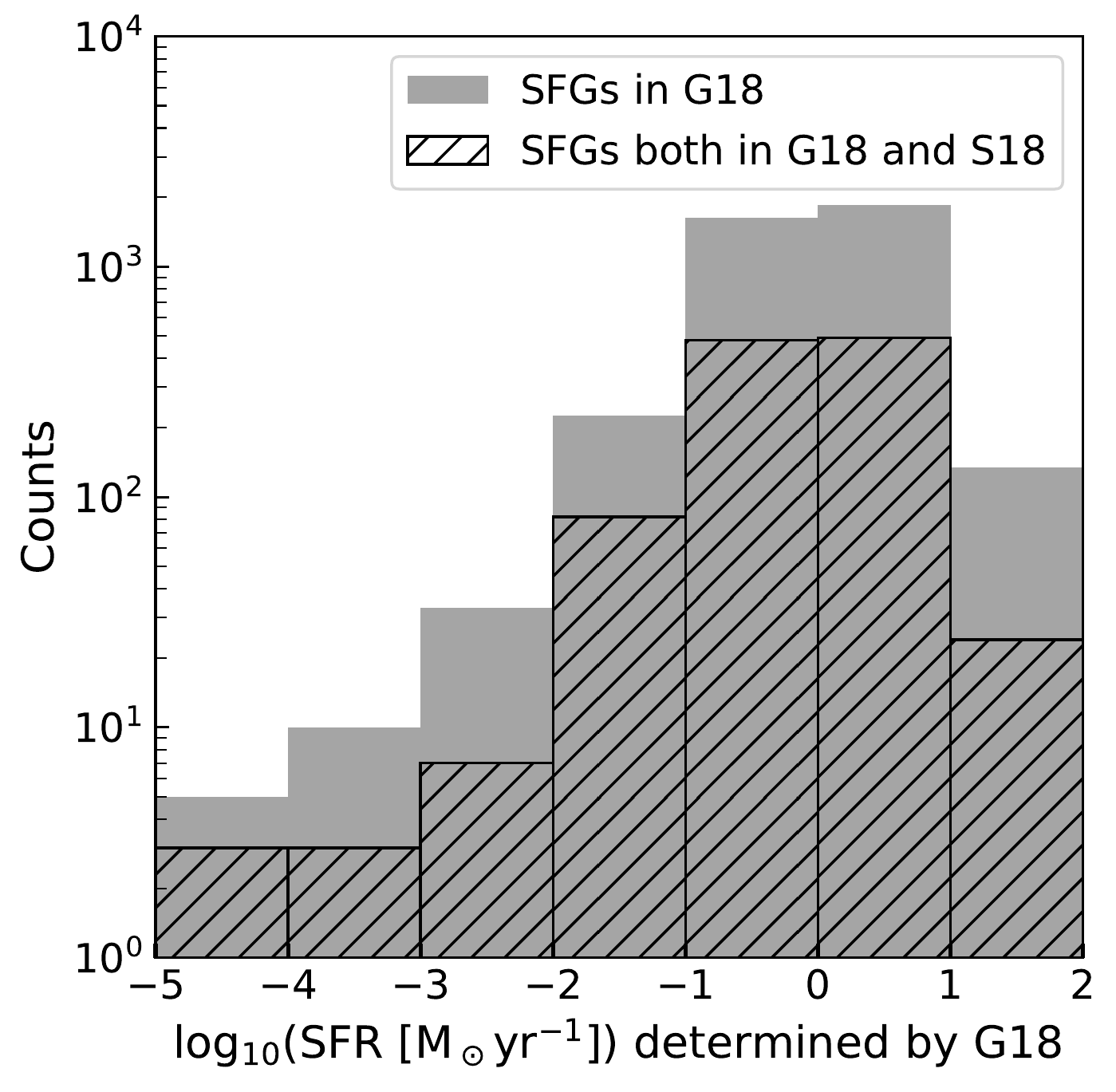}
    \caption{Histogram of SFGs that are included in G18 (gray) and both in G18 and S18 (black, hatched).}
    \label{fig:s18:hist}
\end{figure}

\begin{figure}
    \centering
    \includegraphics[width=0.9\columnwidth]{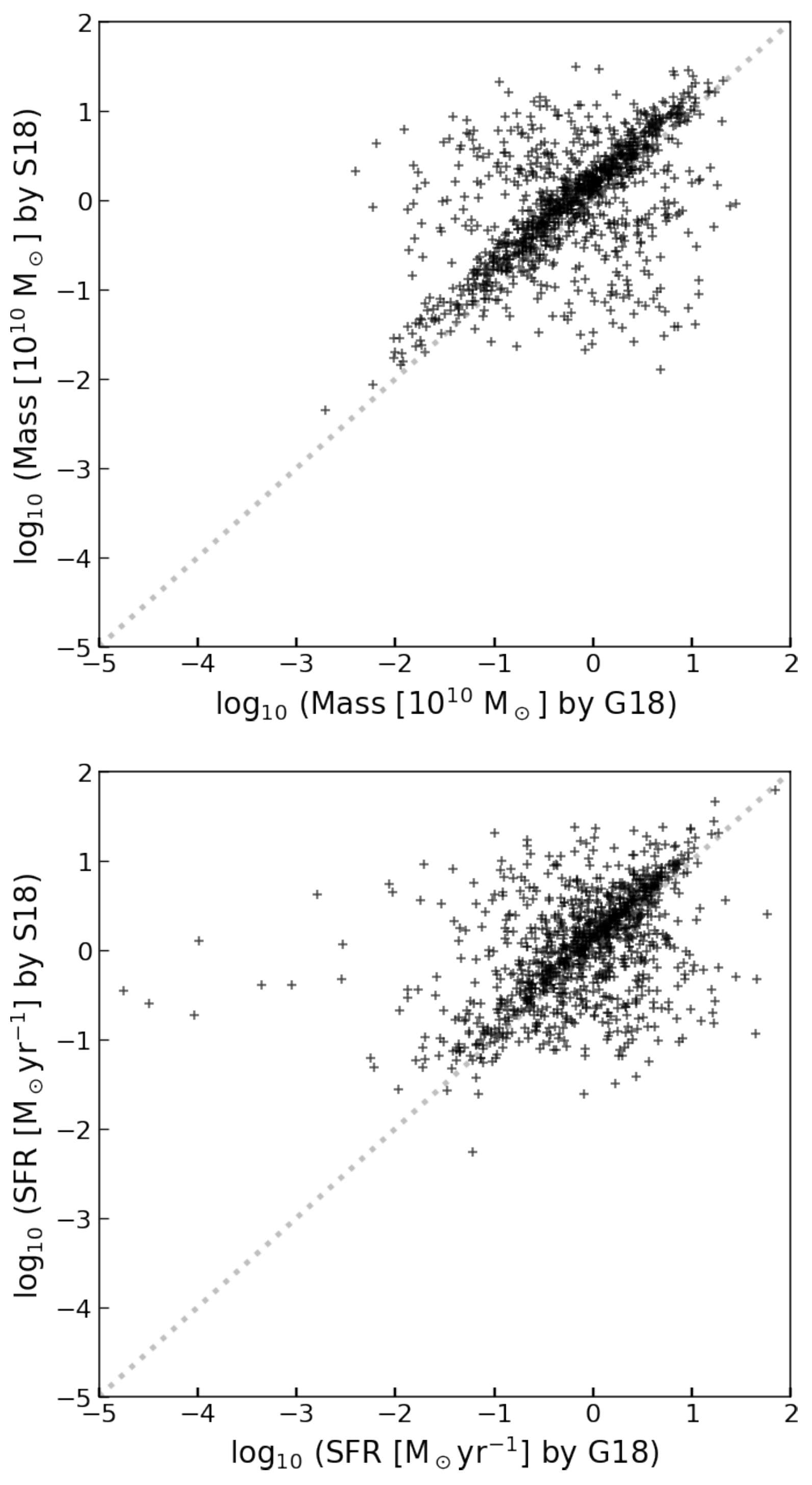}
    \caption{Comparison of stellar masses (top) and SFRs (bottom) determined by \citet{Gurkan18} (G18, x-axis) and \citet{Salim18} (S18, y-axis). Dashed line correspond to the case where these two estimates are identical.}
    \label{fig:s18:sfr}
\end{figure}

We utilize sources from the S18 catalog that are also contained in the study by G18. Since only 1094 out of 3907 SFGs in G18 are included in the S18 catalog, we have to check if this procedure does not induce any bias. Figure~\ref{fig:s18:hist} shows the histogram of sources binned using the SFRs determined by G18. This shows that the cross-correlated catalog is not significantly biased toward high SFR sources. However, we should keep in mind that more than half of low-SFR sources ($<10^{-2}$~M$_\odot$yr$^{-1}$) are not included in the S18 catalog.

Figure~\ref{fig:s18:sfr} compares the masses (top) and SFRs (bottom) determined by each catalog. We find that mass estimations are generally in good agreement, although there are orders of magnitude discrepancies for a small fraction of sources. For SFRs, we find that sources with small ($<10^{-2}$~M$_\odot$yr$^{-1}$) SFRs in G18 systematically have much higher SFRs in the model of S18. This is worrisome, because this suggests that the radio excess in low-SFR sources may be attributed to SFR measurement errors.

Figure~\ref{fig:s18:radio} (in the main text) compares the radio-SFR correlation for different SFR estimations. There are two notable changes. First, the main body of radio-SFR correlation (SFR$>10^{-1}$~M$_\odot$yr$^{-1}$) remain largely unchanged, but the scatter gets significantly larger. Due to this, our method of fitting the correlation with a linear-luminosity model is biased toward bright sources. Therefore, we fit the data using log-luminosity with the method detailed in Appendix~\ref{app:log_luminosity} to derive the slope of radio-SFR correlation. Restricting our analysis to a region with sSFR$>$10$^{-11}$~yr$^{-1}$, where the radio-SFR correlation should hold, we find a flatter slope for S18 SFRs, $L\propto\psi^{0.6}$, which is in significant tension with previous estimates of the radio-SFR correlation at low-frequencies~\cite{1988MNRAS.235.1227C}. 

Second, there are few low-SFR sources when we utilize SFRs from S18. This makes the excess feature in low-SFR sources is much less pronounced. As a result, our mass dependent model is not statistically preferred compared to the SFR-only scaling, contrary to what we observed for G18 SFRs.

However, we note that more than half of the low-SFR sources ($<10^{-2}$~M$_\odot$yr$^{-1}$) observd by G18 are not contained in S18. Therefore, to determine whether the excess feature can be robust against SFR estimates, we need deeper observations and a better determination of SFRs for the low-SFR sources that are not included in the medium- or deep- catalog by S18. 

\section{The Inclusion of Unclassified Sources}
\label{app:unclass}

In the main text, we used 3907 sources that are classified by G18 as SFGs using a BPT-diagram. There are 6370 sources that are not classified due to the absence or weak detection ($<3\sigma$) of emission lines. Although these ``unclassified" sources are not used in the main text, they necessarily include many high-mass and low-SFR sources, which are important for testing our model.

Here, we check whether our model is consistent with LOFAR observations when we include unclassified sources. This analysis should be taken with caution, because there can be sources that are affected by AGN. To avoid biasing our results with the  brightest sources that might be strongly affected by AGN, we fit the data using log-luminosity following the method in Appendix~\ref{app:log_luminosity}.

We find that, if we use the SFRs and masses determined by G18, our SNR+MSP model is preferred over the SNR-only model by $\Delta\rm{LG}(\mathcal{L})$ = 3480. If we replace the SFR and mass determinations by those in S18, the SNR+MSP model is still preferred  by $\Delta\rm{LG}(\mathcal{L})$ = 746. In this cases, and restricting ourselves to sources with \mbox{sSFR$>$10$^{-11}$~yr$^{-1}$} we find a slightly harder value for the radio-SFR correlation, fitting $L\propto\psi^{0.8}$, which is somewhat more consistent with the value obtained in the main text.

\begin{figure*}[h]
 \begin{minipage}[b]{\columnwidth}
  \centering
  \includegraphics[keepaspectratio, scale=0.5]
  {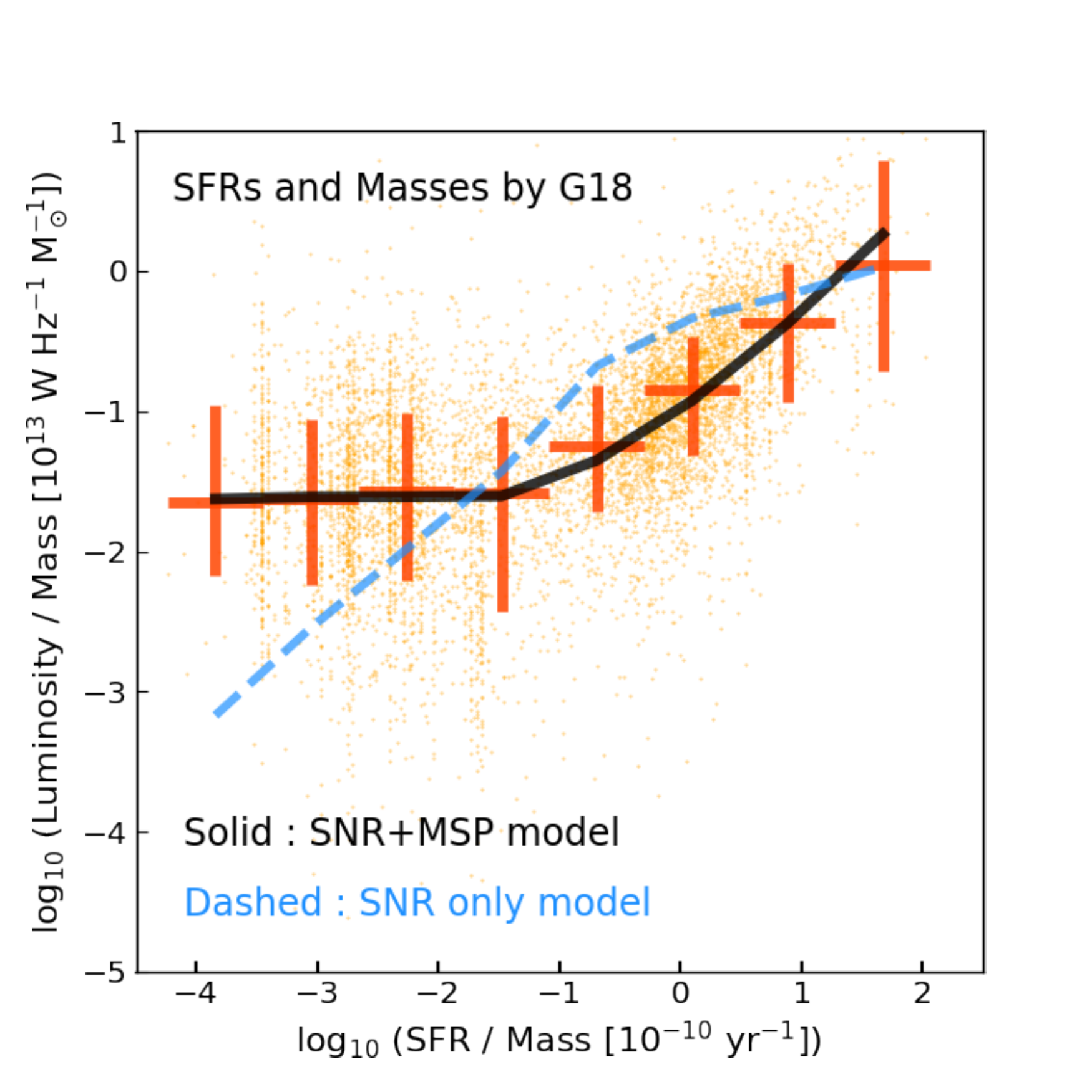}
 \end{minipage}
 \begin{minipage}[b]{\columnwidth}
  \centering
  \includegraphics[keepaspectratio, scale=0.5]
  {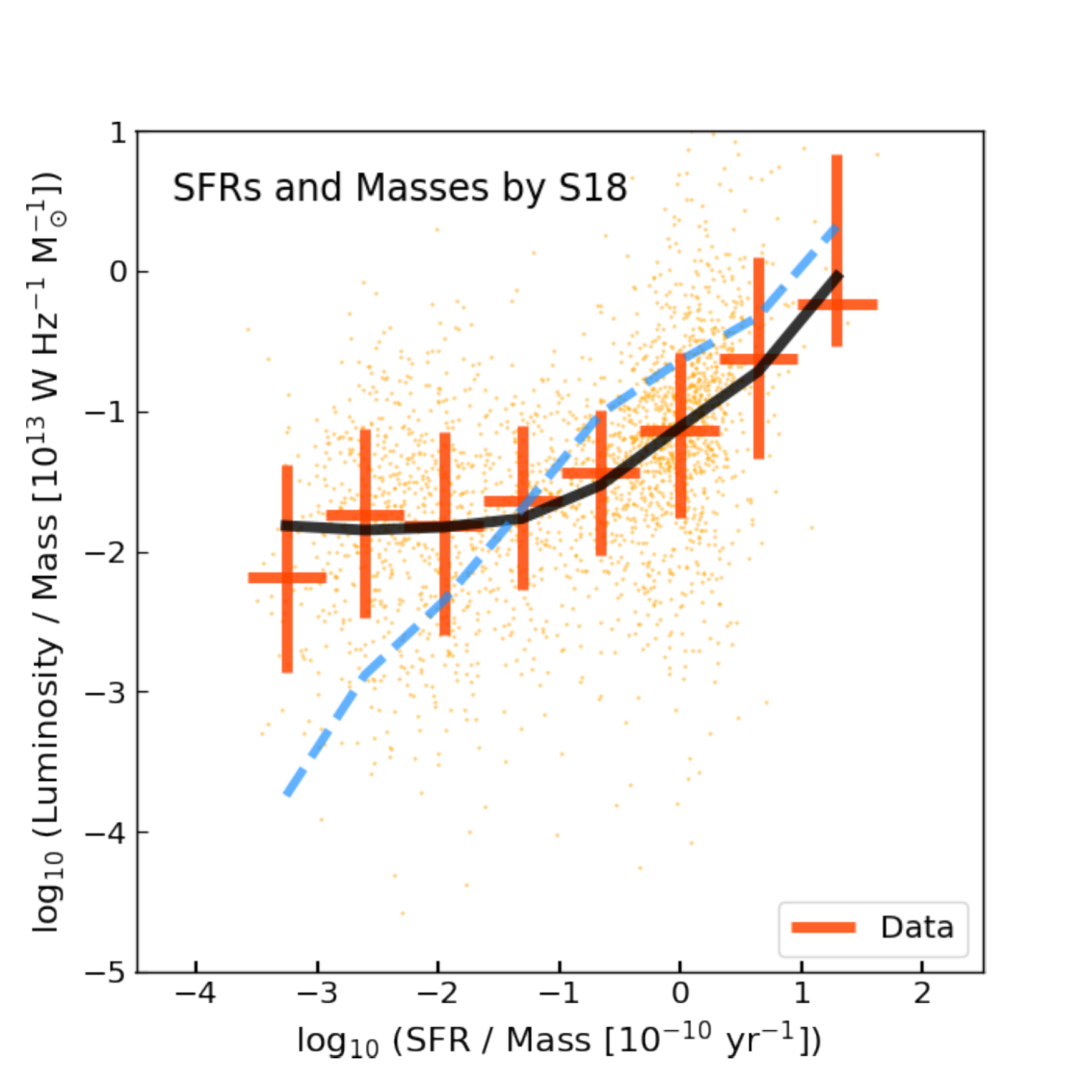}
 \end{minipage}
 \caption{The radio luminosity - SFR correlation scaled by the stellar masses for SFGs and unclassified sources, with galactic parameters determined by G18 (left) and S18 (right). Along with data points for individual galaxies (orange), we show the median and the 16th - 84th percentile range (bars). Lines show theoretical predictions, connecting the median value in each specific-SFR bin that we show with the bars. Theoretically predicted values for individual galaxies are not shown.}\label{fig:app:unclass}
\end{figure*}

Figure~\ref{fig:app:unclass} shows the scaled luminosities vs specific SFRs for two different galactic parameters derived by G18 (left) and S18 (right). In both datasets, we can see a pleateau feature for low specific SFR sources, which is consistent with original findings by G18. This figure clearly illustrates that MSP-based model is significantly favored.

As noted in the main text, this agreement does not prove that MSPs produce the mass-dependent radio emission. In particular, for unclassified sources, we need more careful examination of the contributions from AGN activities. However, it is encouraging that we do see a feature that is expected for MSPs, and the derived parameters are consistent with this interpretation.

\section{Simulation to Test Our Approach}
\label{app:mcsimulation}
We use a Monte-Carlo simulation to produce mock observed data. First, we generate mock galaxies with randomly assigned SFRs ($\psi$) and masses ($M_*$). We assume log-normal distributions for both quantities, with mean and standard deviation obtained from observed data. The number of galaxies is identical to the sample size used in the main analysis. Then, we calculate luminosities for these galaxies, $L^{\rm model}$, with our model equations. Next, we assign measurement errors of luminosities ($L_{\rm err}$) to the simulated galaxies. Observed data show that errors are typically described by $L_{\rm err} \sim 0.1 \sqrt{L}$ (units are in 10$^{23}$ W/Hz) though with large scatter. Here we simply assume $L_{\rm err} = 0.1 \sqrt{L^{\rm model}}$ for all sources. We also assign model errors, denoted as $x$. This is a random factor drawn from a Gaussian distribution with zero mean and standard deviation of $c$. Combining these two error terms, the observed luminosity is calculated by $L = L^{\rm model} + y L_{\rm err} + x L^{\rm model}$, where $y$ is a random variable drawn from Gaussian with mean 0 and standard deviation 1. We run the analysis used in the main text and verified that it can recover injected parameters. If we test a model where we inject a radio luminosity that correlates with only the SNR-term, our analysis shows no statistical preference for the MSP model, which further supports the validity of our analysis.

\color{black}

\clearpage

\bibliography{radio_fir}

\begin{thebibliography}{153}%
\makeatletter
\providecommand \@ifxundefined [1]{%
 \@ifx{#1\undefined}
}%
\providecommand \@ifnum [1]{%
 \ifnum #1\expandafter \@firstoftwo
 \else \expandafter \@secondoftwo
 \fi
}%
\providecommand \@ifx [1]{%
 \ifx #1\expandafter \@firstoftwo
 \else \expandafter \@secondoftwo
 \fi
}%
\providecommand \natexlab [1]{#1}%
\providecommand \enquote  [1]{``#1''}%
\providecommand \bibnamefont  [1]{#1}%
\providecommand \bibfnamefont [1]{#1}%
\providecommand \citenamefont [1]{#1}%
\providecommand \href@noop [0]{\@secondoftwo}%
\providecommand \href [0]{\begingroup \@sanitize@url \@href}%
\providecommand \@href[1]{\@@startlink{#1}\@@href}%
\providecommand \@@href[1]{\endgroup#1\@@endlink}%
\providecommand \@sanitize@url [0]{\catcode `\\12\catcode `\$12\catcode
  `\&12\catcode `\#12\catcode `\^12\catcode `\_12\catcode `\%12\relax}%
\providecommand \@@startlink[1]{}%
\providecommand \@@endlink[0]{}%
\providecommand \url  [0]{\begingroup\@sanitize@url \@url }%
\providecommand \@url [1]{\endgroup\@href {#1}{\urlprefix }}%
\providecommand \urlprefix  [0]{URL }%
\providecommand \Eprint [0]{\href }%
\providecommand \doibase [0]{http://dx.doi.org/}%
\providecommand \selectlanguage [0]{\@gobble}%
\providecommand \bibinfo  [0]{\@secondoftwo}%
\providecommand \bibfield  [0]{\@secondoftwo}%
\providecommand \translation [1]{[#1]}%
\providecommand \BibitemOpen [0]{}%
\providecommand \bibitemStop [0]{}%
\providecommand \bibitemNoStop [0]{.\EOS\space}%
\providecommand \EOS [0]{\spacefactor3000\relax}%
\providecommand \BibitemShut  [1]{\csname bibitem#1\endcsname}%
\let\auto@bib@innerbib\@empty
\bibitem [{\citenamefont {{van der Kruit}}(1973{\natexlab{a}})}]{irc1973a}%
  \BibitemOpen
  \bibfield  {author} {\bibinfo {author} {\bibfnamefont {P.~C.}\ \bibnamefont
  {{van der Kruit}}},\ }\href@noop {} {\bibfield  {journal} {\bibinfo
  {journal} {\aap}\ }\textbf {\bibinfo {volume} {29}},\ \bibinfo {pages} {231}
  (\bibinfo {year} {1973}{\natexlab{a}})}\BibitemShut {NoStop}%
\bibitem [{\citenamefont {{van der Kruit}}(1973{\natexlab{b}})}]{irc1973b}%
  \BibitemOpen
  \bibfield  {author} {\bibinfo {author} {\bibfnamefont {P.~C.}\ \bibnamefont
  {{van der Kruit}}},\ }\href@noop {} {\bibfield  {journal} {\bibinfo
  {journal} {\aap}\ }\textbf {\bibinfo {volume} {29}},\ \bibinfo {pages} {263}
  (\bibinfo {year} {1973}{\natexlab{b}})}\BibitemShut {NoStop}%
\bibitem [{\citenamefont {{Harwit}}\ and\ \citenamefont
  {{Pacini}}(1975)}]{1975ApJ...200L.127H}%
  \BibitemOpen
  \bibfield  {author} {\bibinfo {author} {\bibfnamefont {M.}~\bibnamefont
  {{Harwit}}}\ and\ \bibinfo {author} {\bibfnamefont {F.}~\bibnamefont
  {{Pacini}}},\ }\href {\doibase 10.1086/181913} {\bibfield  {journal}
  {\bibinfo  {journal} {\apjl}\ }\textbf {\bibinfo {volume} {200}},\ \bibinfo
  {pages} {L127} (\bibinfo {year} {1975})}\BibitemShut {NoStop}%
\bibitem [{\citenamefont {{Dickey}}\ and\ \citenamefont
  {{Salpeter}}(1984)}]{irc1984}%
  \BibitemOpen
  \bibfield  {author} {\bibinfo {author} {\bibfnamefont {J.~M.}\ \bibnamefont
  {{Dickey}}}\ and\ \bibinfo {author} {\bibfnamefont {E.~E.}\ \bibnamefont
  {{Salpeter}}},\ }\href {\doibase 10.1086/162428} {\bibfield  {journal}
  {\bibinfo  {journal} {\apj}\ }\textbf {\bibinfo {volume} {284}},\ \bibinfo
  {pages} {461} (\bibinfo {year} {1984})}\BibitemShut {NoStop}%
\bibitem [{\citenamefont {{Rickard}}\ and\ \citenamefont
  {{Harvey}}(1984)}]{1984AJ.....89.1520R}%
  \BibitemOpen
  \bibfield  {author} {\bibinfo {author} {\bibfnamefont {L.~J.}\ \bibnamefont
  {{Rickard}}}\ and\ \bibinfo {author} {\bibfnamefont {P.~M.}\ \bibnamefont
  {{Harvey}}},\ }\href {\doibase 10.1086/113652} {\bibfield  {journal}
  {\bibinfo  {journal} {\aj}\ }\textbf {\bibinfo {volume} {89}},\ \bibinfo
  {pages} {1520} (\bibinfo {year} {1984})}\BibitemShut {NoStop}%
\bibitem [{\citenamefont {{Helou}}\ \emph {et~al.}(1985)\citenamefont
  {{Helou}}, \citenamefont {{Soifer}},\ and\ \citenamefont
  {{Rowan-Robinson}}}]{irc1985}%
  \BibitemOpen
  \bibfield  {author} {\bibinfo {author} {\bibfnamefont {G.}~\bibnamefont
  {{Helou}}}, \bibinfo {author} {\bibfnamefont {B.~T.}\ \bibnamefont
  {{Soifer}}}, \ and\ \bibinfo {author} {\bibfnamefont {M.}~\bibnamefont
  {{Rowan-Robinson}}},\ }\href {\doibase 10.1086/184556} {\bibfield  {journal}
  {\bibinfo  {journal} {\apjl}\ }\textbf {\bibinfo {volume} {298}},\ \bibinfo
  {pages} {L7} (\bibinfo {year} {1985})}\BibitemShut {NoStop}%
\bibitem [{\citenamefont {{de Jong}}\ \emph {et~al.}(1985)\citenamefont {{de
  Jong}}, \citenamefont {{Klein}}, \citenamefont {{Wielebinski}},\ and\
  \citenamefont {{Wunderlich}}}]{1985A&A...147L...6D}%
  \BibitemOpen
  \bibfield  {author} {\bibinfo {author} {\bibfnamefont {T.}~\bibnamefont {{de
  Jong}}}, \bibinfo {author} {\bibfnamefont {U.}~\bibnamefont {{Klein}}},
  \bibinfo {author} {\bibfnamefont {R.}~\bibnamefont {{Wielebinski}}}, \ and\
  \bibinfo {author} {\bibfnamefont {E.}~\bibnamefont {{Wunderlich}}},\
  }\href@noop {} {\bibfield  {journal} {\bibinfo  {journal} {\aap}\ }\textbf
  {\bibinfo {volume} {147}},\ \bibinfo {pages} {L6} (\bibinfo {year}
  {1985})}\BibitemShut {NoStop}%
\bibitem [{\citenamefont {{Hummel}}\ \emph {et~al.}(1988)\citenamefont
  {{Hummel}}, \citenamefont {{Davies}}, \citenamefont {{Wolstencroft}},
  \citenamefont {{van der Hulst}},\ and\ \citenamefont
  {{Pedlar}}}]{1988A&A...199...91H}%
  \BibitemOpen
  \bibfield  {author} {\bibinfo {author} {\bibfnamefont {E.}~\bibnamefont
  {{Hummel}}}, \bibinfo {author} {\bibfnamefont {R.~D.}\ \bibnamefont
  {{Davies}}}, \bibinfo {author} {\bibfnamefont {R.~D.}\ \bibnamefont
  {{Wolstencroft}}}, \bibinfo {author} {\bibfnamefont {J.~M.}\ \bibnamefont
  {{van der Hulst}}}, \ and\ \bibinfo {author} {\bibfnamefont {A.}~\bibnamefont
  {{Pedlar}}},\ }\href@noop {} {\bibfield  {journal} {\bibinfo  {journal}
  {\aap}\ }\textbf {\bibinfo {volume} {199}},\ \bibinfo {pages} {91} (\bibinfo
  {year} {1988})}\BibitemShut {NoStop}%
\bibitem [{\citenamefont {{Condon}}(1992)}]{condon92}%
  \BibitemOpen
  \bibfield  {author} {\bibinfo {author} {\bibfnamefont {J.~J.}\ \bibnamefont
  {{Condon}}},\ }\href {\doibase 10.1146/annurev.aa.30.090192.003043}
  {\bibfield  {journal} {\bibinfo  {journal} {\araa}\ }\textbf {\bibinfo
  {volume} {30}},\ \bibinfo {pages} {575} (\bibinfo {year} {1992})}\BibitemShut
  {NoStop}%
\bibitem [{\citenamefont {{Yun}}\ \emph {et~al.}(2001)\citenamefont {{Yun}},
  \citenamefont {{Reddy}},\ and\ \citenamefont {{Condon}}}]{irc2001}%
  \BibitemOpen
  \bibfield  {author} {\bibinfo {author} {\bibfnamefont {M.~S.}\ \bibnamefont
  {{Yun}}}, \bibinfo {author} {\bibfnamefont {N.~A.}\ \bibnamefont {{Reddy}}},
  \ and\ \bibinfo {author} {\bibfnamefont {J.~J.}\ \bibnamefont {{Condon}}},\
  }\href {\doibase 10.1086/323145} {\bibfield  {journal} {\bibinfo  {journal}
  {\apj}\ }\textbf {\bibinfo {volume} {554}},\ \bibinfo {pages} {803} (\bibinfo
  {year} {2001})},\ \Eprint {http://arxiv.org/abs/astro-ph/0102154}
  {arXiv:astro-ph/0102154 [astro-ph]} \BibitemShut {NoStop}%
\bibitem [{\citenamefont {{Appleton}}\ \emph {et~al.}(2004)\citenamefont
  {{Appleton}} \emph {et~al.}}]{irc2004}%
  \BibitemOpen
  \bibfield  {author} {\bibinfo {author} {\bibfnamefont {P.~N.}\ \bibnamefont
  {{Appleton}}} \emph {et~al.},\ }\href {\doibase 10.1086/422425} {\bibfield
  {journal} {\bibinfo  {journal} {\apjs}\ }\textbf {\bibinfo {volume} {154}},\
  \bibinfo {pages} {147} (\bibinfo {year} {2004})},\ \Eprint
  {http://arxiv.org/abs/astro-ph/0406030} {arXiv:astro-ph/0406030 [astro-ph]}
  \BibitemShut {NoStop}%
\bibitem [{\citenamefont {{Jarvis}}\ \emph {et~al.}(2010)\citenamefont
  {{Jarvis}} \emph {et~al.}}]{irc2010}%
  \BibitemOpen
  \bibfield  {author} {\bibinfo {author} {\bibfnamefont {M.~J.}\ \bibnamefont
  {{Jarvis}}} \emph {et~al.},\ }\href {\doibase
  10.1111/j.1365-2966.2010.17772.x} {\bibfield  {journal} {\bibinfo  {journal}
  {\mnras}\ }\textbf {\bibinfo {volume} {409}},\ \bibinfo {pages} {92}
  (\bibinfo {year} {2010})},\ \Eprint {http://arxiv.org/abs/1009.5390}
  {arXiv:1009.5390 [astro-ph.CO]} \BibitemShut {NoStop}%
\bibitem [{\citenamefont {{Magnelli}}\ \emph {et~al.}(2015)\citenamefont
  {{Magnelli}} \emph {et~al.}}]{irc15M}%
  \BibitemOpen
  \bibfield  {author} {\bibinfo {author} {\bibfnamefont {B.}~\bibnamefont
  {{Magnelli}}} \emph {et~al.},\ }\href {\doibase 10.1051/0004-6361/201424937}
  {\bibfield  {journal} {\bibinfo  {journal} {\aap}\ }\textbf {\bibinfo
  {volume} {573}},\ \bibinfo {eid} {A45} (\bibinfo {year} {2015})},\ \Eprint
  {http://arxiv.org/abs/1410.7412} {arXiv:1410.7412 [astro-ph.GA]} \BibitemShut
  {NoStop}%
\bibitem [{\citenamefont {{Qiu}}\ \emph {et~al.}(2017)\citenamefont {{Qiu}},
  \citenamefont {{Shi}}, \citenamefont {{Wang}}, \citenamefont {{Zhang}},\ and\
  \citenamefont {{Zhou}}}]{irc17Q}%
  \BibitemOpen
  \bibfield  {author} {\bibinfo {author} {\bibfnamefont {J.}~\bibnamefont
  {{Qiu}}}, \bibinfo {author} {\bibfnamefont {Y.}~\bibnamefont {{Shi}}},
  \bibinfo {author} {\bibfnamefont {J.}~\bibnamefont {{Wang}}}, \bibinfo
  {author} {\bibfnamefont {Z.-Y.}\ \bibnamefont {{Zhang}}}, \ and\ \bibinfo
  {author} {\bibfnamefont {L.}~\bibnamefont {{Zhou}}},\ }\href {\doibase
  10.3847/1538-4357/aa832c} {\bibfield  {journal} {\bibinfo  {journal} {\apj}\
  }\textbf {\bibinfo {volume} {846}},\ \bibinfo {eid} {68} (\bibinfo {year}
  {2017})},\ \Eprint {http://arxiv.org/abs/1708.02687} {arXiv:1708.02687
  [astro-ph.GA]} \BibitemShut {NoStop}%
\bibitem [{\citenamefont {{Tabatabaei}}\ \emph {et~al.}(2017)\citenamefont
  {{Tabatabaei}} \emph {et~al.}}]{irc17T}%
  \BibitemOpen
  \bibfield  {author} {\bibinfo {author} {\bibfnamefont {F.~S.}\ \bibnamefont
  {{Tabatabaei}}} \emph {et~al.},\ }\href {\doibase
  10.3847/1538-4357/836/2/185} {\bibfield  {journal} {\bibinfo  {journal}
  {\apj}\ }\textbf {\bibinfo {volume} {836}},\ \bibinfo {eid} {185} (\bibinfo
  {year} {2017})},\ \Eprint {http://arxiv.org/abs/1611.01705} {arXiv:1611.01705
  [astro-ph.GA]} \BibitemShut {NoStop}%
\bibitem [{\citenamefont {{Shao}}\ \emph {et~al.}(2018)\citenamefont {{Shao}},
  \citenamefont {{Koribalski}}, \citenamefont {{Wang}}, \citenamefont {{Ho}},\
  and\ \citenamefont {{Staveley-Smith}}}]{irc18S}%
  \BibitemOpen
  \bibfield  {author} {\bibinfo {author} {\bibfnamefont {L.}~\bibnamefont
  {{Shao}}}, \bibinfo {author} {\bibfnamefont {B.~S.}\ \bibnamefont
  {{Koribalski}}}, \bibinfo {author} {\bibfnamefont {J.}~\bibnamefont
  {{Wang}}}, \bibinfo {author} {\bibfnamefont {L.~C.}\ \bibnamefont {{Ho}}}, \
  and\ \bibinfo {author} {\bibfnamefont {L.}~\bibnamefont {{Staveley-Smith}}},\
  }\href {\doibase 10.1093/mnras/sty1608} {\bibfield  {journal} {\bibinfo
  {journal} {\mnras}\ }\textbf {\bibinfo {volume} {479}},\ \bibinfo {pages}
  {3509} (\bibinfo {year} {2018})},\ \Eprint {http://arxiv.org/abs/1806.05447}
  {arXiv:1806.05447 [astro-ph.GA]} \BibitemShut {NoStop}%
\bibitem [{\citenamefont {{Read}}\ \emph {et~al.}(2018)\citenamefont {{Read}}
  \emph {et~al.}}]{Read18}%
  \BibitemOpen
  \bibfield  {author} {\bibinfo {author} {\bibfnamefont {S.~C.}\ \bibnamefont
  {{Read}}} \emph {et~al.},\ }\href {\doibase 10.1093/mnras/sty2198} {\bibfield
   {journal} {\bibinfo  {journal} {\mnras}\ }\textbf {\bibinfo {volume}
  {480}},\ \bibinfo {pages} {5625} (\bibinfo {year} {2018})},\ \Eprint
  {http://arxiv.org/abs/1808.10452} {arXiv:1808.10452 [astro-ph.GA]}
  \BibitemShut {NoStop}%
\bibitem [{\citenamefont {{Filho}}\ \emph {et~al.}(2019)\citenamefont
  {{Filho}}, \citenamefont {{Tabatabaei}}, \citenamefont {{S{\'a}nchez
  Almeida}}, \citenamefont {{Mu{\~n}oz-Tu{\~n}{\'o}n}},\ and\ \citenamefont
  {{Elmegreen}}}]{irc19F}%
  \BibitemOpen
  \bibfield  {author} {\bibinfo {author} {\bibfnamefont {M.~E.}\ \bibnamefont
  {{Filho}}}, \bibinfo {author} {\bibfnamefont {F.~S.}\ \bibnamefont
  {{Tabatabaei}}}, \bibinfo {author} {\bibfnamefont {J.}~\bibnamefont
  {{S{\'a}nchez Almeida}}}, \bibinfo {author} {\bibfnamefont {C.}~\bibnamefont
  {{Mu{\~n}oz-Tu{\~n}{\'o}n}}}, \ and\ \bibinfo {author} {\bibfnamefont
  {B.~G.}\ \bibnamefont {{Elmegreen}}},\ }\href {\doibase
  10.1093/mnras/sty3199} {\bibfield  {journal} {\bibinfo  {journal} {\mnras}\
  }\textbf {\bibinfo {volume} {484}},\ \bibinfo {pages} {543} (\bibinfo {year}
  {2019})},\ \Eprint {http://arxiv.org/abs/1811.06577} {arXiv:1811.06577
  [astro-ph.GA]} \BibitemShut {NoStop}%
\bibitem [{\citenamefont {{Solarz}}\ \emph {et~al.}(2019)\citenamefont
  {{Solarz}}, \citenamefont {{Pollo}}, \citenamefont {{Bilicki}}, \citenamefont
  {{Pe{\`I}{\textsection}piak}}, \citenamefont {{Takeuchi}},\ and\
  \citenamefont {{Pia{\`I}{\textsection}tek}}}]{irc19S}%
  \BibitemOpen
  \bibfield  {author} {\bibinfo {author} {\bibfnamefont {A.}~\bibnamefont
  {{Solarz}}}, \bibinfo {author} {\bibfnamefont {A.}~\bibnamefont {{Pollo}}},
  \bibinfo {author} {\bibfnamefont {M.}~\bibnamefont {{Bilicki}}}, \bibinfo
  {author} {\bibfnamefont {A.}~\bibnamefont {{Pe{\`I}{\textsection}piak}}},
  \bibinfo {author} {\bibfnamefont {T.~T.}\ \bibnamefont {{Takeuchi}}}, \ and\
  \bibinfo {author} {\bibfnamefont {P.}~\bibnamefont
  {{Pia{\`I}{\textsection}tek}}},\ }\href {\doibase 10.1093/pasj/psz013}
  {\bibfield  {journal} {\bibinfo  {journal} {\pasj}\ }\textbf {\bibinfo
  {volume} {71}},\ \bibinfo {eid} {28} (\bibinfo {year} {2019})},\ \Eprint
  {http://arxiv.org/abs/1901.10410} {arXiv:1901.10410 [astro-ph.GA]}
  \BibitemShut {NoStop}%
\bibitem [{\citenamefont {Ackermann}\ \emph {et~al.}(2012)\citenamefont
  {Ackermann} \emph {et~al.}}]{Ackermann:2012vca}%
  \BibitemOpen
  \bibfield  {author} {\bibinfo {author} {\bibfnamefont {M.}~\bibnamefont
  {Ackermann}} \emph {et~al.} (\bibinfo {collaboration} {Fermi-LAT}),\ }\href
  {\doibase 10.1088/0004-637X/755/2/164} {\bibfield  {journal} {\bibinfo
  {journal} {Astrophys. J.}\ }\textbf {\bibinfo {volume} {755}},\ \bibinfo
  {pages} {164} (\bibinfo {year} {2012})},\ \Eprint
  {http://arxiv.org/abs/1206.1346} {arXiv:1206.1346 [astro-ph.HE]} \BibitemShut
  {NoStop}%
\bibitem [{\citenamefont {Linden}(2017)}]{Linden:2016}%
  \BibitemOpen
  \bibfield  {author} {\bibinfo {author} {\bibfnamefont {T.}~\bibnamefont
  {Linden}},\ }\href {\doibase 10.1103/PhysRevD.96.083001} {\bibfield
  {journal} {\bibinfo  {journal} {Phys. Rev.}\ }\textbf {\bibinfo {volume}
  {D96}},\ \bibinfo {pages} {083001} (\bibinfo {year} {2017})},\ \Eprint
  {http://arxiv.org/abs/1612.03175} {arXiv:1612.03175 [astro-ph.HE]}
  \BibitemShut {NoStop}%
\bibitem [{\citenamefont {{Ajello}}\ \emph {et~al.}(2020)\citenamefont
  {{Ajello}}, \citenamefont {{Di Mauro}}, \citenamefont {{Paliya}},\ and\
  \citenamefont {{Garrappa}}}]{Ajello:2020zna}%
  \BibitemOpen
  \bibfield  {author} {\bibinfo {author} {\bibfnamefont {M.}~\bibnamefont
  {{Ajello}}}, \bibinfo {author} {\bibfnamefont {M.}~\bibnamefont {{Di
  Mauro}}}, \bibinfo {author} {\bibfnamefont {V.~S.}\ \bibnamefont {{Paliya}}},
  \ and\ \bibinfo {author} {\bibfnamefont {S.}~\bibnamefont {{Garrappa}}},\
  }\href {\doibase 10.3847/1538-4357/ab86a6} {\bibfield  {journal} {\bibinfo
  {journal} {\apj}\ }\textbf {\bibinfo {volume} {894}},\ \bibinfo {eid} {88}
  (\bibinfo {year} {2020})},\ \Eprint {http://arxiv.org/abs/2003.05493}
  {arXiv:2003.05493 [astro-ph.GA]} \BibitemShut {NoStop}%
\bibitem [{\citenamefont {Murphy}\ \emph {et~al.}(2009)\citenamefont {Murphy},
  \citenamefont {Kenney}, \citenamefont {Helou}, \citenamefont {Chung},\ and\
  \citenamefont {Howell}}]{Murphy:2008dq}%
  \BibitemOpen
  \bibfield  {author} {\bibinfo {author} {\bibfnamefont {E.~J.}\ \bibnamefont
  {Murphy}}, \bibinfo {author} {\bibfnamefont {J.~D.~P.}\ \bibnamefont
  {Kenney}}, \bibinfo {author} {\bibfnamefont {G.}~\bibnamefont {Helou}},
  \bibinfo {author} {\bibfnamefont {A.}~\bibnamefont {Chung}}, \ and\ \bibinfo
  {author} {\bibfnamefont {J.~H.}\ \bibnamefont {Howell}},\ }\href {\doibase
  10.1088/0004-637X/694/2/1435} {\bibfield  {journal} {\bibinfo  {journal}
  {Astrophys. J.}\ }\textbf {\bibinfo {volume} {694}},\ \bibinfo {pages} {1435}
  (\bibinfo {year} {2009})},\ \Eprint {http://arxiv.org/abs/0812.2922}
  {arXiv:0812.2922 [astro-ph]} \BibitemShut {NoStop}%
\bibitem [{\citenamefont {{Condon}}\ \emph {et~al.}(2002)\citenamefont
  {{Condon}}, \citenamefont {{Cotton}},\ and\ \citenamefont
  {{Broderick}}}]{2002AJ....124..675C}%
  \BibitemOpen
  \bibfield  {author} {\bibinfo {author} {\bibfnamefont {J.~J.}\ \bibnamefont
  {{Condon}}}, \bibinfo {author} {\bibfnamefont {W.~D.}\ \bibnamefont
  {{Cotton}}}, \ and\ \bibinfo {author} {\bibfnamefont {J.~J.}\ \bibnamefont
  {{Broderick}}},\ }\href {\doibase 10.1086/341650} {\bibfield  {journal}
  {\bibinfo  {journal} {\aj}\ }\textbf {\bibinfo {volume} {124}},\ \bibinfo
  {pages} {675} (\bibinfo {year} {2002})}\BibitemShut {NoStop}%
\bibitem [{\citenamefont {{Mori{\'c}}}\ \emph {et~al.}(2010)\citenamefont
  {{Mori{\'c}}}, \citenamefont {{Smol{\v{c}}i{\'c}}}, \citenamefont
  {{Kimball}}, \citenamefont {{Riechers}}, \citenamefont {{Ivezi{\'c}}},\ and\
  \citenamefont {{Scoville}}}]{2010ApJ...724..779M}%
  \BibitemOpen
  \bibfield  {author} {\bibinfo {author} {\bibfnamefont {I.}~\bibnamefont
  {{Mori{\'c}}}}, \bibinfo {author} {\bibfnamefont {V.}~\bibnamefont
  {{Smol{\v{c}}i{\'c}}}}, \bibinfo {author} {\bibfnamefont {A.}~\bibnamefont
  {{Kimball}}}, \bibinfo {author} {\bibfnamefont {D.~A.}\ \bibnamefont
  {{Riechers}}}, \bibinfo {author} {\bibfnamefont {{\v{Z}}.}~\bibnamefont
  {{Ivezi{\'c}}}}, \ and\ \bibinfo {author} {\bibfnamefont {N.}~\bibnamefont
  {{Scoville}}},\ }\href {\doibase 10.1088/0004-637X/724/1/779} {\bibfield
  {journal} {\bibinfo  {journal} {\apj}\ }\textbf {\bibinfo {volume} {724}},\
  \bibinfo {pages} {779} (\bibinfo {year} {2010})},\ \Eprint
  {http://arxiv.org/abs/1010.0435} {arXiv:1010.0435 [astro-ph.GA]} \BibitemShut
  {NoStop}%
\bibitem [{\citenamefont {{Beck}}\ and\ \citenamefont
  {{Golla}}(1988)}]{1988A&A...191L...9B}%
  \BibitemOpen
  \bibfield  {author} {\bibinfo {author} {\bibfnamefont {R.}~\bibnamefont
  {{Beck}}}\ and\ \bibinfo {author} {\bibfnamefont {G.}~\bibnamefont
  {{Golla}}},\ }\href@noop {} {\bibfield  {journal} {\bibinfo  {journal}
  {\aap}\ }\textbf {\bibinfo {volume} {191}},\ \bibinfo {pages} {L9} (\bibinfo
  {year} {1988})}\BibitemShut {NoStop}%
\bibitem [{\citenamefont {{Murphy}}\ \emph {et~al.}(2006)\citenamefont
  {{Murphy}} \emph {et~al.}}]{2006ApJ...638..157M}%
  \BibitemOpen
  \bibfield  {author} {\bibinfo {author} {\bibfnamefont {E.~J.}\ \bibnamefont
  {{Murphy}}} \emph {et~al.},\ }\href {\doibase 10.1086/498636} {\bibfield
  {journal} {\bibinfo  {journal} {\apj}\ }\textbf {\bibinfo {volume} {638}},\
  \bibinfo {pages} {157} (\bibinfo {year} {2006})},\ \Eprint
  {http://arxiv.org/abs/astro-ph/0510227} {arXiv:astro-ph/0510227 [astro-ph]}
  \BibitemShut {NoStop}%
\bibitem [{\citenamefont {{Paladino}}\ \emph {et~al.}(2006)\citenamefont
  {{Paladino}}, \citenamefont {{Murgia}}, \citenamefont {{Helfer}},
  \citenamefont {{Wong}}, \citenamefont {{Ekers}}, \citenamefont {{Blitz}},
  \citenamefont {{Gregorini}},\ and\ \citenamefont
  {{Moscadelli}}}]{Paladino06}%
  \BibitemOpen
  \bibfield  {author} {\bibinfo {author} {\bibfnamefont {R.}~\bibnamefont
  {{Paladino}}}, \bibinfo {author} {\bibfnamefont {M.}~\bibnamefont
  {{Murgia}}}, \bibinfo {author} {\bibfnamefont {T.~T.}\ \bibnamefont
  {{Helfer}}}, \bibinfo {author} {\bibfnamefont {T.}~\bibnamefont {{Wong}}},
  \bibinfo {author} {\bibfnamefont {R.}~\bibnamefont {{Ekers}}}, \bibinfo
  {author} {\bibfnamefont {L.}~\bibnamefont {{Blitz}}}, \bibinfo {author}
  {\bibfnamefont {L.}~\bibnamefont {{Gregorini}}}, \ and\ \bibinfo {author}
  {\bibfnamefont {L.}~\bibnamefont {{Moscadelli}}},\ }\href {\doibase
  10.1051/0004-6361:20065002} {\bibfield  {journal} {\bibinfo  {journal}
  {\aap}\ }\textbf {\bibinfo {volume} {456}},\ \bibinfo {pages} {847} (\bibinfo
  {year} {2006})},\ \Eprint {http://arxiv.org/abs/astro-ph/0606480}
  {arXiv:astro-ph/0606480 [astro-ph]} \BibitemShut {NoStop}%
\bibitem [{\citenamefont {Murphy}\ \emph {et~al.}(2008)\citenamefont {Murphy},
  \citenamefont {Helou}, \citenamefont {Kenney}, \citenamefont {Armus},\ and\
  \citenamefont {Braun}}]{Murphy:2008ma}%
  \BibitemOpen
  \bibfield  {author} {\bibinfo {author} {\bibfnamefont {E.~J.}\ \bibnamefont
  {Murphy}}, \bibinfo {author} {\bibfnamefont {G.}~\bibnamefont {Helou}},
  \bibinfo {author} {\bibfnamefont {J.~D.~P.}\ \bibnamefont {Kenney}}, \bibinfo
  {author} {\bibfnamefont {L.}~\bibnamefont {Armus}}, \ and\ \bibinfo {author}
  {\bibfnamefont {R.}~\bibnamefont {Braun}},\ }\href {\doibase 10.1086/587123}
  {\bibfield  {journal} {\bibinfo  {journal} {Astrophys. J.}\ }\textbf
  {\bibinfo {volume} {678}},\ \bibinfo {pages} {828} (\bibinfo {year}
  {2008})},\ \Eprint {http://arxiv.org/abs/0801.4768} {arXiv:0801.4768
  [astro-ph]} \BibitemShut {NoStop}%
\bibitem [{\citenamefont {{Heesen}}\ \emph {et~al.}(2014)\citenamefont
  {{Heesen}}, \citenamefont {{Brinks}}, \citenamefont {{Leroy}}, \citenamefont
  {{Heald}}, \citenamefont {{Braun}}, \citenamefont {{Bigiel}},\ and\
  \citenamefont {{Beck}}}]{Heesen14}%
  \BibitemOpen
  \bibfield  {author} {\bibinfo {author} {\bibfnamefont {V.}~\bibnamefont
  {{Heesen}}}, \bibinfo {author} {\bibfnamefont {E.}~\bibnamefont {{Brinks}}},
  \bibinfo {author} {\bibfnamefont {A.~K.}\ \bibnamefont {{Leroy}}}, \bibinfo
  {author} {\bibfnamefont {G.}~\bibnamefont {{Heald}}}, \bibinfo {author}
  {\bibfnamefont {R.}~\bibnamefont {{Braun}}}, \bibinfo {author} {\bibfnamefont
  {F.}~\bibnamefont {{Bigiel}}}, \ and\ \bibinfo {author} {\bibfnamefont
  {R.}~\bibnamefont {{Beck}}},\ }\href {\doibase 10.1088/0004-6256/147/5/103}
  {\bibfield  {journal} {\bibinfo  {journal} {\aj}\ }\textbf {\bibinfo {volume}
  {147}},\ \bibinfo {eid} {103} (\bibinfo {year} {2014})},\ \Eprint
  {http://arxiv.org/abs/1402.1711} {arXiv:1402.1711 [astro-ph.GA]} \BibitemShut
  {NoStop}%
\bibitem [{\citenamefont {{Heesen}}\ \emph {et~al.}(2019)\citenamefont
  {{Heesen}} \emph {et~al.}}]{Heesen19}%
  \BibitemOpen
  \bibfield  {author} {\bibinfo {author} {\bibfnamefont {V.}~\bibnamefont
  {{Heesen}}} \emph {et~al.},\ }\href {\doibase 10.1051/0004-6361/201833905}
  {\bibfield  {journal} {\bibinfo  {journal} {\aap}\ }\textbf {\bibinfo
  {volume} {622}},\ \bibinfo {eid} {A8} (\bibinfo {year} {2019})},\ \Eprint
  {http://arxiv.org/abs/1811.07968} {arXiv:1811.07968 [astro-ph.GA]}
  \BibitemShut {NoStop}%
\bibitem [{\citenamefont {Bell}(2003)}]{Bell:2002kg}%
  \BibitemOpen
  \bibfield  {author} {\bibinfo {author} {\bibfnamefont {E.~F.}\ \bibnamefont
  {Bell}},\ }\href {\doibase 10.1086/367829} {\bibfield  {journal} {\bibinfo
  {journal} {Astrophys. J.}\ }\textbf {\bibinfo {volume} {586}},\ \bibinfo
  {pages} {794} (\bibinfo {year} {2003})},\ \Eprint
  {http://arxiv.org/abs/astro-ph/0212121} {arXiv:astro-ph/0212121 [astro-ph]}
  \BibitemShut {NoStop}%
\bibitem [{\citenamefont {Lacki}\ \emph {et~al.}(2010)\citenamefont {Lacki},
  \citenamefont {Thompson},\ and\ \citenamefont {Quataert}}]{Lacki:2009mj}%
  \BibitemOpen
  \bibfield  {author} {\bibinfo {author} {\bibfnamefont {B.~C.}\ \bibnamefont
  {Lacki}}, \bibinfo {author} {\bibfnamefont {T.~A.}\ \bibnamefont {Thompson}},
  \ and\ \bibinfo {author} {\bibfnamefont {E.}~\bibnamefont {Quataert}},\
  }\href {\doibase 10.1088/0004-637X/717/1/1} {\bibfield  {journal} {\bibinfo
  {journal} {Astrophys. J.}\ }\textbf {\bibinfo {volume} {717}},\ \bibinfo
  {pages} {1} (\bibinfo {year} {2010})},\ \Eprint
  {http://arxiv.org/abs/0907.4161} {arXiv:0907.4161 [astro-ph.CO]} \BibitemShut
  {NoStop}%
\bibitem [{\citenamefont {{Hodge}}\ \emph {et~al.}(2008)\citenamefont
  {{Hodge}}, \citenamefont {{Becker}}, \citenamefont {{White}},\ and\
  \citenamefont {{de Vries}}}]{Hodge08}%
  \BibitemOpen
  \bibfield  {author} {\bibinfo {author} {\bibfnamefont {J.~A.}\ \bibnamefont
  {{Hodge}}}, \bibinfo {author} {\bibfnamefont {R.~H.}\ \bibnamefont
  {{Becker}}}, \bibinfo {author} {\bibfnamefont {R.~L.}\ \bibnamefont
  {{White}}}, \ and\ \bibinfo {author} {\bibfnamefont {W.~H.}\ \bibnamefont
  {{de Vries}}},\ }\href {\doibase 10.1088/0004-6256/136/3/1097} {\bibfield
  {journal} {\bibinfo  {journal} {\aj}\ }\textbf {\bibinfo {volume} {136}},\
  \bibinfo {pages} {1097} (\bibinfo {year} {2008})},\ \Eprint
  {http://arxiv.org/abs/0806.3986} {arXiv:0806.3986 [astro-ph]} \BibitemShut
  {NoStop}%
\bibitem [{\citenamefont {{Brown}}\ \emph {et~al.}(2017)\citenamefont {{Brown}}
  \emph {et~al.}}]{Brown17}%
  \BibitemOpen
  \bibfield  {author} {\bibinfo {author} {\bibfnamefont {M.~J.~I.}\
  \bibnamefont {{Brown}}} \emph {et~al.},\ }\href {\doibase
  10.3847/1538-4357/aa8ad2} {\bibfield  {journal} {\bibinfo  {journal} {\apj}\
  }\textbf {\bibinfo {volume} {847}},\ \bibinfo {eid} {136} (\bibinfo {year}
  {2017})},\ \Eprint {http://arxiv.org/abs/1709.00183} {arXiv:1709.00183
  [astro-ph.GA]} \BibitemShut {NoStop}%
\bibitem [{\citenamefont {{Davies}}\ \emph {et~al.}(2017)\citenamefont
  {{Davies}} \emph {et~al.}}]{Davis17}%
  \BibitemOpen
  \bibfield  {author} {\bibinfo {author} {\bibfnamefont {L.~J.~M.}\
  \bibnamefont {{Davies}}} \emph {et~al.},\ }\href {\doibase
  10.1093/mnras/stw3080} {\bibfield  {journal} {\bibinfo  {journal} {\mnras}\
  }\textbf {\bibinfo {volume} {466}},\ \bibinfo {pages} {2312} (\bibinfo {year}
  {2017})},\ \Eprint {http://arxiv.org/abs/1701.06242} {arXiv:1701.06242
  [astro-ph.GA]} \BibitemShut {NoStop}%
\bibitem [{\citenamefont {{Hindson}}\ \emph {et~al.}(2018)\citenamefont
  {{Hindson}}, \citenamefont {{Kitchener}}, \citenamefont {{Brinks}},
  \citenamefont {{Heesen}}, \citenamefont {{Westcott}}, \citenamefont
  {{Hunter}}, \citenamefont {{Zhang}}, \citenamefont {{Rupen}},\ and\
  \citenamefont {{Rau}}}]{irc18H}%
  \BibitemOpen
  \bibfield  {author} {\bibinfo {author} {\bibfnamefont {L.}~\bibnamefont
  {{Hindson}}}, \bibinfo {author} {\bibfnamefont {G.}~\bibnamefont
  {{Kitchener}}}, \bibinfo {author} {\bibfnamefont {E.}~\bibnamefont
  {{Brinks}}}, \bibinfo {author} {\bibfnamefont {V.}~\bibnamefont {{Heesen}}},
  \bibinfo {author} {\bibfnamefont {J.}~\bibnamefont {{Westcott}}}, \bibinfo
  {author} {\bibfnamefont {D.}~\bibnamefont {{Hunter}}}, \bibinfo {author}
  {\bibfnamefont {H.-X.}\ \bibnamefont {{Zhang}}}, \bibinfo {author}
  {\bibfnamefont {M.}~\bibnamefont {{Rupen}}}, \ and\ \bibinfo {author}
  {\bibfnamefont {U.}~\bibnamefont {{Rau}}},\ }\href {\doibase
  10.3847/1538-4365/aaa42c} {\bibfield  {journal} {\bibinfo  {journal} {\apjs}\
  }\textbf {\bibinfo {volume} {234}},\ \bibinfo {eid} {29} (\bibinfo {year}
  {2018})},\ \Eprint {http://arxiv.org/abs/1801.05348} {arXiv:1801.05348
  [astro-ph.GA]} \BibitemShut {NoStop}%
\bibitem [{\citenamefont {{Calistro Rivera}}\ \emph {et~al.}(2017)\citenamefont
  {{Calistro Rivera}} \emph {et~al.}}]{CalistroRivera17}%
  \BibitemOpen
  \bibfield  {author} {\bibinfo {author} {\bibfnamefont {G.}~\bibnamefont
  {{Calistro Rivera}}} \emph {et~al.},\ }\href {\doibase 10.1093/mnras/stx1040}
  {\bibfield  {journal} {\bibinfo  {journal} {\mnras}\ }\textbf {\bibinfo
  {volume} {469}},\ \bibinfo {pages} {3468} (\bibinfo {year} {2017})},\ \Eprint
  {http://arxiv.org/abs/1704.06268} {arXiv:1704.06268 [astro-ph.GA]}
  \BibitemShut {NoStop}%
\bibitem [{\citenamefont {{G{\"u}rkan}}\ \emph {et~al.}(2018)\citenamefont
  {{G{\"u}rkan}} \emph {et~al.}}]{Gurkan18}%
  \BibitemOpen
  \bibfield  {author} {\bibinfo {author} {\bibfnamefont {G.}~\bibnamefont
  {{G{\"u}rkan}}} \emph {et~al.},\ }\href {\doibase 10.1093/mnras/sty016}
  {\bibfield  {journal} {\bibinfo  {journal} {\mnras}\ }\textbf {\bibinfo
  {volume} {475}},\ \bibinfo {pages} {3010} (\bibinfo {year} {2018})},\ \Eprint
  {http://arxiv.org/abs/1801.02629} {arXiv:1801.02629 [astro-ph.GA]}
  \BibitemShut {NoStop}%
\bibitem [{\citenamefont {{Wang}}\ \emph {et~al.}(2019)\citenamefont {{Wang}}
  \emph {et~al.}}]{Wang19}%
  \BibitemOpen
  \bibfield  {author} {\bibinfo {author} {\bibfnamefont {L.}~\bibnamefont
  {{Wang}}} \emph {et~al.},\ }\href {\doibase 10.1051/0004-6361/201935913}
  {\bibfield  {journal} {\bibinfo  {journal} {\aap}\ }\textbf {\bibinfo
  {volume} {631}},\ \bibinfo {eid} {A109} (\bibinfo {year} {2019})},\ \Eprint
  {http://arxiv.org/abs/1909.04489} {arXiv:1909.04489 [astro-ph.GA]}
  \BibitemShut {NoStop}%
\bibitem [{\citenamefont {{Niklas}}\ and\ \citenamefont
  {{Beck}}(1997)}]{Niklas97}%
  \BibitemOpen
  \bibfield  {author} {\bibinfo {author} {\bibfnamefont {S.}~\bibnamefont
  {{Niklas}}}\ and\ \bibinfo {author} {\bibfnamefont {R.}~\bibnamefont
  {{Beck}}},\ }\href@noop {} {\bibfield  {journal} {\bibinfo  {journal} {\aap}\
  }\textbf {\bibinfo {volume} {320}},\ \bibinfo {pages} {54} (\bibinfo {year}
  {1997})}\BibitemShut {NoStop}%
\bibitem [{\citenamefont {{Schleicher}}\ and\ \citenamefont
  {{Beck}}(2016)}]{Schleicher16}%
  \BibitemOpen
  \bibfield  {author} {\bibinfo {author} {\bibfnamefont {D.~R.~G.}\
  \bibnamefont {{Schleicher}}}\ and\ \bibinfo {author} {\bibfnamefont
  {R.}~\bibnamefont {{Beck}}},\ }\href {\doibase 10.1051/0004-6361/201628843}
  {\bibfield  {journal} {\bibinfo  {journal} {\aap}\ }\textbf {\bibinfo
  {volume} {593}},\ \bibinfo {eid} {A77} (\bibinfo {year} {2016})},\ \Eprint
  {http://arxiv.org/abs/1607.00094} {arXiv:1607.00094 [astro-ph.GA]}
  \BibitemShut {NoStop}%
\bibitem [{\citenamefont {{Hooper}}\ and\ \citenamefont
  {{Linden}}(2018)}]{Hooper18}%
  \BibitemOpen
  \bibfield  {author} {\bibinfo {author} {\bibfnamefont {D.}~\bibnamefont
  {{Hooper}}}\ and\ \bibinfo {author} {\bibfnamefont {T.}~\bibnamefont
  {{Linden}}},\ }\href {\doibase 10.1103/PhysRevD.98.043005} {\bibfield
  {journal} {\bibinfo  {journal} {\prd}\ }\textbf {\bibinfo {volume} {98}},\
  \bibinfo {eid} {043005} (\bibinfo {year} {2018})},\ \Eprint
  {http://arxiv.org/abs/1803.08046} {arXiv:1803.08046 [astro-ph.HE]}
  \BibitemShut {NoStop}%
\bibitem [{\citenamefont {{Fragos}}\ \emph
  {et~al.}(2013{\natexlab{a}})\citenamefont {{Fragos}} \emph
  {et~al.}}]{2013ApJ...764...41F}%
  \BibitemOpen
  \bibfield  {author} {\bibinfo {author} {\bibfnamefont {T.}~\bibnamefont
  {{Fragos}}} \emph {et~al.},\ }\href {\doibase 10.1088/0004-637X/764/1/41}
  {\bibfield  {journal} {\bibinfo  {journal} {\apj}\ }\textbf {\bibinfo
  {volume} {764}},\ \bibinfo {eid} {41} (\bibinfo {year}
  {2013}{\natexlab{a}})},\ \Eprint {http://arxiv.org/abs/1206.2395}
  {arXiv:1206.2395 [astro-ph.HE]} \BibitemShut {NoStop}%
\bibitem [{\citenamefont {{Corbet}}(1984)}]{1984A&A...141...91C}%
  \BibitemOpen
  \bibfield  {author} {\bibinfo {author} {\bibfnamefont {R.~H.~D.}\
  \bibnamefont {{Corbet}}},\ }\href@noop {} {\bibfield  {journal} {\bibinfo
  {journal} {\aap}\ }\textbf {\bibinfo {volume} {141}},\ \bibinfo {pages} {91}
  (\bibinfo {year} {1984})}\BibitemShut {NoStop}%
\bibitem [{\citenamefont {{Tauris}}\ and\ \citenamefont {{van den
  Heuvel}}(2006)}]{Tauris06}%
  \BibitemOpen
  \bibfield  {author} {\bibinfo {author} {\bibfnamefont {T.~M.}\ \bibnamefont
  {{Tauris}}}\ and\ \bibinfo {author} {\bibfnamefont {E.~P.~J.}\ \bibnamefont
  {{van den Heuvel}}},\ }\enquote {\bibinfo {title} {{Formation and evolution
  of compact stellar X-ray sources}},}\ in\ \href@noop {} {\emph {\bibinfo
  {booktitle} {Compact stellar X-ray sources}}},\ Vol.~\bibinfo {volume} {39}\
  (\bibinfo {year} {2006})\ pp.\ \bibinfo {pages} {623--665}\BibitemShut
  {NoStop}%
\bibitem [{\citenamefont {{Tatischeff}}(2009)}]{sn1993j}%
  \BibitemOpen
  \bibfield  {author} {\bibinfo {author} {\bibfnamefont {V.}~\bibnamefont
  {{Tatischeff}}},\ }\href {\doibase 10.1051/0004-6361/200811511} {\bibfield
  {journal} {\bibinfo  {journal} {\aap}\ }\textbf {\bibinfo {volume} {499}},\
  \bibinfo {pages} {191} (\bibinfo {year} {2009})},\ \Eprint
  {http://arxiv.org/abs/0903.2944} {arXiv:0903.2944 [astro-ph.HE]} \BibitemShut
  {NoStop}%
\bibitem [{\citenamefont {{Park}}\ \emph {et~al.}(2015)\citenamefont {{Park}},
  \citenamefont {{Caprioli}},\ and\ \citenamefont {{Spitkovsky}}}]{Park15}%
  \BibitemOpen
  \bibfield  {author} {\bibinfo {author} {\bibfnamefont {J.}~\bibnamefont
  {{Park}}}, \bibinfo {author} {\bibfnamefont {D.}~\bibnamefont {{Caprioli}}},
  \ and\ \bibinfo {author} {\bibfnamefont {A.}~\bibnamefont {{Spitkovsky}}},\
  }\href {\doibase 10.1103/PhysRevLett.114.085003} {\bibfield  {journal}
  {\bibinfo  {journal} {\prl}\ }\textbf {\bibinfo {volume} {114}},\ \bibinfo
  {eid} {085003} (\bibinfo {year} {2015})},\ \Eprint
  {http://arxiv.org/abs/1412.0672} {arXiv:1412.0672 [astro-ph.HE]} \BibitemShut
  {NoStop}%
\bibitem [{\citenamefont {{Sarbadhicary}}\ \emph {et~al.}(2017)\citenamefont
  {{Sarbadhicary}}, \citenamefont {{Badenes}}, \citenamefont {{Chomiuk}},
  \citenamefont {{Caprioli}},\ and\ \citenamefont
  {{Huizenga}}}]{Sarbadhicary17}%
  \BibitemOpen
  \bibfield  {author} {\bibinfo {author} {\bibfnamefont {S.~K.}\ \bibnamefont
  {{Sarbadhicary}}}, \bibinfo {author} {\bibfnamefont {C.}~\bibnamefont
  {{Badenes}}}, \bibinfo {author} {\bibfnamefont {L.}~\bibnamefont
  {{Chomiuk}}}, \bibinfo {author} {\bibfnamefont {D.}~\bibnamefont
  {{Caprioli}}}, \ and\ \bibinfo {author} {\bibfnamefont {D.}~\bibnamefont
  {{Huizenga}}},\ }\href {\doibase 10.1093/mnras/stw2566} {\bibfield  {journal}
  {\bibinfo  {journal} {\mnras}\ }\textbf {\bibinfo {volume} {464}},\ \bibinfo
  {pages} {2326} (\bibinfo {year} {2017})},\ \Eprint
  {http://arxiv.org/abs/1605.04923} {arXiv:1605.04923 [astro-ph.HE]}
  \BibitemShut {NoStop}%
\bibitem [{\citenamefont {{Lopez}}\ \emph {et~al.}(2018)\citenamefont
  {{Lopez}}, \citenamefont {{Auchettl}}, \citenamefont {{Linden}},
  \citenamefont {{Bolatto}}, \citenamefont {{Thompson}},\ and\ \citenamefont
  {{Ramirez-Ruiz}}}]{Lopez18}%
  \BibitemOpen
  \bibfield  {author} {\bibinfo {author} {\bibfnamefont {L.~A.}\ \bibnamefont
  {{Lopez}}}, \bibinfo {author} {\bibfnamefont {K.}~\bibnamefont {{Auchettl}}},
  \bibinfo {author} {\bibfnamefont {T.}~\bibnamefont {{Linden}}}, \bibinfo
  {author} {\bibfnamefont {A.~D.}\ \bibnamefont {{Bolatto}}}, \bibinfo {author}
  {\bibfnamefont {T.~A.}\ \bibnamefont {{Thompson}}}, \ and\ \bibinfo {author}
  {\bibfnamefont {E.}~\bibnamefont {{Ramirez-Ruiz}}},\ }\href {\doibase
  10.3847/1538-4357/aae0f8} {\bibfield  {journal} {\bibinfo  {journal} {\apj}\
  }\textbf {\bibinfo {volume} {867}},\ \bibinfo {eid} {44} (\bibinfo {year}
  {2018})},\ \Eprint {http://arxiv.org/abs/1807.06595} {arXiv:1807.06595
  [astro-ph.HE]} \BibitemShut {NoStop}%
\bibitem [{\citenamefont {{Strong}}\ \emph {et~al.}(2010)\citenamefont
  {{Strong}}, \citenamefont {{Porter}}, \citenamefont {{Digel}}, \citenamefont
  {{J{\'o}hannesson}}, \citenamefont {{Martin}}, \citenamefont {{Moskalenko}},
  \citenamefont {{Murphy}},\ and\ \citenamefont {{Orlando}}}]{Strong10}%
  \BibitemOpen
  \bibfield  {author} {\bibinfo {author} {\bibfnamefont {A.~W.}\ \bibnamefont
  {{Strong}}}, \bibinfo {author} {\bibfnamefont {T.~A.}\ \bibnamefont
  {{Porter}}}, \bibinfo {author} {\bibfnamefont {S.~W.}\ \bibnamefont
  {{Digel}}}, \bibinfo {author} {\bibfnamefont {G.}~\bibnamefont
  {{J{\'o}hannesson}}}, \bibinfo {author} {\bibfnamefont {P.}~\bibnamefont
  {{Martin}}}, \bibinfo {author} {\bibfnamefont {I.~V.}\ \bibnamefont
  {{Moskalenko}}}, \bibinfo {author} {\bibfnamefont {E.~J.}\ \bibnamefont
  {{Murphy}}}, \ and\ \bibinfo {author} {\bibfnamefont {E.}~\bibnamefont
  {{Orlando}}},\ }\href {\doibase 10.1088/2041-8205/722/1/L58} {\bibfield
  {journal} {\bibinfo  {journal} {\apjl}\ }\textbf {\bibinfo {volume} {722}},\
  \bibinfo {pages} {L58} (\bibinfo {year} {2010})},\ \Eprint
  {http://arxiv.org/abs/1008.4330} {arXiv:1008.4330 [astro-ph.HE]} \BibitemShut
  {NoStop}%
\bibitem [{\citenamefont {{Thompson}}\ \emph {et~al.}(2007)\citenamefont
  {{Thompson}}, \citenamefont {{Quataert}},\ and\ \citenamefont
  {{Waxman}}}]{Thompson07}%
  \BibitemOpen
  \bibfield  {author} {\bibinfo {author} {\bibfnamefont {T.~A.}\ \bibnamefont
  {{Thompson}}}, \bibinfo {author} {\bibfnamefont {E.}~\bibnamefont
  {{Quataert}}}, \ and\ \bibinfo {author} {\bibfnamefont {E.}~\bibnamefont
  {{Waxman}}},\ }\href {\doibase 10.1086/509068} {\bibfield  {journal}
  {\bibinfo  {journal} {\apj}\ }\textbf {\bibinfo {volume} {654}},\ \bibinfo
  {pages} {219} (\bibinfo {year} {2007})},\ \Eprint
  {http://arxiv.org/abs/astro-ph/0606665} {arXiv:astro-ph/0606665 [astro-ph]}
  \BibitemShut {NoStop}%
\bibitem [{\citenamefont {{Lacki}}\ \emph {et~al.}(2011)\citenamefont
  {{Lacki}}, \citenamefont {{Thompson}}, \citenamefont {{Quataert}},
  \citenamefont {{Loeb}},\ and\ \citenamefont {{Waxman}}}]{Lacki11}%
  \BibitemOpen
  \bibfield  {author} {\bibinfo {author} {\bibfnamefont {B.~C.}\ \bibnamefont
  {{Lacki}}}, \bibinfo {author} {\bibfnamefont {T.~A.}\ \bibnamefont
  {{Thompson}}}, \bibinfo {author} {\bibfnamefont {E.}~\bibnamefont
  {{Quataert}}}, \bibinfo {author} {\bibfnamefont {A.}~\bibnamefont {{Loeb}}},
  \ and\ \bibinfo {author} {\bibfnamefont {E.}~\bibnamefont {{Waxman}}},\
  }\href {\doibase 10.1088/0004-637X/734/2/107} {\bibfield  {journal} {\bibinfo
   {journal} {\apj}\ }\textbf {\bibinfo {volume} {734}},\ \bibinfo {eid} {107}
  (\bibinfo {year} {2011})},\ \Eprint {http://arxiv.org/abs/1003.3257}
  {arXiv:1003.3257 [astro-ph.HE]} \BibitemShut {NoStop}%
\bibitem [{\citenamefont {{Griffin}}\ \emph {et~al.}(2016)\citenamefont
  {{Griffin}}, \citenamefont {{Dai}},\ and\ \citenamefont
  {{Thompson}}}]{Griffin16}%
  \BibitemOpen
  \bibfield  {author} {\bibinfo {author} {\bibfnamefont {R.~D.}\ \bibnamefont
  {{Griffin}}}, \bibinfo {author} {\bibfnamefont {X.}~\bibnamefont {{Dai}}}, \
  and\ \bibinfo {author} {\bibfnamefont {T.~A.}\ \bibnamefont {{Thompson}}},\
  }\href {\doibase 10.3847/2041-8205/823/1/L17} {\bibfield  {journal} {\bibinfo
   {journal} {\apjl}\ }\textbf {\bibinfo {volume} {823}},\ \bibinfo {eid} {L17}
  (\bibinfo {year} {2016})},\ \Eprint {http://arxiv.org/abs/1603.06949}
  {arXiv:1603.06949 [astro-ph.HE]} \BibitemShut {NoStop}%
\bibitem [{\citenamefont {{Lacki}}\ and\ \citenamefont
  {{Beck}}(2013)}]{Lacki13}%
  \BibitemOpen
  \bibfield  {author} {\bibinfo {author} {\bibfnamefont {B.~C.}\ \bibnamefont
  {{Lacki}}}\ and\ \bibinfo {author} {\bibfnamefont {R.}~\bibnamefont
  {{Beck}}},\ }\href {\doibase 10.1093/mnras/stt122} {\bibfield  {journal}
  {\bibinfo  {journal} {\mnras}\ }\textbf {\bibinfo {volume} {430}},\ \bibinfo
  {pages} {3171} (\bibinfo {year} {2013})},\ \Eprint
  {http://arxiv.org/abs/1301.5391} {arXiv:1301.5391 [astro-ph.CO]} \BibitemShut
  {NoStop}%
\bibitem [{\citenamefont {{Hooper}}\ \emph {et~al.}(2017)\citenamefont
  {{Hooper}}, \citenamefont {{Cholis}}, \citenamefont {{Linden}},\ and\
  \citenamefont {{Fang}}}]{Hooper17halo}%
  \BibitemOpen
  \bibfield  {author} {\bibinfo {author} {\bibfnamefont {D.}~\bibnamefont
  {{Hooper}}}, \bibinfo {author} {\bibfnamefont {I.}~\bibnamefont {{Cholis}}},
  \bibinfo {author} {\bibfnamefont {T.}~\bibnamefont {{Linden}}}, \ and\
  \bibinfo {author} {\bibfnamefont {K.}~\bibnamefont {{Fang}}},\ }\href
  {\doibase 10.1103/PhysRevD.96.103013} {\bibfield  {journal} {\bibinfo
  {journal} {\prd}\ }\textbf {\bibinfo {volume} {96}},\ \bibinfo {eid} {103013}
  (\bibinfo {year} {2017})},\ \Eprint {http://arxiv.org/abs/1702.08436}
  {arXiv:1702.08436 [astro-ph.HE]} \BibitemShut {NoStop}%
\bibitem [{\citenamefont {{Linden}}\ \emph {et~al.}(2017)\citenamefont
  {{Linden}}, \citenamefont {{Auchettl}}, \citenamefont {{Bramante}},
  \citenamefont {{Cholis}}, \citenamefont {{Fang}}, \citenamefont {{Hooper}},
  \citenamefont {{Karwal}},\ and\ \citenamefont {{Li}}}]{Linden17halo}%
  \BibitemOpen
  \bibfield  {author} {\bibinfo {author} {\bibfnamefont {T.}~\bibnamefont
  {{Linden}}}, \bibinfo {author} {\bibfnamefont {K.}~\bibnamefont
  {{Auchettl}}}, \bibinfo {author} {\bibfnamefont {J.}~\bibnamefont
  {{Bramante}}}, \bibinfo {author} {\bibfnamefont {I.}~\bibnamefont
  {{Cholis}}}, \bibinfo {author} {\bibfnamefont {K.}~\bibnamefont {{Fang}}},
  \bibinfo {author} {\bibfnamefont {D.}~\bibnamefont {{Hooper}}}, \bibinfo
  {author} {\bibfnamefont {T.}~\bibnamefont {{Karwal}}}, \ and\ \bibinfo
  {author} {\bibfnamefont {S.~W.}\ \bibnamefont {{Li}}},\ }\href {\doibase
  10.1103/PhysRevD.96.103016} {\bibfield  {journal} {\bibinfo  {journal}
  {\prd}\ }\textbf {\bibinfo {volume} {96}},\ \bibinfo {eid} {103016} (\bibinfo
  {year} {2017})},\ \Eprint {http://arxiv.org/abs/1703.09704} {arXiv:1703.09704
  [astro-ph.HE]} \BibitemShut {NoStop}%
\bibitem [{\citenamefont {Faucher-Giguere}\ and\ \citenamefont
  {Kaspi}(2006)}]{FaucherGiguere:2005ny}%
  \BibitemOpen
  \bibfield  {author} {\bibinfo {author} {\bibfnamefont {C.-A.}\ \bibnamefont
  {Faucher-Giguere}}\ and\ \bibinfo {author} {\bibfnamefont {V.~M.}\
  \bibnamefont {Kaspi}},\ }\href {\doibase 10.1086/501516} {\bibfield
  {journal} {\bibinfo  {journal} {Astrophys. J.}\ }\textbf {\bibinfo {volume}
  {643}},\ \bibinfo {pages} {332} (\bibinfo {year} {2006})},\ \Eprint
  {http://arxiv.org/abs/astro-ph/0512585} {arXiv:astro-ph/0512585 [astro-ph]}
  \BibitemShut {NoStop}%
\bibitem [{\citenamefont {de~Jager}(2008)}]{deJager:2008ke}%
  \BibitemOpen
  \bibfield  {author} {\bibinfo {author} {\bibfnamefont {O.~C.}\ \bibnamefont
  {de~Jager}},\ }\href {\doibase 10.1086/588283} {\bibfield  {journal}
  {\bibinfo  {journal} {Astrophys. J.}\ }\textbf {\bibinfo {volume} {678}},\
  \bibinfo {pages} {L113} (\bibinfo {year} {2008})},\ \Eprint
  {http://arxiv.org/abs/0803.2104} {arXiv:0803.2104 [astro-ph]} \BibitemShut
  {NoStop}%
\bibitem [{\citenamefont {{Gaensler}}\ and\ \citenamefont
  {{Slane}}(2006)}]{Gaensler06}%
  \BibitemOpen
  \bibfield  {author} {\bibinfo {author} {\bibfnamefont {B.~M.}\ \bibnamefont
  {{Gaensler}}}\ and\ \bibinfo {author} {\bibfnamefont {P.~O.}\ \bibnamefont
  {{Slane}}},\ }\href {\doibase 10.1146/annurev.astro.44.051905.092528}
  {\bibfield  {journal} {\bibinfo  {journal} {\araa}\ }\textbf {\bibinfo
  {volume} {44}},\ \bibinfo {pages} {17} (\bibinfo {year} {2006})},\ \Eprint
  {http://arxiv.org/abs/astro-ph/0601081} {arXiv:astro-ph/0601081 [astro-ph]}
  \BibitemShut {NoStop}%
\bibitem [{\citenamefont {{Reynolds}}\ \emph {et~al.}(2017)\citenamefont
  {{Reynolds}}, \citenamefont {{Pavlov}}, \citenamefont {{Kargaltsev}},
  \citenamefont {{Klingler}}, \citenamefont {{Renaud}},\ and\ \citenamefont
  {{Mereghetti}}}]{Reynolds17}%
  \BibitemOpen
  \bibfield  {author} {\bibinfo {author} {\bibfnamefont {S.~P.}\ \bibnamefont
  {{Reynolds}}}, \bibinfo {author} {\bibfnamefont {G.~G.}\ \bibnamefont
  {{Pavlov}}}, \bibinfo {author} {\bibfnamefont {O.}~\bibnamefont
  {{Kargaltsev}}}, \bibinfo {author} {\bibfnamefont {N.}~\bibnamefont
  {{Klingler}}}, \bibinfo {author} {\bibfnamefont {M.}~\bibnamefont
  {{Renaud}}}, \ and\ \bibinfo {author} {\bibfnamefont {S.}~\bibnamefont
  {{Mereghetti}}},\ }\href {\doibase 10.1007/s11214-017-0356-6} {\bibfield
  {journal} {\bibinfo  {journal} {\ssr}\ }\textbf {\bibinfo {volume} {207}},\
  \bibinfo {pages} {175} (\bibinfo {year} {2017})},\ \Eprint
  {http://arxiv.org/abs/1705.08897} {arXiv:1705.08897 [astro-ph.HE]}
  \BibitemShut {NoStop}%
\bibitem [{\citenamefont {{Szary}}\ \emph {et~al.}(2014)\citenamefont
  {{Szary}}, \citenamefont {{Zhang}}, \citenamefont {{Melikidze}},
  \citenamefont {{Gil}},\ and\ \citenamefont {{Xu}}}]{Szary14}%
  \BibitemOpen
  \bibfield  {author} {\bibinfo {author} {\bibfnamefont {A.}~\bibnamefont
  {{Szary}}}, \bibinfo {author} {\bibfnamefont {B.}~\bibnamefont {{Zhang}}},
  \bibinfo {author} {\bibfnamefont {G.~I.}\ \bibnamefont {{Melikidze}}},
  \bibinfo {author} {\bibfnamefont {J.}~\bibnamefont {{Gil}}}, \ and\ \bibinfo
  {author} {\bibfnamefont {R.-X.}\ \bibnamefont {{Xu}}},\ }\href {\doibase
  10.1088/0004-637X/784/1/59} {\bibfield  {journal} {\bibinfo  {journal}
  {\apj}\ }\textbf {\bibinfo {volume} {784}},\ \bibinfo {eid} {59} (\bibinfo
  {year} {2014})},\ \Eprint {http://arxiv.org/abs/1402.0228} {arXiv:1402.0228
  [astro-ph.HE]} \BibitemShut {NoStop}%
\bibitem [{\citenamefont {{Lorimer}}(2013)}]{Lorimer13}%
  \BibitemOpen
  \bibfield  {author} {\bibinfo {author} {\bibfnamefont {D.~R.}\ \bibnamefont
  {{Lorimer}}},\ }in\ \href {\doibase 10.1017/S1743921312023769} {\emph
  {\bibinfo {booktitle} {Neutron Stars and Pulsars: Challenges and
  Opportunities after 80 years}}},\ \bibinfo {series} {IAU Symposium}, Vol.\
  \bibinfo {volume} {291},\ \bibinfo {editor} {edited by\ \bibinfo {editor}
  {\bibfnamefont {J.}~\bibnamefont {{van Leeuwen}}}}\ (\bibinfo {year} {2013})\
  pp.\ \bibinfo {pages} {237--242},\ \Eprint {http://arxiv.org/abs/1210.2746}
  {arXiv:1210.2746 [astro-ph.GA]} \BibitemShut {NoStop}%
\bibitem [{\citenamefont {{Gonthier}}\ \emph {et~al.}(2018)\citenamefont
  {{Gonthier}}, \citenamefont {{Harding}}, \citenamefont {{Ferrara}},
  \citenamefont {{Frederick}}, \citenamefont {{Mohr}},\ and\ \citenamefont
  {{Koh}}}]{Gonthier18}%
  \BibitemOpen
  \bibfield  {author} {\bibinfo {author} {\bibfnamefont {P.~L.}\ \bibnamefont
  {{Gonthier}}}, \bibinfo {author} {\bibfnamefont {A.~K.}\ \bibnamefont
  {{Harding}}}, \bibinfo {author} {\bibfnamefont {E.~C.}\ \bibnamefont
  {{Ferrara}}}, \bibinfo {author} {\bibfnamefont {S.~E.}\ \bibnamefont
  {{Frederick}}}, \bibinfo {author} {\bibfnamefont {V.~E.}\ \bibnamefont
  {{Mohr}}}, \ and\ \bibinfo {author} {\bibfnamefont {Y.-M.}\ \bibnamefont
  {{Koh}}},\ }\href {\doibase 10.3847/1538-4357/aad08d} {\bibfield  {journal}
  {\bibinfo  {journal} {\apj}\ }\textbf {\bibinfo {volume} {863}},\ \bibinfo
  {eid} {199} (\bibinfo {year} {2018})},\ \Eprint
  {http://arxiv.org/abs/1806.11215} {arXiv:1806.11215 [astro-ph.HE]}
  \BibitemShut {NoStop}%
\bibitem [{\citenamefont {{Abdo}}\ \emph {et~al.}(2009)\citenamefont {{Abdo}}
  \emph {et~al.}}]{Abdo09gc}%
  \BibitemOpen
  \bibfield  {author} {\bibinfo {author} {\bibfnamefont {A.~A.}\ \bibnamefont
  {{Abdo}}} \emph {et~al.},\ }\href {\doibase 10.1126/science.1177023}
  {\bibfield  {journal} {\bibinfo  {journal} {Science}\ }\textbf {\bibinfo
  {volume} {325}},\ \bibinfo {pages} {845} (\bibinfo {year}
  {2009})}\BibitemShut {NoStop}%
\bibitem [{\citenamefont {{Hooper}}\ and\ \citenamefont
  {{Linden}}(2016)}]{Hooper16gc}%
  \BibitemOpen
  \bibfield  {author} {\bibinfo {author} {\bibfnamefont {D.}~\bibnamefont
  {{Hooper}}}\ and\ \bibinfo {author} {\bibfnamefont {T.}~\bibnamefont
  {{Linden}}},\ }\href {\doibase 10.1088/1475-7516/2016/08/018} {\bibfield
  {journal} {\bibinfo  {journal} {\jcap}\ }\textbf {\bibinfo {volume} {2016}},\
  \bibinfo {eid} {018} (\bibinfo {year} {2016})},\ \Eprint
  {http://arxiv.org/abs/1606.09250} {arXiv:1606.09250 [astro-ph.HE]}
  \BibitemShut {NoStop}%
\bibitem [{\citenamefont {{Macias}}\ \emph {et~al.}(2019)\citenamefont
  {{Macias}}, \citenamefont {{Horiuchi}}, \citenamefont {{Kaplinghat}},
  \citenamefont {{Gordon}}, \citenamefont {{Crocker}},\ and\ \citenamefont
  {{Nataf}}}]{Macias19}%
  \BibitemOpen
  \bibfield  {author} {\bibinfo {author} {\bibfnamefont {O.}~\bibnamefont
  {{Macias}}}, \bibinfo {author} {\bibfnamefont {S.}~\bibnamefont
  {{Horiuchi}}}, \bibinfo {author} {\bibfnamefont {M.}~\bibnamefont
  {{Kaplinghat}}}, \bibinfo {author} {\bibfnamefont {C.}~\bibnamefont
  {{Gordon}}}, \bibinfo {author} {\bibfnamefont {R.~M.}\ \bibnamefont
  {{Crocker}}}, \ and\ \bibinfo {author} {\bibfnamefont {D.~M.}\ \bibnamefont
  {{Nataf}}},\ }\href {\doibase 10.1088/1475-7516/2019/09/042} {\bibfield
  {journal} {\bibinfo  {journal} {\jcap}\ }\textbf {\bibinfo {volume} {2019}},\
  \bibinfo {eid} {042} (\bibinfo {year} {2019})},\ \Eprint
  {http://arxiv.org/abs/1901.03822} {arXiv:1901.03822 [astro-ph.HE]}
  \BibitemShut {NoStop}%
\bibitem [{\citenamefont {{Gilfanov}}(2004)}]{Gilfanov04}%
  \BibitemOpen
  \bibfield  {author} {\bibinfo {author} {\bibfnamefont {M.}~\bibnamefont
  {{Gilfanov}}},\ }\href {\doibase 10.1111/j.1365-2966.2004.07473.x} {\bibfield
   {journal} {\bibinfo  {journal} {\mnras}\ }\textbf {\bibinfo {volume}
  {349}},\ \bibinfo {pages} {146} (\bibinfo {year} {2004})},\ \Eprint
  {http://arxiv.org/abs/astro-ph/0309454} {arXiv:astro-ph/0309454 [astro-ph]}
  \BibitemShut {NoStop}%
\bibitem [{\citenamefont {{Lehmer}}\ \emph {et~al.}(2010)\citenamefont
  {{Lehmer}}, \citenamefont {{Alexander}}, \citenamefont {{Bauer}},
  \citenamefont {{Brand t}}, \citenamefont {{Goulding}}, \citenamefont
  {{Jenkins}}, \citenamefont {{Ptak}},\ and\ \citenamefont
  {{Roberts}}}]{Lehmer10}%
  \BibitemOpen
  \bibfield  {author} {\bibinfo {author} {\bibfnamefont {B.~D.}\ \bibnamefont
  {{Lehmer}}}, \bibinfo {author} {\bibfnamefont {D.~M.}\ \bibnamefont
  {{Alexander}}}, \bibinfo {author} {\bibfnamefont {F.~E.}\ \bibnamefont
  {{Bauer}}}, \bibinfo {author} {\bibfnamefont {W.~N.}\ \bibnamefont {{Brand
  t}}}, \bibinfo {author} {\bibfnamefont {A.~D.}\ \bibnamefont {{Goulding}}},
  \bibinfo {author} {\bibfnamefont {L.~P.}\ \bibnamefont {{Jenkins}}}, \bibinfo
  {author} {\bibfnamefont {A.}~\bibnamefont {{Ptak}}}, \ and\ \bibinfo {author}
  {\bibfnamefont {T.~P.}\ \bibnamefont {{Roberts}}},\ }\href {\doibase
  10.1088/0004-637X/724/1/559} {\bibfield  {journal} {\bibinfo  {journal}
  {\apj}\ }\textbf {\bibinfo {volume} {724}},\ \bibinfo {pages} {559} (\bibinfo
  {year} {2010})},\ \Eprint {http://arxiv.org/abs/1009.3943} {arXiv:1009.3943
  [astro-ph.CO]} \BibitemShut {NoStop}%
\bibitem [{\citenamefont {{Boroson}}\ \emph {et~al.}(2011)\citenamefont
  {{Boroson}}, \citenamefont {{Kim}},\ and\ \citenamefont
  {{Fabbiano}}}]{Boroson11}%
  \BibitemOpen
  \bibfield  {author} {\bibinfo {author} {\bibfnamefont {B.}~\bibnamefont
  {{Boroson}}}, \bibinfo {author} {\bibfnamefont {D.-W.}\ \bibnamefont
  {{Kim}}}, \ and\ \bibinfo {author} {\bibfnamefont {G.}~\bibnamefont
  {{Fabbiano}}},\ }\href {\doibase 10.1088/0004-637X/729/1/12} {\bibfield
  {journal} {\bibinfo  {journal} {\apj}\ }\textbf {\bibinfo {volume} {729}},\
  \bibinfo {eid} {12} (\bibinfo {year} {2011})},\ \Eprint
  {http://arxiv.org/abs/1011.2529} {arXiv:1011.2529 [astro-ph.HE]} \BibitemShut
  {NoStop}%
\bibitem [{\citenamefont {{Fragos}}\ \emph
  {et~al.}(2013{\natexlab{b}})\citenamefont {{Fragos}} \emph
  {et~al.}}]{Fragos13}%
  \BibitemOpen
  \bibfield  {author} {\bibinfo {author} {\bibfnamefont {T.}~\bibnamefont
  {{Fragos}}} \emph {et~al.},\ }\href {\doibase 10.1088/0004-637X/764/1/41}
  {\bibfield  {journal} {\bibinfo  {journal} {\apj}\ }\textbf {\bibinfo
  {volume} {764}},\ \bibinfo {eid} {41} (\bibinfo {year}
  {2013}{\natexlab{b}})},\ \Eprint {http://arxiv.org/abs/1206.2395}
  {arXiv:1206.2395 [astro-ph.HE]} \BibitemShut {NoStop}%
\bibitem [{\citenamefont {{Eckner}}\ \emph {et~al.}(2018)\citenamefont
  {{Eckner}} \emph {et~al.}}]{Eckner2018}%
  \BibitemOpen
  \bibfield  {author} {\bibinfo {author} {\bibfnamefont {C.}~\bibnamefont
  {{Eckner}}} \emph {et~al.},\ }\href {\doibase 10.3847/1538-4357/aac029}
  {\bibfield  {journal} {\bibinfo  {journal} {\apj}\ }\textbf {\bibinfo
  {volume} {862}},\ \bibinfo {eid} {79} (\bibinfo {year} {2018})},\ \Eprint
  {http://arxiv.org/abs/1711.05127} {arXiv:1711.05127 [astro-ph.HE]}
  \BibitemShut {NoStop}%
\bibitem [{\citenamefont {{Winter}}\ \emph {et~al.}(2016)\citenamefont
  {{Winter}}, \citenamefont {{Zaharijas}}, \citenamefont {{Bechtol}},\ and\
  \citenamefont {{Vand enbroucke}}}]{Winter16}%
  \BibitemOpen
  \bibfield  {author} {\bibinfo {author} {\bibfnamefont {M.}~\bibnamefont
  {{Winter}}}, \bibinfo {author} {\bibfnamefont {G.}~\bibnamefont
  {{Zaharijas}}}, \bibinfo {author} {\bibfnamefont {K.}~\bibnamefont
  {{Bechtol}}}, \ and\ \bibinfo {author} {\bibfnamefont {J.}~\bibnamefont
  {{Vand enbroucke}}},\ }\href {\doibase 10.3847/2041-8205/832/1/L6} {\bibfield
   {journal} {\bibinfo  {journal} {\apjl}\ }\textbf {\bibinfo {volume} {832}},\
  \bibinfo {eid} {L6} (\bibinfo {year} {2016})},\ \Eprint
  {http://arxiv.org/abs/1607.06390} {arXiv:1607.06390 [astro-ph.HE]}
  \BibitemShut {NoStop}%
\bibitem [{\citenamefont {{Ploeg}}\ \emph {et~al.}(2017)\citenamefont
  {{Ploeg}}, \citenamefont {{Gordon}}, \citenamefont {{Crocker}},\ and\
  \citenamefont {{Macias}}}]{Ploeg17}%
  \BibitemOpen
  \bibfield  {author} {\bibinfo {author} {\bibfnamefont {H.}~\bibnamefont
  {{Ploeg}}}, \bibinfo {author} {\bibfnamefont {C.}~\bibnamefont {{Gordon}}},
  \bibinfo {author} {\bibfnamefont {R.}~\bibnamefont {{Crocker}}}, \ and\
  \bibinfo {author} {\bibfnamefont {O.}~\bibnamefont {{Macias}}},\ }\href
  {\doibase 10.1088/1475-7516/2017/08/015} {\bibfield  {journal} {\bibinfo
  {journal} {\jcap}\ }\textbf {\bibinfo {volume} {2017}},\ \bibinfo {eid} {015}
  (\bibinfo {year} {2017})},\ \Eprint {http://arxiv.org/abs/1705.00806}
  {arXiv:1705.00806 [astro-ph.HE]} \BibitemShut {NoStop}%
\bibitem [{\citenamefont {{Bartels}}\ \emph
  {et~al.}(2018{\natexlab{a}})\citenamefont {{Bartels}}, \citenamefont
  {{Storm}}, \citenamefont {{Weniger}},\ and\ \citenamefont
  {{Calore}}}]{Bartels18}%
  \BibitemOpen
  \bibfield  {author} {\bibinfo {author} {\bibfnamefont {R.}~\bibnamefont
  {{Bartels}}}, \bibinfo {author} {\bibfnamefont {E.}~\bibnamefont {{Storm}}},
  \bibinfo {author} {\bibfnamefont {C.}~\bibnamefont {{Weniger}}}, \ and\
  \bibinfo {author} {\bibfnamefont {F.}~\bibnamefont {{Calore}}},\ }\href
  {\doibase 10.1038/s41550-018-0531-z} {\bibfield  {journal} {\bibinfo
  {journal} {Nature Astronomy}\ }\textbf {\bibinfo {volume} {2}},\ \bibinfo
  {pages} {819} (\bibinfo {year} {2018}{\natexlab{a}})},\ \Eprint
  {http://arxiv.org/abs/1711.04778} {arXiv:1711.04778 [astro-ph.HE]}
  \BibitemShut {NoStop}%
\bibitem [{\citenamefont {{Bartels}}\ \emph
  {et~al.}(2018{\natexlab{b}})\citenamefont {{Bartels}}, \citenamefont
  {{Edwards}},\ and\ \citenamefont {{Weniger}}}]{Bartels18b}%
  \BibitemOpen
  \bibfield  {author} {\bibinfo {author} {\bibfnamefont {R.~T.}\ \bibnamefont
  {{Bartels}}}, \bibinfo {author} {\bibfnamefont {T.~D.~P.}\ \bibnamefont
  {{Edwards}}}, \ and\ \bibinfo {author} {\bibfnamefont {C.}~\bibnamefont
  {{Weniger}}},\ }\href {\doibase 10.1093/mnras/sty2529} {\bibfield  {journal}
  {\bibinfo  {journal} {\mnras}\ }\textbf {\bibinfo {volume} {481}},\ \bibinfo
  {pages} {3966} (\bibinfo {year} {2018}{\natexlab{b}})},\ \Eprint
  {http://arxiv.org/abs/1805.11097} {arXiv:1805.11097 [astro-ph.HE]}
  \BibitemShut {NoStop}%
\bibitem [{\citenamefont {{Licquia}}\ and\ \citenamefont
  {{Newman}}(2015)}]{Licquia15}%
  \BibitemOpen
  \bibfield  {author} {\bibinfo {author} {\bibfnamefont {T.~C.}\ \bibnamefont
  {{Licquia}}}\ and\ \bibinfo {author} {\bibfnamefont {J.~A.}\ \bibnamefont
  {{Newman}}},\ }\href {\doibase 10.1088/0004-637X/806/1/96} {\bibfield
  {journal} {\bibinfo  {journal} {\apj}\ }\textbf {\bibinfo {volume} {806}},\
  \bibinfo {eid} {96} (\bibinfo {year} {2015})},\ \Eprint
  {http://arxiv.org/abs/1407.1078} {arXiv:1407.1078 [astro-ph.GA]} \BibitemShut
  {NoStop}%
\bibitem [{\citenamefont {{Abdo}}\ \emph {et~al.}(2013)\citenamefont {{Abdo}}
  \emph {et~al.}}]{2013ApJS..208...17A}%
  \BibitemOpen
  \bibfield  {author} {\bibinfo {author} {\bibfnamefont {A.~A.}\ \bibnamefont
  {{Abdo}}} \emph {et~al.},\ }\href {\doibase 10.1088/0067-0049/208/2/17}
  {\bibfield  {journal} {\bibinfo  {journal} {\apjs}\ }\textbf {\bibinfo
  {volume} {208}},\ \bibinfo {eid} {17} (\bibinfo {year} {2013})},\ \Eprint
  {http://arxiv.org/abs/1305.4385} {arXiv:1305.4385 [astro-ph.HE]} \BibitemShut
  {NoStop}%
\bibitem [{\citenamefont {Gilfanov}(2004)}]{Gilfanov:2003th}%
  \BibitemOpen
  \bibfield  {author} {\bibinfo {author} {\bibfnamefont {M.}~\bibnamefont
  {Gilfanov}},\ }\href {\doibase 10.1111/j.1365-2966.2004.07473.x} {\bibfield
  {journal} {\bibinfo  {journal} {Mon. Not. Roy. Astron. Soc.}\ }\textbf
  {\bibinfo {volume} {349}},\ \bibinfo {pages} {146} (\bibinfo {year}
  {2004})},\ \Eprint {http://arxiv.org/abs/astro-ph/0309454}
  {arXiv:astro-ph/0309454} \BibitemShut {NoStop}%
\bibitem [{\citenamefont {{Ackermann}}\ \emph {et~al.}(2017)\citenamefont
  {{Ackermann}} \emph {et~al.}}]{Ackermann17m31}%
  \BibitemOpen
  \bibfield  {author} {\bibinfo {author} {\bibfnamefont {M.}~\bibnamefont
  {{Ackermann}}} \emph {et~al.},\ }\href {\doibase 10.3847/1538-4357/aa5c3d}
  {\bibfield  {journal} {\bibinfo  {journal} {\apj}\ }\textbf {\bibinfo
  {volume} {836}},\ \bibinfo {eid} {208} (\bibinfo {year} {2017})},\ \Eprint
  {http://arxiv.org/abs/1702.08602} {arXiv:1702.08602 [astro-ph.HE]}
  \BibitemShut {NoStop}%
\bibitem [{\citenamefont {{Stappers}}\ \emph {et~al.}(2003)\citenamefont
  {{Stappers}}, \citenamefont {{Gaensler}}, \citenamefont {{Kaspi}},
  \citenamefont {{van der Klis}},\ and\ \citenamefont {{Lewin}}}]{Stappers03}%
  \BibitemOpen
  \bibfield  {author} {\bibinfo {author} {\bibfnamefont {B.~W.}\ \bibnamefont
  {{Stappers}}}, \bibinfo {author} {\bibfnamefont {B.~M.}\ \bibnamefont
  {{Gaensler}}}, \bibinfo {author} {\bibfnamefont {V.~M.}\ \bibnamefont
  {{Kaspi}}}, \bibinfo {author} {\bibfnamefont {M.}~\bibnamefont {{van der
  Klis}}}, \ and\ \bibinfo {author} {\bibfnamefont {W.~H.~G.}\ \bibnamefont
  {{Lewin}}},\ }\href {\doibase 10.1126/science.1079841} {\bibfield  {journal}
  {\bibinfo  {journal} {Science}\ }\textbf {\bibinfo {volume} {299}},\ \bibinfo
  {pages} {1372} (\bibinfo {year} {2003})},\ \Eprint
  {http://arxiv.org/abs/astro-ph/0302588} {arXiv:astro-ph/0302588 [astro-ph]}
  \BibitemShut {NoStop}%
\bibitem [{\citenamefont {{Hui}}\ and\ \citenamefont {{Becker}}(2006)}]{Hui06}%
  \BibitemOpen
  \bibfield  {author} {\bibinfo {author} {\bibfnamefont {C.~Y.}\ \bibnamefont
  {{Hui}}}\ and\ \bibinfo {author} {\bibfnamefont {W.}~\bibnamefont
  {{Becker}}},\ }\href {\doibase 10.1051/0004-6361:200600008} {\bibfield
  {journal} {\bibinfo  {journal} {\aap}\ }\textbf {\bibinfo {volume} {448}},\
  \bibinfo {pages} {L13} (\bibinfo {year} {2006})},\ \Eprint
  {http://arxiv.org/abs/astro-ph/0601189} {arXiv:astro-ph/0601189 [astro-ph]}
  \BibitemShut {NoStop}%
\bibitem [{\citenamefont {{Lee}}\ \emph {et~al.}(2018)\citenamefont {{Lee}},
  \citenamefont {{Hui}}, \citenamefont {{Takata}},\ and\ \citenamefont
  {{Lin}}}]{Lee18}%
  \BibitemOpen
  \bibfield  {author} {\bibinfo {author} {\bibfnamefont {J.}~\bibnamefont
  {{Lee}}}, \bibinfo {author} {\bibfnamefont {C.~Y.}\ \bibnamefont {{Hui}}},
  \bibinfo {author} {\bibfnamefont {J.}~\bibnamefont {{Takata}}}, \ and\
  \bibinfo {author} {\bibfnamefont {L.~C.~C.}\ \bibnamefont {{Lin}}},\ }\href
  {\doibase 10.1051/0004-6361/201833760} {\bibfield  {journal} {\bibinfo
  {journal} {\aap}\ }\textbf {\bibinfo {volume} {620}},\ \bibinfo {eid} {L14}
  (\bibinfo {year} {2018})},\ \Eprint {http://arxiv.org/abs/1811.03284}
  {arXiv:1811.03284 [astro-ph.HE]} \BibitemShut {NoStop}%
\bibitem [{\citenamefont {{MAGIC Collaboration}}(2019)}]{MAGIC19gc}%
  \BibitemOpen
  \bibfield  {author} {\bibinfo {author} {\bibnamefont {{MAGIC
  Collaboration}}},\ }\href {\doibase 10.1093/mnras/stz179} {\bibfield
  {journal} {\bibinfo  {journal} {\mnras}\ }\textbf {\bibinfo {volume} {484}},\
  \bibinfo {pages} {2876} (\bibinfo {year} {2019})},\ \Eprint
  {http://arxiv.org/abs/1901.04367} {arXiv:1901.04367 [astro-ph.HE]}
  \BibitemShut {NoStop}%
\bibitem [{\citenamefont {{Cheng}}\ \emph {et~al.}(2010)\citenamefont
  {{Cheng}}, \citenamefont {{Chernyshov}}, \citenamefont {{Dogiel}},
  \citenamefont {{Hui}},\ and\ \citenamefont {{Kong}}}]{Cheng10}%
  \BibitemOpen
  \bibfield  {author} {\bibinfo {author} {\bibfnamefont {K.~S.}\ \bibnamefont
  {{Cheng}}}, \bibinfo {author} {\bibfnamefont {D.~O.}\ \bibnamefont
  {{Chernyshov}}}, \bibinfo {author} {\bibfnamefont {V.~A.}\ \bibnamefont
  {{Dogiel}}}, \bibinfo {author} {\bibfnamefont {C.~Y.}\ \bibnamefont {{Hui}}},
  \ and\ \bibinfo {author} {\bibfnamefont {A.~K.~H.}\ \bibnamefont {{Kong}}},\
  }\href {\doibase 10.1088/0004-637X/723/2/1219} {\bibfield  {journal}
  {\bibinfo  {journal} {\apj}\ }\textbf {\bibinfo {volume} {723}},\ \bibinfo
  {pages} {1219} (\bibinfo {year} {2010})},\ \Eprint
  {http://arxiv.org/abs/1009.2278} {arXiv:1009.2278 [astro-ph.HE]} \BibitemShut
  {NoStop}%
\bibitem [{\citenamefont {{Bednarek}}\ and\ \citenamefont
  {{Sitarek}}(2007)}]{Bednarek07}%
  \BibitemOpen
  \bibfield  {author} {\bibinfo {author} {\bibfnamefont {W.}~\bibnamefont
  {{Bednarek}}}\ and\ \bibinfo {author} {\bibfnamefont {J.}~\bibnamefont
  {{Sitarek}}},\ }\href {\doibase 10.1111/j.1365-2966.2007.11664.x} {\bibfield
  {journal} {\bibinfo  {journal} {\mnras}\ }\textbf {\bibinfo {volume} {377}},\
  \bibinfo {pages} {920} (\bibinfo {year} {2007})},\ \Eprint
  {http://arxiv.org/abs/astro-ph/0701522} {arXiv:astro-ph/0701522 [astro-ph]}
  \BibitemShut {NoStop}%
\bibitem [{\citenamefont {{Harding}}\ and\ \citenamefont
  {{Muslimov}}(2011)}]{Harding11}%
  \BibitemOpen
  \bibfield  {author} {\bibinfo {author} {\bibfnamefont {A.~K.}\ \bibnamefont
  {{Harding}}}\ and\ \bibinfo {author} {\bibfnamefont {A.~G.}\ \bibnamefont
  {{Muslimov}}},\ }\href {\doibase 10.1088/0004-637X/743/2/181} {\bibfield
  {journal} {\bibinfo  {journal} {\apj}\ }\textbf {\bibinfo {volume} {743}},\
  \bibinfo {eid} {181} (\bibinfo {year} {2011})},\ \Eprint
  {http://arxiv.org/abs/1111.1668} {arXiv:1111.1668 [astro-ph.HE]} \BibitemShut
  {NoStop}%
\bibitem [{\citenamefont {{Kisaka}}\ and\ \citenamefont
  {{Kawanaka}}(2012)}]{Kisaka12}%
  \BibitemOpen
  \bibfield  {author} {\bibinfo {author} {\bibfnamefont {S.}~\bibnamefont
  {{Kisaka}}}\ and\ \bibinfo {author} {\bibfnamefont {N.}~\bibnamefont
  {{Kawanaka}}},\ }\href {\doibase 10.1111/j.1365-2966.2012.20576.x} {\bibfield
   {journal} {\bibinfo  {journal} {\mnras}\ }\textbf {\bibinfo {volume}
  {421}},\ \bibinfo {pages} {3543} (\bibinfo {year} {2012})},\ \Eprint
  {http://arxiv.org/abs/1112.5312} {arXiv:1112.5312 [astro-ph.HE]} \BibitemShut
  {NoStop}%
\bibitem [{\citenamefont {Venter}\ \emph {et~al.}(2015)\citenamefont {Venter},
  \citenamefont {Kopp}, \citenamefont {Harding}, \citenamefont {Gonthier},\
  and\ \citenamefont {Büsching}}]{Venter:2015gga}%
  \BibitemOpen
  \bibfield  {author} {\bibinfo {author} {\bibfnamefont {C.}~\bibnamefont
  {Venter}}, \bibinfo {author} {\bibfnamefont {A.}~\bibnamefont {Kopp}},
  \bibinfo {author} {\bibfnamefont {A.}~\bibnamefont {Harding}}, \bibinfo
  {author} {\bibfnamefont {P.}~\bibnamefont {Gonthier}}, \ and\ \bibinfo
  {author} {\bibfnamefont {I.}~\bibnamefont {Büsching}},\ }\href {\doibase
  10.1088/0004-637X/807/2/130} {\bibfield  {journal} {\bibinfo  {journal}
  {Astrophys. J.}\ }\textbf {\bibinfo {volume} {807}},\ \bibinfo {pages} {130}
  (\bibinfo {year} {2015})},\ \Eprint {http://arxiv.org/abs/1506.01211}
  {arXiv:1506.01211 [astro-ph.HE]} \BibitemShut {NoStop}%
\bibitem [{\citenamefont {{Yuan}}\ and\ \citenamefont {{Ioka}}(2015)}]{Yuan15}%
  \BibitemOpen
  \bibfield  {author} {\bibinfo {author} {\bibfnamefont {Q.}~\bibnamefont
  {{Yuan}}}\ and\ \bibinfo {author} {\bibfnamefont {K.}~\bibnamefont
  {{Ioka}}},\ }\href {\doibase 10.1088/0004-637X/802/2/124} {\bibfield
  {journal} {\bibinfo  {journal} {\apj}\ }\textbf {\bibinfo {volume} {802}},\
  \bibinfo {eid} {124} (\bibinfo {year} {2015})},\ \Eprint
  {http://arxiv.org/abs/1411.4363} {arXiv:1411.4363 [astro-ph.HE]} \BibitemShut
  {NoStop}%
\bibitem [{\citenamefont {{Petrovi{\'c}}}\ \emph {et~al.}(2015)\citenamefont
  {{Petrovi{\'c}}}, \citenamefont {{Serpico}},\ and\ \citenamefont
  {{Zaharijas}}}]{Petrovic15}%
  \BibitemOpen
  \bibfield  {author} {\bibinfo {author} {\bibfnamefont {J.}~\bibnamefont
  {{Petrovi{\'c}}}}, \bibinfo {author} {\bibfnamefont {P.~D.}\ \bibnamefont
  {{Serpico}}}, \ and\ \bibinfo {author} {\bibfnamefont {G.}~\bibnamefont
  {{Zaharijas}}},\ }\href {\doibase 10.1088/1475-7516/2015/02/023} {\bibfield
  {journal} {\bibinfo  {journal} {\jcap}\ }\textbf {\bibinfo {volume} {2015}},\
  \bibinfo {eid} {023} (\bibinfo {year} {2015})},\ \Eprint
  {http://arxiv.org/abs/1411.2980} {arXiv:1411.2980 [astro-ph.HE]} \BibitemShut
  {NoStop}%
\bibitem [{\citenamefont {{Bednarek}}\ \emph {et~al.}(2016)\citenamefont
  {{Bednarek}}, \citenamefont {{Sitarek}},\ and\ \citenamefont
  {{Sobczak}}}]{Bednarek16}%
  \BibitemOpen
  \bibfield  {author} {\bibinfo {author} {\bibfnamefont {W.}~\bibnamefont
  {{Bednarek}}}, \bibinfo {author} {\bibfnamefont {J.}~\bibnamefont
  {{Sitarek}}}, \ and\ \bibinfo {author} {\bibfnamefont {T.}~\bibnamefont
  {{Sobczak}}},\ }\href {\doibase 10.1093/mnras/stw367} {\bibfield  {journal}
  {\bibinfo  {journal} {\mnras}\ }\textbf {\bibinfo {volume} {458}},\ \bibinfo
  {pages} {1083} (\bibinfo {year} {2016})},\ \Eprint
  {http://arxiv.org/abs/1602.03629} {arXiv:1602.03629 [astro-ph.HE]}
  \BibitemShut {NoStop}%
\bibitem [{\citenamefont {{Song}}\ \emph {et~al.}(2019)\citenamefont {{Song}},
  \citenamefont {{Macias}},\ and\ \citenamefont {{Horiuchi}}}]{Song19}%
  \BibitemOpen
  \bibfield  {author} {\bibinfo {author} {\bibfnamefont {D.}~\bibnamefont
  {{Song}}}, \bibinfo {author} {\bibfnamefont {O.}~\bibnamefont {{Macias}}}, \
  and\ \bibinfo {author} {\bibfnamefont {S.}~\bibnamefont {{Horiuchi}}},\
  }\href {\doibase 10.1103/PhysRevD.99.123020} {\bibfield  {journal} {\bibinfo
  {journal} {\prd}\ }\textbf {\bibinfo {volume} {99}},\ \bibinfo {eid} {123020}
  (\bibinfo {year} {2019})},\ \Eprint {http://arxiv.org/abs/1901.07025}
  {arXiv:1901.07025 [astro-ph.HE]} \BibitemShut {NoStop}%
\bibitem [{\citenamefont {{Ndiyavala}}\ \emph {et~al.}(2019)\citenamefont
  {{Ndiyavala}}, \citenamefont {{Venter}}, \citenamefont {{Johnson}},
  \citenamefont {{Harding}}, \citenamefont {{Smith}}, \citenamefont {{Eger}},
  \citenamefont {{Kopp}},\ and\ \citenamefont {{van der Walt}}}]{Ndiyavala19}%
  \BibitemOpen
  \bibfield  {author} {\bibinfo {author} {\bibfnamefont {H.}~\bibnamefont
  {{Ndiyavala}}}, \bibinfo {author} {\bibfnamefont {C.}~\bibnamefont
  {{Venter}}}, \bibinfo {author} {\bibfnamefont {T.~J.}\ \bibnamefont
  {{Johnson}}}, \bibinfo {author} {\bibfnamefont {A.~K.}\ \bibnamefont
  {{Harding}}}, \bibinfo {author} {\bibfnamefont {D.~A.}\ \bibnamefont
  {{Smith}}}, \bibinfo {author} {\bibfnamefont {P.}~\bibnamefont {{Eger}}},
  \bibinfo {author} {\bibfnamefont {A.}~\bibnamefont {{Kopp}}}, \ and\ \bibinfo
  {author} {\bibfnamefont {D.~J.}\ \bibnamefont {{van der Walt}}},\ }\href
  {\doibase 10.3847/1538-4357/ab24ca} {\bibfield  {journal} {\bibinfo
  {journal} {\apj}\ }\textbf {\bibinfo {volume} {880}},\ \bibinfo {eid} {53}
  (\bibinfo {year} {2019})},\ \Eprint {http://arxiv.org/abs/1905.10229}
  {arXiv:1905.10229 [astro-ph.HE]} \BibitemShut {NoStop}%
\bibitem [{\citenamefont {{Bykov}}\ \emph {et~al.}(2019)\citenamefont
  {{Bykov}}, \citenamefont {{Petrov}}, \citenamefont {{Krassilchtchikov}},
  \citenamefont {{Levenfish}}, \citenamefont {{Osipov}},\ and\ \citenamefont
  {{Pavlov}}}]{Bykov19}%
  \BibitemOpen
  \bibfield  {author} {\bibinfo {author} {\bibfnamefont {A.~M.}\ \bibnamefont
  {{Bykov}}}, \bibinfo {author} {\bibfnamefont {A.~E.}\ \bibnamefont
  {{Petrov}}}, \bibinfo {author} {\bibfnamefont {A.~M.}\ \bibnamefont
  {{Krassilchtchikov}}}, \bibinfo {author} {\bibfnamefont {K.~P.}\ \bibnamefont
  {{Levenfish}}}, \bibinfo {author} {\bibfnamefont {S.~M.}\ \bibnamefont
  {{Osipov}}}, \ and\ \bibinfo {author} {\bibfnamefont {G.~G.}\ \bibnamefont
  {{Pavlov}}},\ }\href {\doibase 10.3847/2041-8213/ab1922} {\bibfield
  {journal} {\bibinfo  {journal} {\apjl}\ }\textbf {\bibinfo {volume} {876}},\
  \bibinfo {eid} {L8} (\bibinfo {year} {2019})},\ \Eprint
  {http://arxiv.org/abs/1904.09430} {arXiv:1904.09430 [astro-ph.HE]}
  \BibitemShut {NoStop}%
\bibitem [{\citenamefont {{Atoyan}}\ \emph {et~al.}(1995)\citenamefont
  {{Atoyan}}, \citenamefont {{Aharonian}},\ and\ \citenamefont
  {{V{\"o}lk}}}]{Atoyan95}%
  \BibitemOpen
  \bibfield  {author} {\bibinfo {author} {\bibfnamefont {A.~M.}\ \bibnamefont
  {{Atoyan}}}, \bibinfo {author} {\bibfnamefont {F.~A.}\ \bibnamefont
  {{Aharonian}}}, \ and\ \bibinfo {author} {\bibfnamefont {H.~J.}\ \bibnamefont
  {{V{\"o}lk}}},\ }\href {\doibase 10.1103/PhysRevD.52.3265} {\bibfield
  {journal} {\bibinfo  {journal} {\prd}\ }\textbf {\bibinfo {volume} {52}},\
  \bibinfo {pages} {3265} (\bibinfo {year} {1995})}\BibitemShut {NoStop}%
\bibitem [{\citenamefont {{Beck}}\ \emph {et~al.}(2019)\citenamefont {{Beck}},
  \citenamefont {{Chamandy}}, \citenamefont {{Elson}},\ and\ \citenamefont
  {{Blackman}}}]{Beck19}%
  \BibitemOpen
  \bibfield  {author} {\bibinfo {author} {\bibfnamefont {R.}~\bibnamefont
  {{Beck}}}, \bibinfo {author} {\bibfnamefont {L.}~\bibnamefont {{Chamandy}}},
  \bibinfo {author} {\bibfnamefont {E.}~\bibnamefont {{Elson}}}, \ and\
  \bibinfo {author} {\bibfnamefont {E.~G.}\ \bibnamefont {{Blackman}}},\ }\href
  {\doibase 10.3390/galaxies8010004} {\bibfield  {journal} {\bibinfo  {journal}
  {Galaxies}\ }\textbf {\bibinfo {volume} {8}},\ \bibinfo {pages} {4} (\bibinfo
  {year} {2019})},\ \Eprint {http://arxiv.org/abs/1912.08962} {arXiv:1912.08962
  [astro-ph.GA]} \BibitemShut {NoStop}%
\bibitem [{\citenamefont {{Thompson}}\ \emph {et~al.}(2006)\citenamefont
  {{Thompson}}, \citenamefont {{Quataert}}, \citenamefont {{Waxman}},
  \citenamefont {{Murray}},\ and\ \citenamefont {{Martin}}}]{Thompson06}%
  \BibitemOpen
  \bibfield  {author} {\bibinfo {author} {\bibfnamefont {T.~A.}\ \bibnamefont
  {{Thompson}}}, \bibinfo {author} {\bibfnamefont {E.}~\bibnamefont
  {{Quataert}}}, \bibinfo {author} {\bibfnamefont {E.}~\bibnamefont
  {{Waxman}}}, \bibinfo {author} {\bibfnamefont {N.}~\bibnamefont {{Murray}}},
  \ and\ \bibinfo {author} {\bibfnamefont {C.~L.}\ \bibnamefont {{Martin}}},\
  }\href {\doibase 10.1086/504035} {\bibfield  {journal} {\bibinfo  {journal}
  {\apj}\ }\textbf {\bibinfo {volume} {645}},\ \bibinfo {pages} {186} (\bibinfo
  {year} {2006})},\ \Eprint {http://arxiv.org/abs/astro-ph/0601626}
  {arXiv:astro-ph/0601626 [astro-ph]} \BibitemShut {NoStop}%
\bibitem [{\citenamefont {{Mathews}}\ and\ \citenamefont
  {{Brighenti}}(2003)}]{Mathews03}%
  \BibitemOpen
  \bibfield  {author} {\bibinfo {author} {\bibfnamefont {W.~G.}\ \bibnamefont
  {{Mathews}}}\ and\ \bibinfo {author} {\bibfnamefont {F.}~\bibnamefont
  {{Brighenti}}},\ }\href {\doibase 10.1146/annurev.astro.41.090401.094542}
  {\bibfield  {journal} {\bibinfo  {journal} {\araa}\ }\textbf {\bibinfo
  {volume} {41}},\ \bibinfo {pages} {191} (\bibinfo {year} {2003})},\ \Eprint
  {http://arxiv.org/abs/astro-ph/0309553} {arXiv:astro-ph/0309553 [astro-ph]}
  \BibitemShut {NoStop}%
\bibitem [{\citenamefont {{Evoli}}\ \emph {et~al.}(2020)\citenamefont
  {{Evoli}}, \citenamefont {{Morlino}}, \citenamefont {{Blasi}},\ and\
  \citenamefont {{Aloisio}}}]{Evoli20}%
  \BibitemOpen
  \bibfield  {author} {\bibinfo {author} {\bibfnamefont {C.}~\bibnamefont
  {{Evoli}}}, \bibinfo {author} {\bibfnamefont {G.}~\bibnamefont {{Morlino}}},
  \bibinfo {author} {\bibfnamefont {P.}~\bibnamefont {{Blasi}}}, \ and\
  \bibinfo {author} {\bibfnamefont {R.}~\bibnamefont {{Aloisio}}},\ }\href
  {\doibase 10.1103/PhysRevD.101.023013} {\bibfield  {journal} {\bibinfo
  {journal} {\prd}\ }\textbf {\bibinfo {volume} {101}},\ \bibinfo {eid}
  {023013} (\bibinfo {year} {2020})},\ \Eprint
  {http://arxiv.org/abs/1910.04113} {arXiv:1910.04113 [astro-ph.HE]}
  \BibitemShut {NoStop}%
\bibitem [{\citenamefont {{Morlino}}\ and\ \citenamefont
  {{Amato}}(2020)}]{Morlino20}%
  \BibitemOpen
  \bibfield  {author} {\bibinfo {author} {\bibfnamefont {G.}~\bibnamefont
  {{Morlino}}}\ and\ \bibinfo {author} {\bibfnamefont {E.}~\bibnamefont
  {{Amato}}},\ }\href {\doibase 10.1103/PhysRevD.101.083017} {\bibfield
  {journal} {\bibinfo  {journal} {\prd}\ }\textbf {\bibinfo {volume} {101}},\
  \bibinfo {eid} {083017} (\bibinfo {year} {2020})},\ \Eprint
  {http://arxiv.org/abs/2003.04700} {arXiv:2003.04700 [astro-ph.HE]}
  \BibitemShut {NoStop}%
\bibitem [{\citenamefont {{Cowsik}}\ and\ \citenamefont
  {{Madziwa-Nussinov}}(2016)}]{Cowsik16}%
  \BibitemOpen
  \bibfield  {author} {\bibinfo {author} {\bibfnamefont {R.}~\bibnamefont
  {{Cowsik}}}\ and\ \bibinfo {author} {\bibfnamefont {T.}~\bibnamefont
  {{Madziwa-Nussinov}}},\ }\href {\doibase 10.3847/0004-637X/827/2/119}
  {\bibfield  {journal} {\bibinfo  {journal} {\apj}\ }\textbf {\bibinfo
  {volume} {827}},\ \bibinfo {eid} {119} (\bibinfo {year} {2016})},\ \Eprint
  {http://arxiv.org/abs/1505.00305} {arXiv:1505.00305 [astro-ph.HE]}
  \BibitemShut {NoStop}%
\bibitem [{\citenamefont {{Lipari}}(2017)}]{Lipari17}%
  \BibitemOpen
  \bibfield  {author} {\bibinfo {author} {\bibfnamefont {P.}~\bibnamefont
  {{Lipari}}},\ }\href {\doibase 10.1103/PhysRevD.95.063009} {\bibfield
  {journal} {\bibinfo  {journal} {\prd}\ }\textbf {\bibinfo {volume} {95}},\
  \bibinfo {eid} {063009} (\bibinfo {year} {2017})},\ \Eprint
  {http://arxiv.org/abs/1608.02018} {arXiv:1608.02018 [astro-ph.HE]}
  \BibitemShut {NoStop}%
\bibitem [{\citenamefont {{Fujita}}\ \emph {et~al.}(2010)\citenamefont
  {{Fujita}}, \citenamefont {{Ohira}},\ and\ \citenamefont
  {{Takahara}}}]{Fujita10}%
  \BibitemOpen
  \bibfield  {author} {\bibinfo {author} {\bibfnamefont {Y.}~\bibnamefont
  {{Fujita}}}, \bibinfo {author} {\bibfnamefont {Y.}~\bibnamefont {{Ohira}}}, \
  and\ \bibinfo {author} {\bibfnamefont {F.}~\bibnamefont {{Takahara}}},\
  }\href {\doibase 10.1088/2041-8205/712/2/L153} {\bibfield  {journal}
  {\bibinfo  {journal} {\apjl}\ }\textbf {\bibinfo {volume} {712}},\ \bibinfo
  {pages} {L153} (\bibinfo {year} {2010})},\ \Eprint
  {http://arxiv.org/abs/1002.4871} {arXiv:1002.4871 [astro-ph.HE]} \BibitemShut
  {NoStop}%
\bibitem [{\citenamefont {{Fujita}}\ \emph {et~al.}(2011)\citenamefont
  {{Fujita}}, \citenamefont {{Takahara}}, \citenamefont {{Ohira}},\ and\
  \citenamefont {{Iwasaki}}}]{Fujita11}%
  \BibitemOpen
  \bibfield  {author} {\bibinfo {author} {\bibfnamefont {Y.}~\bibnamefont
  {{Fujita}}}, \bibinfo {author} {\bibfnamefont {F.}~\bibnamefont
  {{Takahara}}}, \bibinfo {author} {\bibfnamefont {Y.}~\bibnamefont {{Ohira}}},
  \ and\ \bibinfo {author} {\bibfnamefont {K.}~\bibnamefont {{Iwasaki}}},\
  }\href {\doibase 10.1111/j.1365-2966.2011.18980.x} {\bibfield  {journal}
  {\bibinfo  {journal} {\mnras}\ }\textbf {\bibinfo {volume} {415}},\ \bibinfo
  {pages} {3434} (\bibinfo {year} {2011})},\ \Eprint
  {http://arxiv.org/abs/1105.0683} {arXiv:1105.0683 [astro-ph.HE]} \BibitemShut
  {NoStop}%
\bibitem [{\citenamefont {{Malkov}}\ \emph {et~al.}(2013)\citenamefont
  {{Malkov}}, \citenamefont {{Diamond}}, \citenamefont {{Sagdeev}},
  \citenamefont {{Aharonian}},\ and\ \citenamefont {{Moskalenko}}}]{Malkov13}%
  \BibitemOpen
  \bibfield  {author} {\bibinfo {author} {\bibfnamefont {M.~A.}\ \bibnamefont
  {{Malkov}}}, \bibinfo {author} {\bibfnamefont {P.~H.}\ \bibnamefont
  {{Diamond}}}, \bibinfo {author} {\bibfnamefont {R.~Z.}\ \bibnamefont
  {{Sagdeev}}}, \bibinfo {author} {\bibfnamefont {F.~A.}\ \bibnamefont
  {{Aharonian}}}, \ and\ \bibinfo {author} {\bibfnamefont {I.~V.}\ \bibnamefont
  {{Moskalenko}}},\ }\href {\doibase 10.1088/0004-637X/768/1/73} {\bibfield
  {journal} {\bibinfo  {journal} {\apj}\ }\textbf {\bibinfo {volume} {768}},\
  \bibinfo {eid} {73} (\bibinfo {year} {2013})},\ \Eprint
  {http://arxiv.org/abs/1207.4728} {arXiv:1207.4728 [astro-ph.HE]} \BibitemShut
  {NoStop}%
\bibitem [{\citenamefont {{Nava}}\ \emph {et~al.}(2016)\citenamefont {{Nava}},
  \citenamefont {{Gabici}}, \citenamefont {{Marcowith}}, \citenamefont
  {{Morlino}},\ and\ \citenamefont {{Ptuskin}}}]{Nava16}%
  \BibitemOpen
  \bibfield  {author} {\bibinfo {author} {\bibfnamefont {L.}~\bibnamefont
  {{Nava}}}, \bibinfo {author} {\bibfnamefont {S.}~\bibnamefont {{Gabici}}},
  \bibinfo {author} {\bibfnamefont {A.}~\bibnamefont {{Marcowith}}}, \bibinfo
  {author} {\bibfnamefont {G.}~\bibnamefont {{Morlino}}}, \ and\ \bibinfo
  {author} {\bibfnamefont {V.~S.}\ \bibnamefont {{Ptuskin}}},\ }\href {\doibase
  10.1093/mnras/stw1592} {\bibfield  {journal} {\bibinfo  {journal} {\mnras}\
  }\textbf {\bibinfo {volume} {461}},\ \bibinfo {pages} {3552} (\bibinfo {year}
  {2016})},\ \Eprint {http://arxiv.org/abs/1606.06902} {arXiv:1606.06902
  [astro-ph.HE]} \BibitemShut {NoStop}%
\bibitem [{\citenamefont {D'Angelo}\ \emph {et~al.}(2018)\citenamefont
  {D'Angelo}, \citenamefont {Morlino}, \citenamefont {Amato},\ and\
  \citenamefont {Blasi}}]{DAngelo:2017rou}%
  \BibitemOpen
  \bibfield  {author} {\bibinfo {author} {\bibfnamefont {M.}~\bibnamefont
  {D'Angelo}}, \bibinfo {author} {\bibfnamefont {G.}~\bibnamefont {Morlino}},
  \bibinfo {author} {\bibfnamefont {E.}~\bibnamefont {Amato}}, \ and\ \bibinfo
  {author} {\bibfnamefont {P.}~\bibnamefont {Blasi}},\ }\href {\doibase
  10.1093/mnras/stx2828} {\bibfield  {journal} {\bibinfo  {journal} {Mon. Not.
  Roy. Astron. Soc.}\ }\textbf {\bibinfo {volume} {474}},\ \bibinfo {pages}
  {1944} (\bibinfo {year} {2018})},\ \Eprint {http://arxiv.org/abs/1710.10937}
  {arXiv:1710.10937 [astro-ph.HE]} \BibitemShut {NoStop}%
\bibitem [{\citenamefont {Evoli}\ \emph {et~al.}(2018)\citenamefont {Evoli},
  \citenamefont {Linden},\ and\ \citenamefont {Morlino}}]{Evoli:2018aza}%
  \BibitemOpen
  \bibfield  {author} {\bibinfo {author} {\bibfnamefont {C.}~\bibnamefont
  {Evoli}}, \bibinfo {author} {\bibfnamefont {T.}~\bibnamefont {Linden}}, \
  and\ \bibinfo {author} {\bibfnamefont {G.}~\bibnamefont {Morlino}},\ }\href
  {\doibase 10.1103/PhysRevD.98.063017} {\bibfield  {journal} {\bibinfo
  {journal} {Phys. Rev. D}\ }\textbf {\bibinfo {volume} {98}},\ \bibinfo
  {pages} {063017} (\bibinfo {year} {2018})},\ \Eprint
  {http://arxiv.org/abs/1807.09263} {arXiv:1807.09263 [astro-ph.HE]}
  \BibitemShut {NoStop}%
\bibitem [{\citenamefont {{Fang}}\ \emph {et~al.}(2019)\citenamefont {{Fang}},
  \citenamefont {{Bi}},\ and\ \citenamefont {{Yin}}}]{Fang19}%
  \BibitemOpen
  \bibfield  {author} {\bibinfo {author} {\bibfnamefont {K.}~\bibnamefont
  {{Fang}}}, \bibinfo {author} {\bibfnamefont {X.-J.}\ \bibnamefont {{Bi}}}, \
  and\ \bibinfo {author} {\bibfnamefont {P.-F.}\ \bibnamefont {{Yin}}},\ }\href
  {\doibase 10.1093/mnras/stz1974} {\bibfield  {journal} {\bibinfo  {journal}
  {\mnras}\ }\textbf {\bibinfo {volume} {488}},\ \bibinfo {pages} {4074}
  (\bibinfo {year} {2019})},\ \Eprint {http://arxiv.org/abs/1903.06421}
  {arXiv:1903.06421 [astro-ph.HE]} \BibitemShut {NoStop}%
\bibitem [{\citenamefont {{Israel}}\ and\ \citenamefont
  {{Mahoney}}(1990)}]{Israel90}%
  \BibitemOpen
  \bibfield  {author} {\bibinfo {author} {\bibfnamefont {F.~P.}\ \bibnamefont
  {{Israel}}}\ and\ \bibinfo {author} {\bibfnamefont {M.~J.}\ \bibnamefont
  {{Mahoney}}},\ }\href {\doibase 10.1086/168513} {\bibfield  {journal}
  {\bibinfo  {journal} {\apj}\ }\textbf {\bibinfo {volume} {352}},\ \bibinfo
  {pages} {30} (\bibinfo {year} {1990})}\BibitemShut {NoStop}%
\bibitem [{\citenamefont {{Hummel}}(1991)}]{Hummel91}%
  \BibitemOpen
  \bibfield  {author} {\bibinfo {author} {\bibfnamefont {E.}~\bibnamefont
  {{Hummel}}},\ }\href@noop {} {\bibfield  {journal} {\bibinfo  {journal}
  {\aap}\ }\textbf {\bibinfo {volume} {251}},\ \bibinfo {pages} {442} (\bibinfo
  {year} {1991})}\BibitemShut {NoStop}%
\bibitem [{\citenamefont {{Basu}}\ \emph {et~al.}(2015)\citenamefont {{Basu}},
  \citenamefont {{Beck}}, \citenamefont {{Schmidt}},\ and\ \citenamefont
  {{Roy}}}]{Basu15}%
  \BibitemOpen
  \bibfield  {author} {\bibinfo {author} {\bibfnamefont {A.}~\bibnamefont
  {{Basu}}}, \bibinfo {author} {\bibfnamefont {R.}~\bibnamefont {{Beck}}},
  \bibinfo {author} {\bibfnamefont {P.}~\bibnamefont {{Schmidt}}}, \ and\
  \bibinfo {author} {\bibfnamefont {S.}~\bibnamefont {{Roy}}},\ }\href
  {\doibase 10.1093/mnras/stv510} {\bibfield  {journal} {\bibinfo  {journal}
  {\mnras}\ }\textbf {\bibinfo {volume} {449}},\ \bibinfo {pages} {3879}
  (\bibinfo {year} {2015})},\ \Eprint {http://arxiv.org/abs/1503.02420}
  {arXiv:1503.02420 [astro-ph.GA]} \BibitemShut {NoStop}%
\bibitem [{\citenamefont {{Marvil}}\ \emph {et~al.}(2015)\citenamefont
  {{Marvil}}, \citenamefont {{Owen}},\ and\ \citenamefont
  {{Eilek}}}]{Marvil15}%
  \BibitemOpen
  \bibfield  {author} {\bibinfo {author} {\bibfnamefont {J.}~\bibnamefont
  {{Marvil}}}, \bibinfo {author} {\bibfnamefont {F.}~\bibnamefont {{Owen}}}, \
  and\ \bibinfo {author} {\bibfnamefont {J.}~\bibnamefont {{Eilek}}},\ }\href
  {\doibase 10.1088/0004-6256/149/1/32} {\bibfield  {journal} {\bibinfo
  {journal} {\aj}\ }\textbf {\bibinfo {volume} {149}},\ \bibinfo {eid} {32}
  (\bibinfo {year} {2015})},\ \Eprint {http://arxiv.org/abs/1408.6296}
  {arXiv:1408.6296 [astro-ph.GA]} \BibitemShut {NoStop}%
\bibitem [{\citenamefont {{Chy{\.z}y}}\ \emph {et~al.}(2018)\citenamefont
  {{Chy{\.z}y}} \emph {et~al.}}]{Chyzy18}%
  \BibitemOpen
  \bibfield  {author} {\bibinfo {author} {\bibfnamefont {K.~T.}\ \bibnamefont
  {{Chy{\.z}y}}} \emph {et~al.},\ }\href {\doibase 10.1051/0004-6361/201833133}
  {\bibfield  {journal} {\bibinfo  {journal} {\aap}\ }\textbf {\bibinfo
  {volume} {619}},\ \bibinfo {eid} {A36} (\bibinfo {year} {2018})},\ \Eprint
  {http://arxiv.org/abs/1808.10374} {arXiv:1808.10374 [astro-ph.GA]}
  \BibitemShut {NoStop}%
\bibitem [{\citenamefont {{Torres}}(2004)}]{Torres04}%
  \BibitemOpen
  \bibfield  {author} {\bibinfo {author} {\bibfnamefont {D.~F.}\ \bibnamefont
  {{Torres}}},\ }\href {\doibase 10.1086/425415} {\bibfield  {journal}
  {\bibinfo  {journal} {\apj}\ }\textbf {\bibinfo {volume} {617}},\ \bibinfo
  {pages} {966} (\bibinfo {year} {2004})},\ \Eprint
  {http://arxiv.org/abs/astro-ph/0407240} {arXiv:astro-ph/0407240 [astro-ph]}
  \BibitemShut {NoStop}%
\bibitem [{\citenamefont {{Clemens}}\ \emph {et~al.}(2010)\citenamefont
  {{Clemens}}, \citenamefont {{Scaife}}, \citenamefont {{Vega}},\ and\
  \citenamefont {{Bressan}}}]{Clemens10}%
  \BibitemOpen
  \bibfield  {author} {\bibinfo {author} {\bibfnamefont {M.~S.}\ \bibnamefont
  {{Clemens}}}, \bibinfo {author} {\bibfnamefont {A.}~\bibnamefont {{Scaife}}},
  \bibinfo {author} {\bibfnamefont {O.}~\bibnamefont {{Vega}}}, \ and\ \bibinfo
  {author} {\bibfnamefont {A.}~\bibnamefont {{Bressan}}},\ }\href {\doibase
  10.1111/j.1365-2966.2010.16534.x} {\bibfield  {journal} {\bibinfo  {journal}
  {\mnras}\ }\textbf {\bibinfo {volume} {405}},\ \bibinfo {pages} {887}
  (\bibinfo {year} {2010})},\ \Eprint {http://arxiv.org/abs/1002.3334}
  {arXiv:1002.3334 [astro-ph.GA]} \BibitemShut {NoStop}%
\bibitem [{\citenamefont {{Best}}\ and\ \citenamefont
  {{Heckman}}(2012)}]{BH12}%
  \BibitemOpen
  \bibfield  {author} {\bibinfo {author} {\bibfnamefont {P.~N.}\ \bibnamefont
  {{Best}}}\ and\ \bibinfo {author} {\bibfnamefont {T.~M.}\ \bibnamefont
  {{Heckman}}},\ }\href {\doibase 10.1111/j.1365-2966.2012.20414.x} {\bibfield
  {journal} {\bibinfo  {journal} {\mnras}\ }\textbf {\bibinfo {volume} {421}},\
  \bibinfo {pages} {1569} (\bibinfo {year} {2012})},\ \Eprint
  {http://arxiv.org/abs/1201.2397} {arXiv:1201.2397 [astro-ph.CO]} \BibitemShut
  {NoStop}%
\bibitem [{\citenamefont {{James}}\ and\ \citenamefont
  {{Roos}}(1975)}]{minuit}%
  \BibitemOpen
  \bibfield  {author} {\bibinfo {author} {\bibfnamefont {F.}~\bibnamefont
  {{James}}}\ and\ \bibinfo {author} {\bibfnamefont {M.}~\bibnamefont
  {{Roos}}},\ }\href {\doibase 10.1016/0010-4655(75)90039-9} {\bibfield
  {journal} {\bibinfo  {journal} {Computer Physics Communications}\ }\textbf
  {\bibinfo {volume} {10}},\ \bibinfo {pages} {343} (\bibinfo {year}
  {1975})}\BibitemShut {NoStop}%
\bibitem [{\citenamefont {{da Cunha}}\ \emph {et~al.}(2008)\citenamefont {{da
  Cunha}}, \citenamefont {{Charlot}},\ and\ \citenamefont {{Elbaz}}}]{magphys}%
  \BibitemOpen
  \bibfield  {author} {\bibinfo {author} {\bibfnamefont {E.}~\bibnamefont {{da
  Cunha}}}, \bibinfo {author} {\bibfnamefont {S.}~\bibnamefont {{Charlot}}}, \
  and\ \bibinfo {author} {\bibfnamefont {D.}~\bibnamefont {{Elbaz}}},\ }\href
  {\doibase 10.1111/j.1365-2966.2008.13535.x} {\bibfield  {journal} {\bibinfo
  {journal} {\mnras}\ }\textbf {\bibinfo {volume} {388}},\ \bibinfo {pages}
  {1595} (\bibinfo {year} {2008})},\ \Eprint {http://arxiv.org/abs/0806.1020}
  {arXiv:0806.1020 [astro-ph]} \BibitemShut {NoStop}%
\bibitem [{\citenamefont {{Yin}}\ \emph {et~al.}(2009)\citenamefont {{Yin}},
  \citenamefont {{Hou}}, \citenamefont {{Prantzos}}, \citenamefont
  {{Boissier}}, \citenamefont {{Chang}}, \citenamefont {{Shen}},\ and\
  \citenamefont {{Zhang}}}]{Yin09}%
  \BibitemOpen
  \bibfield  {author} {\bibinfo {author} {\bibfnamefont {J.}~\bibnamefont
  {{Yin}}}, \bibinfo {author} {\bibfnamefont {J.~L.}\ \bibnamefont {{Hou}}},
  \bibinfo {author} {\bibfnamefont {N.}~\bibnamefont {{Prantzos}}}, \bibinfo
  {author} {\bibfnamefont {S.}~\bibnamefont {{Boissier}}}, \bibinfo {author}
  {\bibfnamefont {R.~X.}\ \bibnamefont {{Chang}}}, \bibinfo {author}
  {\bibfnamefont {S.~Y.}\ \bibnamefont {{Shen}}}, \ and\ \bibinfo {author}
  {\bibfnamefont {B.}~\bibnamefont {{Zhang}}},\ }\href {\doibase
  10.1051/0004-6361/200912316} {\bibfield  {journal} {\bibinfo  {journal}
  {\aap}\ }\textbf {\bibinfo {volume} {505}},\ \bibinfo {pages} {497} (\bibinfo
  {year} {2009})},\ \Eprint {http://arxiv.org/abs/0906.4821} {arXiv:0906.4821
  [astro-ph.GA]} \BibitemShut {NoStop}%
\bibitem [{\citenamefont {{Sick}}\ \emph {et~al.}(2015)\citenamefont {{Sick}},
  \citenamefont {{Courteau}}, \citenamefont {{Cuilland re}}, \citenamefont
  {{Dalcanton}}, \citenamefont {{de Jong}}, \citenamefont {{McDonald}},
  \citenamefont {{Simard}},\ and\ \citenamefont {{Tully}}}]{Sick15}%
  \BibitemOpen
  \bibfield  {author} {\bibinfo {author} {\bibfnamefont {J.}~\bibnamefont
  {{Sick}}}, \bibinfo {author} {\bibfnamefont {S.}~\bibnamefont {{Courteau}}},
  \bibinfo {author} {\bibfnamefont {J.-C.}\ \bibnamefont {{Cuilland re}}},
  \bibinfo {author} {\bibfnamefont {J.}~\bibnamefont {{Dalcanton}}}, \bibinfo
  {author} {\bibfnamefont {R.}~\bibnamefont {{de Jong}}}, \bibinfo {author}
  {\bibfnamefont {M.}~\bibnamefont {{McDonald}}}, \bibinfo {author}
  {\bibfnamefont {D.}~\bibnamefont {{Simard}}}, \ and\ \bibinfo {author}
  {\bibfnamefont {R.~B.}\ \bibnamefont {{Tully}}},\ }in\ \href {\doibase
  10.1017/S1743921315003440} {\emph {\bibinfo {booktitle} {Galaxy Masses as
  Constraints of Formation Models}}},\ \bibinfo {series} {IAU Symposium}, Vol.\
  \bibinfo {volume} {311},\ \bibinfo {editor} {edited by\ \bibinfo {editor}
  {\bibfnamefont {M.}~\bibnamefont {{Cappellari}}}\ and\ \bibinfo {editor}
  {\bibfnamefont {S.}~\bibnamefont {{Courteau}}}}\ (\bibinfo {year} {2015})\
  pp.\ \bibinfo {pages} {82--85},\ \Eprint {http://arxiv.org/abs/1410.0017}
  {arXiv:1410.0017 [astro-ph.GA]} \BibitemShut {NoStop}%
\bibitem [{\citenamefont {{Johnson}}\ \emph {et~al.}(2014)\citenamefont
  {{Johnson}}, \citenamefont {{Venter}}, \citenamefont {{Harding}},
  \citenamefont {{Guillemot}}, \citenamefont {{Smith}}, \citenamefont
  {{Kramer}}, \citenamefont {{{\c{C}}elik}}, \citenamefont {{den Hartog}},
  \citenamefont {{Ferrara}}, \citenamefont {{Hou}}, \citenamefont {{Lande}},\
  and\ \citenamefont {{Ray}}}]{Johnson14}%
  \BibitemOpen
  \bibfield  {author} {\bibinfo {author} {\bibfnamefont {T.~J.}\ \bibnamefont
  {{Johnson}}}, \bibinfo {author} {\bibfnamefont {C.}~\bibnamefont {{Venter}}},
  \bibinfo {author} {\bibfnamefont {A.~K.}\ \bibnamefont {{Harding}}}, \bibinfo
  {author} {\bibfnamefont {L.}~\bibnamefont {{Guillemot}}}, \bibinfo {author}
  {\bibfnamefont {D.~A.}\ \bibnamefont {{Smith}}}, \bibinfo {author}
  {\bibfnamefont {M.}~\bibnamefont {{Kramer}}}, \bibinfo {author}
  {\bibfnamefont {{\"O}.}~\bibnamefont {{{\c{C}}elik}}}, \bibinfo {author}
  {\bibfnamefont {P.~R.}\ \bibnamefont {{den Hartog}}}, \bibinfo {author}
  {\bibfnamefont {E.~C.}\ \bibnamefont {{Ferrara}}}, \bibinfo {author}
  {\bibfnamefont {X.}~\bibnamefont {{Hou}}}, \bibinfo {author} {\bibfnamefont
  {J.}~\bibnamefont {{Lande}}}, \ and\ \bibinfo {author} {\bibfnamefont
  {P.~S.}\ \bibnamefont {{Ray}}},\ }\href {\doibase 10.1088/0067-0049/213/1/6}
  {\bibfield  {journal} {\bibinfo  {journal} {\apjs}\ }\textbf {\bibinfo
  {volume} {213}},\ \bibinfo {eid} {6} (\bibinfo {year} {2014})},\ \Eprint
  {http://arxiv.org/abs/1404.2264} {arXiv:1404.2264 [astro-ph.HE]} \BibitemShut
  {NoStop}%
\bibitem [{\citenamefont {{Coroniti}}(1990)}]{Coroniti90}%
  \BibitemOpen
  \bibfield  {author} {\bibinfo {author} {\bibfnamefont {F.~V.}\ \bibnamefont
  {{Coroniti}}},\ }\href {\doibase 10.1086/168340} {\bibfield  {journal}
  {\bibinfo  {journal} {\apj}\ }\textbf {\bibinfo {volume} {349}},\ \bibinfo
  {pages} {538} (\bibinfo {year} {1990})}\BibitemShut {NoStop}%
\bibitem [{\citenamefont {Ajello}\ \emph {et~al.}(2016)\citenamefont {Ajello}
  \emph {et~al.}}]{TheFermi-LAT:2015kwa}%
  \BibitemOpen
  \bibfield  {author} {\bibinfo {author} {\bibfnamefont {M.}~\bibnamefont
  {Ajello}} \emph {et~al.} (\bibinfo {collaboration} {Fermi-LAT}),\ }\href
  {\doibase 10.3847/0004-637X/819/1/44} {\bibfield  {journal} {\bibinfo
  {journal} {Astrophys. J.}\ }\textbf {\bibinfo {volume} {819}},\ \bibinfo
  {pages} {44} (\bibinfo {year} {2016})},\ \Eprint
  {http://arxiv.org/abs/1511.02938} {arXiv:1511.02938 [astro-ph.HE]}
  \BibitemShut {NoStop}%
\bibitem [{\citenamefont {{Salim}}\ \emph {et~al.}(2018)\citenamefont
  {{Salim}}, \citenamefont {{Boquien}},\ and\ \citenamefont {{Lee}}}]{Salim18}%
  \BibitemOpen
  \bibfield  {author} {\bibinfo {author} {\bibfnamefont {S.}~\bibnamefont
  {{Salim}}}, \bibinfo {author} {\bibfnamefont {M.}~\bibnamefont {{Boquien}}},
  \ and\ \bibinfo {author} {\bibfnamefont {J.~C.}\ \bibnamefont {{Lee}}},\
  }\href {\doibase 10.3847/1538-4357/aabf3c} {\bibfield  {journal} {\bibinfo
  {journal} {\apj}\ }\textbf {\bibinfo {volume} {859}},\ \bibinfo {eid} {11}
  (\bibinfo {year} {2018})},\ \Eprint {http://arxiv.org/abs/1804.05850}
  {arXiv:1804.05850 [astro-ph.GA]} \BibitemShut {NoStop}%
\bibitem [{\citenamefont {{Salim}}\ \emph {et~al.}(2016)\citenamefont
  {{Salim}}, \citenamefont {{Lee}}, \citenamefont {{Janowiecki}}, \citenamefont
  {{da Cunha}}, \citenamefont {{Dickinson}}, \citenamefont {{Boquien}},
  \citenamefont {{Burgarella}}, \citenamefont {{Salzer}},\ and\ \citenamefont
  {{Charlot}}}]{Salim16}%
  \BibitemOpen
  \bibfield  {author} {\bibinfo {author} {\bibfnamefont {S.}~\bibnamefont
  {{Salim}}}, \bibinfo {author} {\bibfnamefont {J.~C.}\ \bibnamefont {{Lee}}},
  \bibinfo {author} {\bibfnamefont {S.}~\bibnamefont {{Janowiecki}}}, \bibinfo
  {author} {\bibfnamefont {E.}~\bibnamefont {{da Cunha}}}, \bibinfo {author}
  {\bibfnamefont {M.}~\bibnamefont {{Dickinson}}}, \bibinfo {author}
  {\bibfnamefont {M.}~\bibnamefont {{Boquien}}}, \bibinfo {author}
  {\bibfnamefont {D.}~\bibnamefont {{Burgarella}}}, \bibinfo {author}
  {\bibfnamefont {J.~J.}\ \bibnamefont {{Salzer}}}, \ and\ \bibinfo {author}
  {\bibfnamefont {S.}~\bibnamefont {{Charlot}}},\ }\href {\doibase
  10.3847/0067-0049/227/1/2} {\bibfield  {journal} {\bibinfo  {journal}
  {\apjs}\ }\textbf {\bibinfo {volume} {227}},\ \bibinfo {eid} {2} (\bibinfo
  {year} {2016})},\ \Eprint {http://arxiv.org/abs/1610.00712} {arXiv:1610.00712
  [astro-ph.GA]} \BibitemShut {NoStop}%
\bibitem [{\citenamefont {{Noll}}\ \emph {et~al.}(2009)\citenamefont {{Noll}},
  \citenamefont {{Burgarella}}, \citenamefont {{Giovannoli}}, \citenamefont
  {{Buat}}, \citenamefont {{Marcillac}},\ and\ \citenamefont
  {{Mu{\~n}oz-Mateos}}}]{cigale}%
  \BibitemOpen
  \bibfield  {author} {\bibinfo {author} {\bibfnamefont {S.}~\bibnamefont
  {{Noll}}}, \bibinfo {author} {\bibfnamefont {D.}~\bibnamefont
  {{Burgarella}}}, \bibinfo {author} {\bibfnamefont {E.}~\bibnamefont
  {{Giovannoli}}}, \bibinfo {author} {\bibfnamefont {V.}~\bibnamefont
  {{Buat}}}, \bibinfo {author} {\bibfnamefont {D.}~\bibnamefont {{Marcillac}}},
  \ and\ \bibinfo {author} {\bibfnamefont {J.~C.}\ \bibnamefont
  {{Mu{\~n}oz-Mateos}}},\ }\href {\doibase 10.1051/0004-6361/200912497}
  {\bibfield  {journal} {\bibinfo  {journal} {\aap}\ }\textbf {\bibinfo
  {volume} {507}},\ \bibinfo {pages} {1793} (\bibinfo {year} {2009})},\ \Eprint
  {http://arxiv.org/abs/0909.5439} {arXiv:0909.5439 [astro-ph.CO]} \BibitemShut
  {NoStop}%
\bibitem [{\citenamefont {{Cox}}\ \emph {et~al.}(1988)\citenamefont {{Cox}},
  \citenamefont {{Eales}}, \citenamefont {{Alexander}},\ and\ \citenamefont
  {{Fitt}}}]{1988MNRAS.235.1227C}%
  \BibitemOpen
  \bibfield  {author} {\bibinfo {author} {\bibfnamefont {M.~J.}\ \bibnamefont
  {{Cox}}}, \bibinfo {author} {\bibfnamefont {S.~A.~E.}\ \bibnamefont
  {{Eales}}}, \bibinfo {author} {\bibfnamefont {P.}~\bibnamefont
  {{Alexander}}}, \ and\ \bibinfo {author} {\bibfnamefont {A.~J.}\ \bibnamefont
  {{Fitt}}},\ }\href {\doibase 10.1093/mnras/235.4.1227} {\bibfield  {journal}
  {\bibinfo  {journal} {\mnras}\ }\textbf {\bibinfo {volume} {235}},\ \bibinfo
  {pages} {1227} (\bibinfo {year} {1988})}\BibitemShut {NoStop}%
\bibitem [{\citenamefont {{Salim}}\ \emph {et~al.}(2009)\citenamefont {{Salim}}
  \emph {et~al.}}]{Salim09}%
  \BibitemOpen
  \bibfield  {author} {\bibinfo {author} {\bibfnamefont {S.}~\bibnamefont
  {{Salim}}} \emph {et~al.},\ }\href {\doibase 10.1088/0004-637X/700/1/161}
  {\bibfield  {journal} {\bibinfo  {journal} {\apj}\ }\textbf {\bibinfo
  {volume} {700}},\ \bibinfo {pages} {161} (\bibinfo {year} {2009})},\ \Eprint
  {http://arxiv.org/abs/0905.0162} {arXiv:0905.0162 [astro-ph.CO]} \BibitemShut
  {NoStop}%
\bibitem [{\citenamefont {{Calzetti}}\ \emph {et~al.}(2010)\citenamefont
  {{Calzetti}} \emph {et~al.}}]{Calzetti10}%
  \BibitemOpen
  \bibfield  {author} {\bibinfo {author} {\bibfnamefont {D.}~\bibnamefont
  {{Calzetti}}} \emph {et~al.},\ }\href {\doibase 10.1088/0004-637X/714/2/1256}
  {\bibfield  {journal} {\bibinfo  {journal} {\apj}\ }\textbf {\bibinfo
  {volume} {714}},\ \bibinfo {pages} {1256} (\bibinfo {year} {2010})},\ \Eprint
  {http://arxiv.org/abs/1003.0961} {arXiv:1003.0961 [astro-ph.CO]} \BibitemShut
  {NoStop}%
\bibitem [{\citenamefont {{Magorrian}}\ \emph {et~al.}(1998)\citenamefont
  {{Magorrian}} \emph {et~al.}}]{1998AJ....115.2285M}%
  \BibitemOpen
  \bibfield  {author} {\bibinfo {author} {\bibfnamefont {J.}~\bibnamefont
  {{Magorrian}}} \emph {et~al.},\ }\href {\doibase 10.1086/300353} {\bibfield
  {journal} {\bibinfo  {journal} {\aj}\ }\textbf {\bibinfo {volume} {115}},\
  \bibinfo {pages} {2285} (\bibinfo {year} {1998})},\ \Eprint
  {http://arxiv.org/abs/astro-ph/9708072} {arXiv:astro-ph/9708072 [astro-ph]}
  \BibitemShut {NoStop}%
\bibitem [{\citenamefont {{Voss}}\ and\ \citenamefont
  {{Gilfanov}}(2007)}]{2007A&A...468...49V}%
  \BibitemOpen
  \bibfield  {author} {\bibinfo {author} {\bibfnamefont {R.}~\bibnamefont
  {{Voss}}}\ and\ \bibinfo {author} {\bibfnamefont {M.}~\bibnamefont
  {{Gilfanov}}},\ }\href {\doibase 10.1051/0004-6361:20066614} {\bibfield
  {journal} {\bibinfo  {journal} {\aap}\ }\textbf {\bibinfo {volume} {468}},\
  \bibinfo {pages} {49} (\bibinfo {year} {2007})},\ \Eprint
  {http://arxiv.org/abs/astro-ph/0610649} {arXiv:astro-ph/0610649 [astro-ph]}
  \BibitemShut {NoStop}%
\bibitem [{\citenamefont {{McDaniel}}\ \emph {et~al.}(2019)\citenamefont
  {{McDaniel}}, \citenamefont {{Jeltema}},\ and\ \citenamefont
  {{Profumo}}}]{McDaniel19}%
  \BibitemOpen
  \bibfield  {author} {\bibinfo {author} {\bibfnamefont {A.}~\bibnamefont
  {{McDaniel}}}, \bibinfo {author} {\bibfnamefont {T.}~\bibnamefont
  {{Jeltema}}}, \ and\ \bibinfo {author} {\bibfnamefont {S.}~\bibnamefont
  {{Profumo}}},\ }\href {\doibase 10.1103/PhysRevD.100.023014} {\bibfield
  {journal} {\bibinfo  {journal} {\prd}\ }\textbf {\bibinfo {volume} {100}},\
  \bibinfo {eid} {023014} (\bibinfo {year} {2019})},\ \Eprint
  {http://arxiv.org/abs/1903.06833} {arXiv:1903.06833 [astro-ph.HE]}
  \BibitemShut {NoStop}%
\bibitem [{\citenamefont {Gordon}\ \emph {et~al.}(2004)\citenamefont {Gordon}
  \emph {et~al.}}]{Gordon:2004wn}%
  \BibitemOpen
  \bibfield  {author} {\bibinfo {author} {\bibfnamefont {K.~D.}\ \bibnamefont
  {Gordon}} \emph {et~al.},\ }\href {\doibase 10.1086/422714} {\bibfield
  {journal} {\bibinfo  {journal} {Astrophys. J. Suppl.}\ }\textbf {\bibinfo
  {volume} {154}},\ \bibinfo {pages} {215} (\bibinfo {year} {2004})},\ \Eprint
  {http://arxiv.org/abs/astro-ph/0406064} {arXiv:astro-ph/0406064} \BibitemShut
  {NoStop}%
\bibitem [{\citenamefont {{Grindlay}}(1984)}]{1984AdSpR...3...19G}%
  \BibitemOpen
  \bibfield  {author} {\bibinfo {author} {\bibfnamefont {J.~E.}\ \bibnamefont
  {{Grindlay}}},\ }\href {\doibase 10.1016/0273-1177(84)90054-1} {\bibfield
  {journal} {\bibinfo  {journal} {Advances in Space Research}\ }\textbf
  {\bibinfo {volume} {3}},\ \bibinfo {pages} {19} (\bibinfo {year}
  {1984})}\BibitemShut {NoStop}%
\bibitem [{\citenamefont {{Song}}\ \emph {et~al.}(2021)\citenamefont {{Song}},
  \citenamefont {{Macias}}, \citenamefont {{Horiuchi}}, \citenamefont
  {{Crocker}},\ and\ \citenamefont {{Nataf}}}]{2021arXiv210200061S}%
  \BibitemOpen
  \bibfield  {author} {\bibinfo {author} {\bibfnamefont {D.}~\bibnamefont
  {{Song}}}, \bibinfo {author} {\bibfnamefont {O.}~\bibnamefont {{Macias}}},
  \bibinfo {author} {\bibfnamefont {S.}~\bibnamefont {{Horiuchi}}}, \bibinfo
  {author} {\bibfnamefont {R.~M.}\ \bibnamefont {{Crocker}}}, \ and\ \bibinfo
  {author} {\bibfnamefont {D.~M.}\ \bibnamefont {{Nataf}}},\ }\href@noop {}
  {\bibfield  {journal} {\bibinfo  {journal} {arXiv e-prints}\ ,\ \bibinfo
  {eid} {arXiv:2102.00061}} (\bibinfo {year} {2021})},\ \Eprint
  {http://arxiv.org/abs/2102.00061} {arXiv:2102.00061 [astro-ph.HE]}
  \BibitemShut {NoStop}%
\bibitem [{\citenamefont {{Macias}}\ \emph {et~al.}(2021)\citenamefont
  {{Macias}}, \citenamefont {{van Leijen}}, \citenamefont {{Song}},
  \citenamefont {{Ando}}, \citenamefont {{Horiuchi}},\ and\ \citenamefont
  {{Crocker}}}]{2021arXiv210205648M}%
  \BibitemOpen
  \bibfield  {author} {\bibinfo {author} {\bibfnamefont {O.}~\bibnamefont
  {{Macias}}}, \bibinfo {author} {\bibfnamefont {H.}~\bibnamefont {{van
  Leijen}}}, \bibinfo {author} {\bibfnamefont {D.}~\bibnamefont {{Song}}},
  \bibinfo {author} {\bibfnamefont {S.}~\bibnamefont {{Ando}}}, \bibinfo
  {author} {\bibfnamefont {S.}~\bibnamefont {{Horiuchi}}}, \ and\ \bibinfo
  {author} {\bibfnamefont {R.~M.}\ \bibnamefont {{Crocker}}},\ }\href@noop {}
  {\bibfield  {journal} {\bibinfo  {journal} {arXiv e-prints}\ ,\ \bibinfo
  {eid} {arXiv:2102.05648}} (\bibinfo {year} {2021})},\ \Eprint
  {http://arxiv.org/abs/2102.05648} {arXiv:2102.05648 [astro-ph.HE]}
  \BibitemShut {NoStop}%
\bibitem [{\citenamefont {{Goodenough}}\ and\ \citenamefont
  {{Hooper}}(2009)}]{Goodenough:2009gk}%
  \BibitemOpen
  \bibfield  {author} {\bibinfo {author} {\bibfnamefont {L.}~\bibnamefont
  {{Goodenough}}}\ and\ \bibinfo {author} {\bibfnamefont {D.}~\bibnamefont
  {{Hooper}}},\ }\href@noop {} {\bibfield  {journal} {\bibinfo  {journal}
  {arXiv e-prints}\ ,\ \bibinfo {eid} {arXiv:0910.2998}} (\bibinfo {year}
  {2009})},\ \Eprint {http://arxiv.org/abs/0910.2998} {arXiv:0910.2998
  [hep-ph]} \BibitemShut {NoStop}%
\bibitem [{\citenamefont {Daylan}\ \emph {et~al.}(2016)\citenamefont {Daylan},
  \citenamefont {Finkbeiner}, \citenamefont {Hooper}, \citenamefont {Linden},
  \citenamefont {Portillo}, \citenamefont {Rodd},\ and\ \citenamefont
  {Slatyer}}]{Daylan:2014rsa}%
  \BibitemOpen
  \bibfield  {author} {\bibinfo {author} {\bibfnamefont {T.}~\bibnamefont
  {Daylan}}, \bibinfo {author} {\bibfnamefont {D.~P.}\ \bibnamefont
  {Finkbeiner}}, \bibinfo {author} {\bibfnamefont {D.}~\bibnamefont {Hooper}},
  \bibinfo {author} {\bibfnamefont {T.}~\bibnamefont {Linden}}, \bibinfo
  {author} {\bibfnamefont {S.~K.~N.}\ \bibnamefont {Portillo}}, \bibinfo
  {author} {\bibfnamefont {N.~L.}\ \bibnamefont {Rodd}}, \ and\ \bibinfo
  {author} {\bibfnamefont {T.~R.}\ \bibnamefont {Slatyer}},\ }\href {\doibase
  10.1016/j.dark.2015.12.005} {\bibfield  {journal} {\bibinfo  {journal} {Phys.
  Dark Univ.}\ }\textbf {\bibinfo {volume} {12}},\ \bibinfo {pages} {1}
  (\bibinfo {year} {2016})},\ \Eprint {http://arxiv.org/abs/1402.6703}
  {arXiv:1402.6703 [astro-ph.HE]} \BibitemShut {NoStop}%
\bibitem [{\citenamefont {Abazajian}(2011)}]{Abazajian:2010zy}%
  \BibitemOpen
  \bibfield  {author} {\bibinfo {author} {\bibfnamefont {K.~N.}\ \bibnamefont
  {Abazajian}},\ }\href {\doibase 10.1088/1475-7516/2011/03/010} {\bibfield
  {journal} {\bibinfo  {journal} {JCAP}\ }\textbf {\bibinfo {volume} {03}},\
  \bibinfo {pages} {010} (\bibinfo {year} {2011})},\ \Eprint
  {http://arxiv.org/abs/1011.4275} {arXiv:1011.4275 [astro-ph.HE]} \BibitemShut
  {NoStop}%
\bibitem [{\citenamefont {Bartels}\ \emph {et~al.}(2016)\citenamefont
  {Bartels}, \citenamefont {Krishnamurthy},\ and\ \citenamefont
  {Weniger}}]{Bartels:2015aea}%
  \BibitemOpen
  \bibfield  {author} {\bibinfo {author} {\bibfnamefont {R.}~\bibnamefont
  {Bartels}}, \bibinfo {author} {\bibfnamefont {S.}~\bibnamefont
  {Krishnamurthy}}, \ and\ \bibinfo {author} {\bibfnamefont {C.}~\bibnamefont
  {Weniger}},\ }\href {\doibase 10.1103/PhysRevLett.116.051102} {\bibfield
  {journal} {\bibinfo  {journal} {Phys. Rev. Lett.}\ }\textbf {\bibinfo
  {volume} {116}},\ \bibinfo {pages} {051102} (\bibinfo {year} {2016})},\
  \Eprint {http://arxiv.org/abs/1506.05104} {arXiv:1506.05104 [astro-ph.HE]}
  \BibitemShut {NoStop}%
\bibitem [{\citenamefont {Lee}\ \emph {et~al.}(2016)\citenamefont {Lee},
  \citenamefont {Lisanti}, \citenamefont {Safdi}, \citenamefont {Slatyer},\
  and\ \citenamefont {Xue}}]{Lee:2015fea}%
  \BibitemOpen
  \bibfield  {author} {\bibinfo {author} {\bibfnamefont {S.~K.}\ \bibnamefont
  {Lee}}, \bibinfo {author} {\bibfnamefont {M.}~\bibnamefont {Lisanti}},
  \bibinfo {author} {\bibfnamefont {B.~R.}\ \bibnamefont {Safdi}}, \bibinfo
  {author} {\bibfnamefont {T.~R.}\ \bibnamefont {Slatyer}}, \ and\ \bibinfo
  {author} {\bibfnamefont {W.}~\bibnamefont {Xue}},\ }\href {\doibase
  10.1103/PhysRevLett.116.051103} {\bibfield  {journal} {\bibinfo  {journal}
  {Phys. Rev. Lett.}\ }\textbf {\bibinfo {volume} {116}},\ \bibinfo {pages}
  {051103} (\bibinfo {year} {2016})},\ \Eprint
  {http://arxiv.org/abs/1506.05124} {arXiv:1506.05124 [astro-ph.HE]}
  \BibitemShut {NoStop}%
\bibitem [{\citenamefont {Carlson}\ \emph {et~al.}(2016)\citenamefont
  {Carlson}, \citenamefont {Linden},\ and\ \citenamefont
  {Profumo}}]{Carlson:2016iis}%
  \BibitemOpen
  \bibfield  {author} {\bibinfo {author} {\bibfnamefont {E.}~\bibnamefont
  {Carlson}}, \bibinfo {author} {\bibfnamefont {T.}~\bibnamefont {Linden}}, \
  and\ \bibinfo {author} {\bibfnamefont {S.}~\bibnamefont {Profumo}},\ }\href
  {\doibase 10.1103/PhysRevD.94.063504} {\bibfield  {journal} {\bibinfo
  {journal} {Phys. Rev. D}\ }\textbf {\bibinfo {volume} {94}},\ \bibinfo
  {pages} {063504} (\bibinfo {year} {2016})},\ \Eprint
  {http://arxiv.org/abs/1603.06584} {arXiv:1603.06584 [astro-ph.HE]}
  \BibitemShut {NoStop}%
\bibitem [{\citenamefont {{Abeysekara}}\ \emph {et~al.}(2017)\citenamefont
  {{Abeysekara}} \emph {et~al.}}]{TeVhalo:HAWC17}%
  \BibitemOpen
  \bibfield  {author} {\bibinfo {author} {\bibfnamefont {A.~U.}\ \bibnamefont
  {{Abeysekara}}} \emph {et~al.},\ }\href {\doibase 10.1126/science.aan4880}
  {\bibfield  {journal} {\bibinfo  {journal} {Science}\ }\textbf {\bibinfo
  {volume} {358}},\ \bibinfo {pages} {911} (\bibinfo {year} {2017})},\ \Eprint
  {http://arxiv.org/abs/1711.06223} {arXiv:1711.06223 [astro-ph.HE]}
  \BibitemShut {NoStop}%
\bibitem [{\citenamefont {{Di Mauro}}\ \emph {et~al.}(2019)\citenamefont {{Di
  Mauro}}, \citenamefont {{Manconi}},\ and\ \citenamefont
  {{Donato}}}]{DiMauro19}%
  \BibitemOpen
  \bibfield  {author} {\bibinfo {author} {\bibfnamefont {M.}~\bibnamefont {{Di
  Mauro}}}, \bibinfo {author} {\bibfnamefont {S.}~\bibnamefont {{Manconi}}}, \
  and\ \bibinfo {author} {\bibfnamefont {F.}~\bibnamefont {{Donato}}},\ }\href
  {\doibase 10.1103/PhysRevD.100.123015} {\bibfield  {journal} {\bibinfo
  {journal} {\prd}\ }\textbf {\bibinfo {volume} {100}},\ \bibinfo {eid}
  {123015} (\bibinfo {year} {2019})},\ \Eprint
  {http://arxiv.org/abs/1903.05647} {arXiv:1903.05647 [astro-ph.HE]}
  \BibitemShut {NoStop}%
\bibitem [{\citenamefont {{Adriani}}\ \emph {et~al.}(2010)\citenamefont
  {{Adriani}}, \citenamefont {{Barbarino}}, \citenamefont {{Bazilevskaya}},
  \citenamefont {{Bellotti}}, \citenamefont {{Boezio}}, \citenamefont
  {{Bogomolov}}, \citenamefont {{Bonechi}},\ and\ \citenamefont {{PAMELA
  Collaboration}}}]{2010PhRvL.105l1101A}%
  \BibitemOpen
  \bibfield  {author} {\bibinfo {author} {\bibfnamefont {O.}~\bibnamefont
  {{Adriani}}}, \bibinfo {author} {\bibfnamefont {G.~C.}\ \bibnamefont
  {{Barbarino}}}, \bibinfo {author} {\bibfnamefont {G.~A.}\ \bibnamefont
  {{Bazilevskaya}}}, \bibinfo {author} {\bibfnamefont {R.}~\bibnamefont
  {{Bellotti}}}, \bibinfo {author} {\bibfnamefont {M.}~\bibnamefont
  {{Boezio}}}, \bibinfo {author} {\bibfnamefont {E.~A.}\ \bibnamefont
  {{Bogomolov}}}, \bibinfo {author} {\bibfnamefont {L.}~\bibnamefont
  {{Bonechi}}}, \ and\ \bibinfo {author} {\bibnamefont {{PAMELA
  Collaboration}}},\ }\href {\doibase 10.1103/PhysRevLett.105.121101}
  {\bibfield  {journal} {\bibinfo  {journal} {\prl}\ }\textbf {\bibinfo
  {volume} {105}},\ \bibinfo {eid} {121101} (\bibinfo {year} {2010})},\ \Eprint
  {http://arxiv.org/abs/1007.0821} {arXiv:1007.0821 [astro-ph.HE]} \BibitemShut
  {NoStop}%
\bibitem [{\citenamefont {Aguilar}\ \emph {et~al.}(2013)\citenamefont {Aguilar}
  \emph {et~al.}}]{PhysRevLett.110.141102}%
  \BibitemOpen
  \bibfield  {author} {\bibinfo {author} {\bibfnamefont {M.}~\bibnamefont
  {Aguilar}} \emph {et~al.} (\bibinfo {collaboration} {AMS Collaboration}),\
  }\href {\doibase 10.1103/PhysRevLett.110.141102} {\bibfield  {journal}
  {\bibinfo  {journal} {Phys. Rev. Lett.}\ }\textbf {\bibinfo {volume} {110}},\
  \bibinfo {pages} {141102} (\bibinfo {year} {2013})}\BibitemShut {NoStop}%
\bibitem [{\citenamefont {{Venter}}\ \emph {et~al.}(2015)\citenamefont
  {{Venter}}, \citenamefont {{Kopp}}, \citenamefont {{Harding}}, \citenamefont
  {{Gonthier}},\ and\ \citenamefont {{Buesching}}}]{2015ICRC...34..462V}%
  \BibitemOpen
  \bibfield  {author} {\bibinfo {author} {\bibfnamefont {C.}~\bibnamefont
  {{Venter}}}, \bibinfo {author} {\bibfnamefont {A.}~\bibnamefont {{Kopp}}},
  \bibinfo {author} {\bibfnamefont {A.~K.}\ \bibnamefont {{Harding}}}, \bibinfo
  {author} {\bibfnamefont {P.~L.}\ \bibnamefont {{Gonthier}}}, \ and\ \bibinfo
  {author} {\bibfnamefont {I.}~\bibnamefont {{Buesching}}},\ }in\ \href@noop {}
  {\emph {\bibinfo {booktitle} {34th International Cosmic Ray Conference
  (ICRC2015)}}},\ \bibinfo {series} {International Cosmic Ray Conference},
  Vol.~\bibinfo {volume} {34}\ (\bibinfo {year} {2015})\ p.\ \bibinfo {pages}
  {462},\ \Eprint {http://arxiv.org/abs/1508.04676} {arXiv:1508.04676
  [astro-ph.HE]} \BibitemShut {NoStop}%
\bibitem [{\citenamefont {{Astropy Collaboration}}(2013)}]{astropy0}%
  \BibitemOpen
  \bibfield  {author} {\bibinfo {author} {\bibnamefont {{Astropy
  Collaboration}}},\ }\href {\doibase 10.1051/0004-6361/201322068} {\bibfield
  {journal} {\bibinfo  {journal} {\aap}\ }\textbf {\bibinfo {volume} {558}},\
  \bibinfo {eid} {A33} (\bibinfo {year} {2013})},\ \Eprint
  {http://arxiv.org/abs/1307.6212} {arXiv:1307.6212 [astro-ph.IM]} \BibitemShut
  {NoStop}%
\bibitem [{\citenamefont {{Astropy Collaboration}}(2018)}]{astropy}%
  \BibitemOpen
  \bibfield  {author} {\bibinfo {author} {\bibnamefont {{Astropy
  Collaboration}}},\ }\href {\doibase 10.3847/1538-3881/aabc4f} {\bibfield
  {journal} {\bibinfo  {journal} {\aj}\ }\textbf {\bibinfo {volume} {156}},\
  \bibinfo {eid} {123} (\bibinfo {year} {2018})},\ \Eprint
  {http://arxiv.org/abs/1801.02634} {arXiv:1801.02634 [astro-ph.IM]}
  \BibitemShut {NoStop}%
\bibitem [{\citenamefont {Hunter}(2007)}]{matplotlib}%
  \BibitemOpen
  \bibfield  {author} {\bibinfo {author} {\bibfnamefont {J.~D.}\ \bibnamefont
  {Hunter}},\ }\href {\doibase 10.1109/MCSE.2007.55} {\bibfield  {journal}
  {\bibinfo  {journal} {Computing in Science \& Engineering}\ }\textbf
  {\bibinfo {volume} {9}},\ \bibinfo {pages} {90} (\bibinfo {year}
  {2007})}\BibitemShut {NoStop}%
\bibitem [{\citenamefont {{van der Walt}}\ \emph {et~al.}(2011)\citenamefont
  {{van der Walt}}, \citenamefont {{Colbert}},\ and\ \citenamefont
  {{Varoquaux}}}]{numpy}%
  \BibitemOpen
  \bibfield  {author} {\bibinfo {author} {\bibfnamefont {S.}~\bibnamefont {{van
  der Walt}}}, \bibinfo {author} {\bibfnamefont {S.~C.}\ \bibnamefont
  {{Colbert}}}, \ and\ \bibinfo {author} {\bibfnamefont {G.}~\bibnamefont
  {{Varoquaux}}},\ }\href {\doibase 10.1109/MCSE.2011.37} {\bibfield  {journal}
  {\bibinfo  {journal} {Computing in Science Engineering}\ }\textbf {\bibinfo
  {volume} {13}},\ \bibinfo {pages} {22} (\bibinfo {year} {2011})}\BibitemShut
  {NoStop}%
\end{thebibliography}%

\end{document}